\documentclass[journal]{llncs}
\pdfoutput=1
\usepackage[ruled,noend,linesnumbered]{algorithm2e}
\usepackage{graphicx} 
\usepackage{comment}
\usepackage{array}
\usepackage{color}
\usepackage{cite}
\usepackage{paralist}

\usepackage{psfrag}
\usepackage{url}

\usepackage[utf8]{inputenc}

\usepackage{subfig} 
\usepackage{multirow} %

\usepackage{amsmath}
\usepackage{amssymb}

\usepackage{latexsym}

\usepackage{boxedminipage}

\usepackage{listings}

\usepackage{times}

\usepackage{xspace}

\usepackage{ifpdf}

\newcommand{\figspa}{\vspace{-15pt}}
\newcommand{\figspb}{\vspace{-9pt}}
\newcommand{\figspc}{\vspace{-5pt}}

\ifpdf
\DeclareGraphicsExtensions{.pdf, .jpg, .tif}
\else
\DeclareGraphicsExtensions{.eps, .jpg}
\fi

\makeatletter
\def\url@leostyle{%
  \@ifundefined{selectfont}{\def\UrlFont{\sf}}{\def\UrlFont{\small\ttfamily}}}
\makeatother
\title{A Secure Location-based Alert System with Tunable Privacy-Performance Trade-off}

\author{
Gabriel Ghinita\inst{1} \and
Kien Nguyen \inst{2} \and
Mihai Maruseac \inst{1} \and
Cyrus Shahabi \inst{2}
}

\institute{University of Massachusetts, Boston, USA\\
	\email{gghinita@cs.umb.edu, mmarusea@cs.umb.edu} \and
	University of Southern California, Los Angeles, USA\\
	\email{kien.nguyen@usc.edu,shahabi@usc.edu}}

\newfont{\mycrnotice}{ptmr8t at 7pt}
\newfont{\myconfname}{ptmri8t at 7pt}

\begin{document}

\maketitle

\begin{abstract}
Monitoring location updates from mobile users has important applications in many areas, ranging from public safety and national security to social networks and advertising. However, sensitive information can be derived from movement patterns, thus protecting the privacy of mobile users is a major concern. Users may only be willing to disclose their locations when some condition is met, for instance in proximity of a disaster area or an event of interest. Currently, such functionality can be achieved using {\em searchable encryption}. Such cryptographic primitives provide provable guarantees for privacy, and allow decryption only when the location satisfies some predicate. Nevertheless, they rely on expensive {\em pairing-based cryptography (PBC)}, of which direct application to the domain of location updates leads to impractical solutions.

We propose secure and efficient techniques for private processing of location updates that complement the use of PBC and lead to significant gains in performance by reducing the amount of required pairing operations. We implement two optimizations that further improve performance: materialization of results to expensive mathematical operations, and parallelization. 
We also propose an heuristic that brings down the computational overhead through enlarging an alert zone by a small factor (given as system parameter), therefore trading off a small and controlled amount of privacy for significant performance gains. 
Extensive experimental results show that the proposed techniques significantly improve performance compared to the baseline, and reduce the searchable encryption overhead to a level that is practical in a computing environment with reasonable resources, such as the cloud.

\end{abstract}

\setcounter{footnote}{0}

\sloppypar


\figspb
\section{Introduction}\label{sec:intro}

Modern mobile devices with positioning capabilities (e.g., GPS) allow users to be informed about events that occur in their proximity. Many classes of applications benefit from the large-scale availability of location data, ranging from public safety and national security to social networks and advertising. One particular scenario of interest is that of {\em location-based alert systems}, where mobile users wish to be immediately notified when their current location satisfies some conditions, expressed as a spatial {\em search predicate}. For instance, in a public safety scenario, users want to be notified when they are getting close to a dangerous accident area. Alternatively, in the commercial domain, a user may want to be alerted when a nearby sale event is underway. 

The typical architecture of such a system uses a server that collects location updates from the users and checks whether the alert condition is met. Such a service is often provided by a commercial entity that is not fully trusted. The collection of user trajectories at a commercial site introduces serious privacy concerns, as sensitive personal information may be derived from a person's whereabouts \cite{gruteser03, GKKST08}. Therefore, protecting the privacy of users is a necessary feature of such a system, and the users must not report their exact locations to the server. Ideally, the only information that the server should be able to derive from the user updates is whether the conditions that the users subscribe to are satisfied or not. Syntactic privacy models~\cite{gruteser03, gedik08, mca06, gks07} that perform generalization of locations before sharing have been proven vulnerable, especially in the presence of background knowledge \cite{GKKST08}. Furthermore, semantic privacy models such as differential privacy \cite{dmns06, D10, cpssy12} are only suitable for releasing statistics, but not for processing privately individual updates. 

Recently, several advanced encryption functions that allow evaluation of predicates on ciphertexts have been proposed \cite{Boneh07, Liu12, Blundo09}. These functions are broadly referred to as {\em searchable encryption (SE)} functions, since they allow the evaluation of certain types of queries without requiring decryption. Some of these encryption systems are asymmetric, i.e., they employ a secret key {\em SK} and a public key {\em PK} pair. 

\begin{figure}[b]
\figspa
\centering
\includegraphics[scale=0.35]{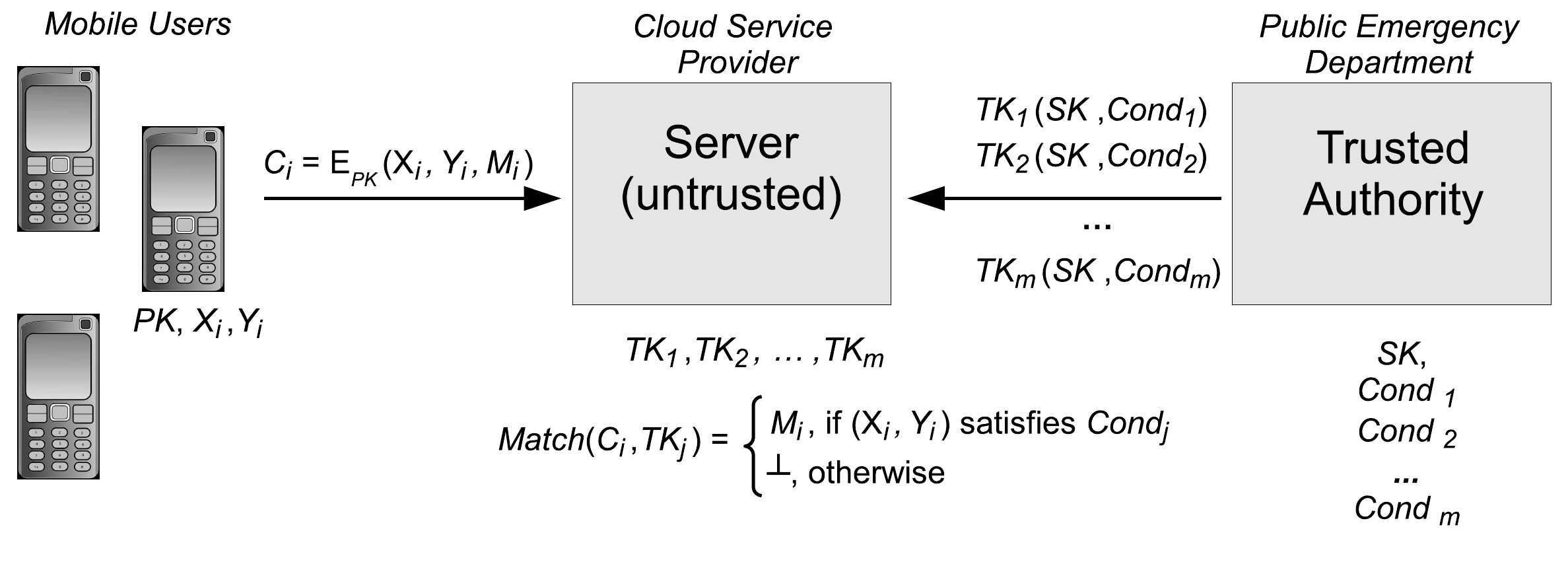}
\caption{Location-based Alert System}\label{fig:Intro}
\end{figure}

Figure~\ref{fig:Intro} shows the envisioned system architecture, with three types of entities: {\em (i)} users who send encrypted location updates using the {\em PK} of a trusted authority; {\em (ii)} a {\em trusted authority (TA)} who generates {\em tokens} for spatial search predicates using secret key $SK$; and {\em (iii)} {\em the server (S)} that collects location updates from users and evaluates the predicates on the ciphertexts using the tokens. In practice, the TA may represent the public emergency department of a city, which is responsible for the safety of the citizens. The TA is trusted, but it does not have the necessary infrastructure to support a large-scale alert system, hence it outsources this service to $S$. 

However, $S$ is a commercial entity that cannot be trusted with user locations, so the TA sets up a {\em SK/PK} pair, and distributes {\em PK} to the users. When an emergency occurs in a region, the TA creates a search token which is sent to $S$ to be matched against the ciphertexts received from users. The properties of SE guarantee that $S$ is able to evaluate the predicate on the ciphertext (e.g., whether the user location is enclosed in the region encoded by the search token) and learns only if the ciphertext matches or not, but no other information about user location.

To understand search on encrypted data, it helps to consider each ciphertext as being composed of two parts: an encrypted index $I$ and encrypted message $M$. $M$ is the payload of the ciphertext, just the same as in the case of conventional encryption. The novel part about searchable encryption is the presence of the index $I$, which is used for search, and can be seen as a parameter of the encryption function $E_{PK}(I,M)$. When a user $u_i$ constructs its update, she uses her current coordinates $(X_i, Y_i)$ as index, and performs encryption as $E_{PK}((X_i, Y_i),M_i)$. If the index satisfies the predicate specified by a token, then the server is able to recover the message $M_i$ from the user. However, this does not imply that $S$ can find the exact user location, as $M_i$ may contain information that is of other nature (e.g., an emergency contact number).

One prominent approach to searchable encryption called {\em Hidden Vector Encryption (HVE)} was proposed in \cite{Boneh07}. HVE can evaluate exact match, range and subset queries on ciphertexts. HVE uses bilinear maps on groups of composite order \cite{crypto04} as mathematical foundation and makes extensive use of expensive operations such as {\em bilinear map pairings}.
As a result, HVE is very expensive and scales poorly. Later in Section~\ref{sec:exps}, we show that in order to process the update from a single user only, it may take up to $100$ seconds. Clearly, direct application of HVE for alert systems is not suitable.

In this paper, we propose secure and efficient techniques to support private location-based alert systems using searchable encryption. To the best of our knowledge, this is the first study of applying asymmetric searchable encryption to the domain of private search with spatial predicates. Our specific contributions are%
\footnote{This submission is an extended version of~\cite{codaspy14}. Additional contributions consist of the technique for relaxation of alert zones presented in Section~\ref{sec:queryexpansion}, and its evaluation in Section~\ref{sec:exp:expansion}.}:
\renewcommand{\theenumi}{\roman{enumi}}
\begin{enumerate}
\figspc
\item We devise specific constructions that allow application of HVE to the problem of location-based alert systems with a reduced number of {\em bilinear pairing} operations, thus lowering the computational overhead of HVE. 
\item We develop optimizations based on reuse of expensive mathematical operation results and parallelization, which further reduce the HVE performance overhead.
\item We introduce a novel heuristic algorithm that provides effective means to tune the privacy-performance trade-off of the system, by allowing enlargement of alert zones by a small factor. By carefully enlarging the alert zone, one can obtain search tokens that require significantly smaller computation time to process.
\item We perform an extensive experimental evaluation which shows that the proposed approach brings the overhead of searchable encryption to acceptable levels in a computing environment such as the cloud.
\figspc
\end{enumerate}

Section~\ref{sec:background} overviews the proposed system and HVE. Section~\ref{sec:contribution} presents the encoding techniques for efficient application of HVE, whereas Section~\ref{sec:optimizations} outlines the optimizations to reduce execution time. 
In Section~\ref{sec:queryexpansion}, we introduce the heuristic for privacy-performance trade-off tuning through alert zone enlargement.
Section~\ref{sec:exps} contains the experimental evaluation results, followed by a survey of related research in Section~\ref{sec:related}. Finally, Section~\ref{sec:conclusion} concludes the paper and highlights directions for future work.

\figspb
\section{Background}
\label{sec:background}

\figspc
\subsection{System and Privacy Model}
\label{sec:model}

\begin{figure}[t]
\centering
\includegraphics[scale=0.35]{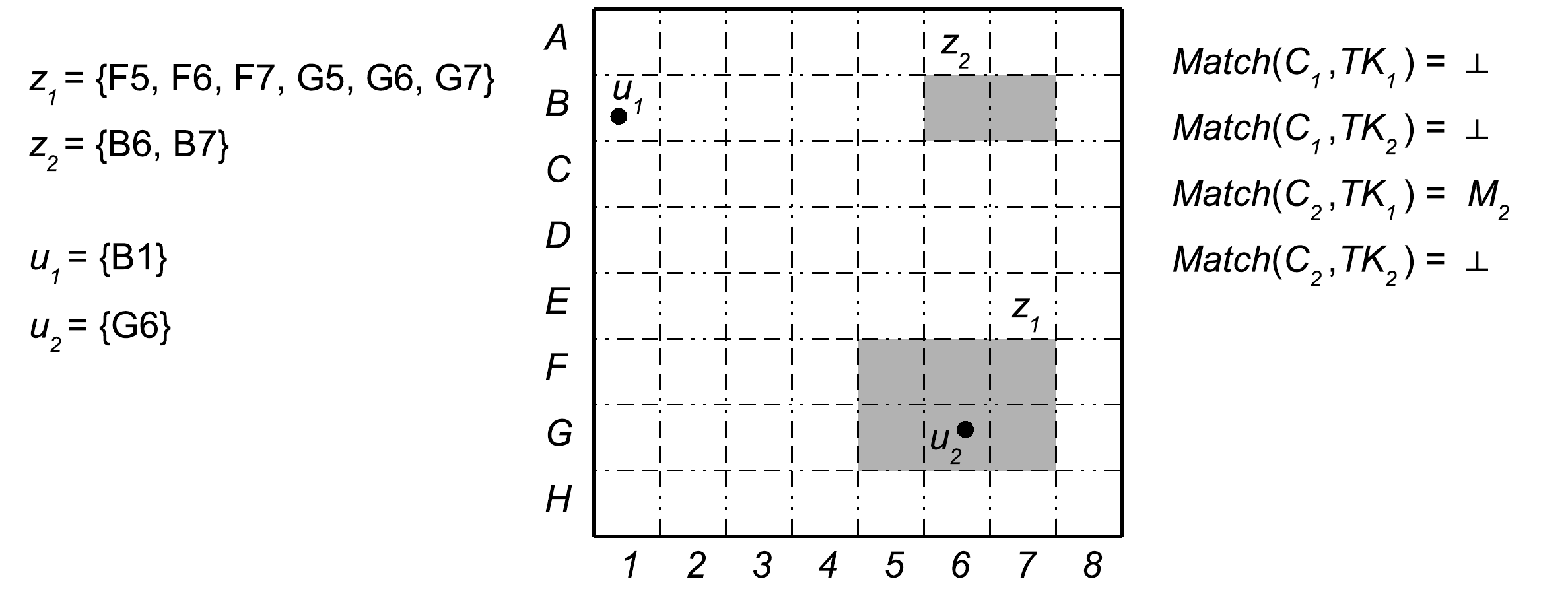}\figspb
\caption{System Model ($n=2$, $m=2$, $d=8$)}\label{fig:sys_model}
\figspb
\figspc
\end{figure}

Figure~\ref{fig:sys_model} illustrates the location-based alert system model, where $n$ users $\{u_1, \ldots, u_n\}$ move within a two-dimensional domain. Users continuously report their coordinates and wish to be notified when their location falls within any of $m$ {\em alert zones} $\{z_1, \ldots, z_m\}$. Alert zones (or simply {\em zones}) are defined by a trusted authority, as detailed later in this section. For simplicity, we assume that the space is partitioned by a regular grid of size $d\times d$, and each alert zone covers a number of grid cells. To facilitate presentation, we assume a square data domain, but our techniques can be immediately extended to a rectangular one, by adjusting the grid cell shape. The functional requirement of the system follows the spatial range query semantics, i.e., a user $u$ must receive an alert corresponding to zone $z$ if its location is {\em enclosed} by zone $z$.

The system (represented in Figure~\ref{fig:Intro}) consists of three types of entities:
\begin{enumerate}
\item 
{\bf Mobile Users} subscribe to the alert system and periodically submit encrypted location updates.
\item
The {\bf Trusted Authority (TA)} is a trusted entity that decides which are the alert zones, and creates for each zone a search {\em token} that allows to check privately if a user location falls within the alert zone or not.
\item
The {\bf Server (S)} is the provider of the alert system. It receives encrypted updates from users and search tokens from TA, and performs the predicate evaluation $Match$ to decide whether encrypted location $C_i (1 \le i \le n)$ falls within alert zone $j$ represented by token $TK_j (1 \le j \le m)$. If the predicate holds, $Match$ returns message $M_i$ encrypted by the user, otherwise it returns a void message ($\bot$).
\end{enumerate}

The {\em privacy requirement} dictates that the server must not learn any information about the user locations, other than what can be derived from the match outcome, i.e., whether the user is in a particular alert zone or not. In case of a successful match, the server $S$ learns that user $u$ is enclosed by zone $z$. In case of a non-match, the server $S$ learns only that the user is outside the zone $z$, but no additional location information. Note that, this model is applicable to many real-life scenarios, such as our motivating example in Section~\ref{sec:intro}. For instance, users wish to keep their location private most of the time, but they want to be immediately notified if they enter a zone where their personal safety may be threatened. Furthermore, the extent of alert zones is typically small compared to the entire data domain, so the fact that $S$ learns that $u$ is {\em not} within the set of alert zones does not disclose significant information about $u$'s location.

In practice, the TA role is played by an organization such as a city's public emergency department. Such an actor is trusted not to disclose $SK$ and compromise user privacy, but at the same time does not have the technological infrastructure to monitor a large user population. Hence, the alert service is outsourced to a commercial entity, e.g., a cloud provider that plays the role of the server. The TA will issue alert zones to signal that certain areas of the city are affected by an emergency. %

A private location-based alert system is also useful in social networks. A social network user $u$ can create a $SK/PK$ pair and distribute $PK$ to its buddies. Next, $u$ creates a token that represents his/her current location, e.g., a downtown restaurant. The network provider (e.g., Facebook), plays the role of the server: it privately monitors users, and sends the identifiers of buddies in the downtown area back to $u$. No information is gained by the server about locations of non-matching users.

\subsection{Searchable Encryption with HVE}
\label{sec:hve}

\begin{figure}
\figspa
\centering
        \subfloat[Match]{
                \centering
                \includegraphics[width=0.25\textwidth]{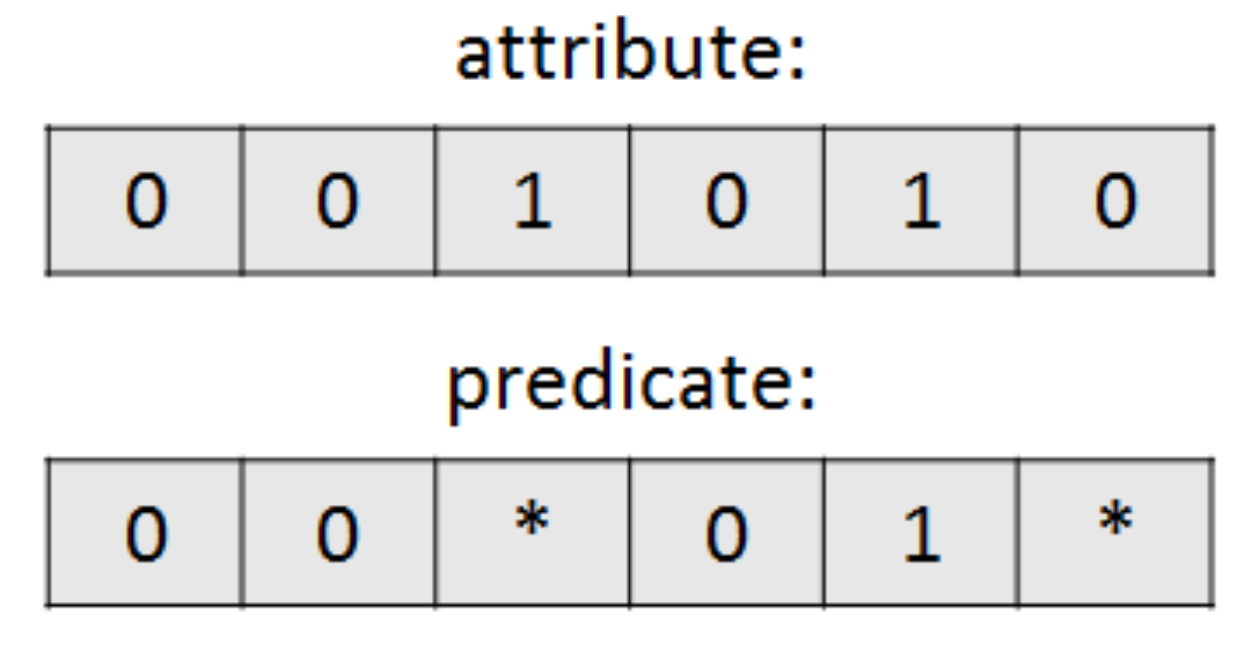}
                \label{fig:hvesucc}
	 }
         \subfloat[Non-Match]{
                \centering
                \includegraphics[width=0.25\textwidth]{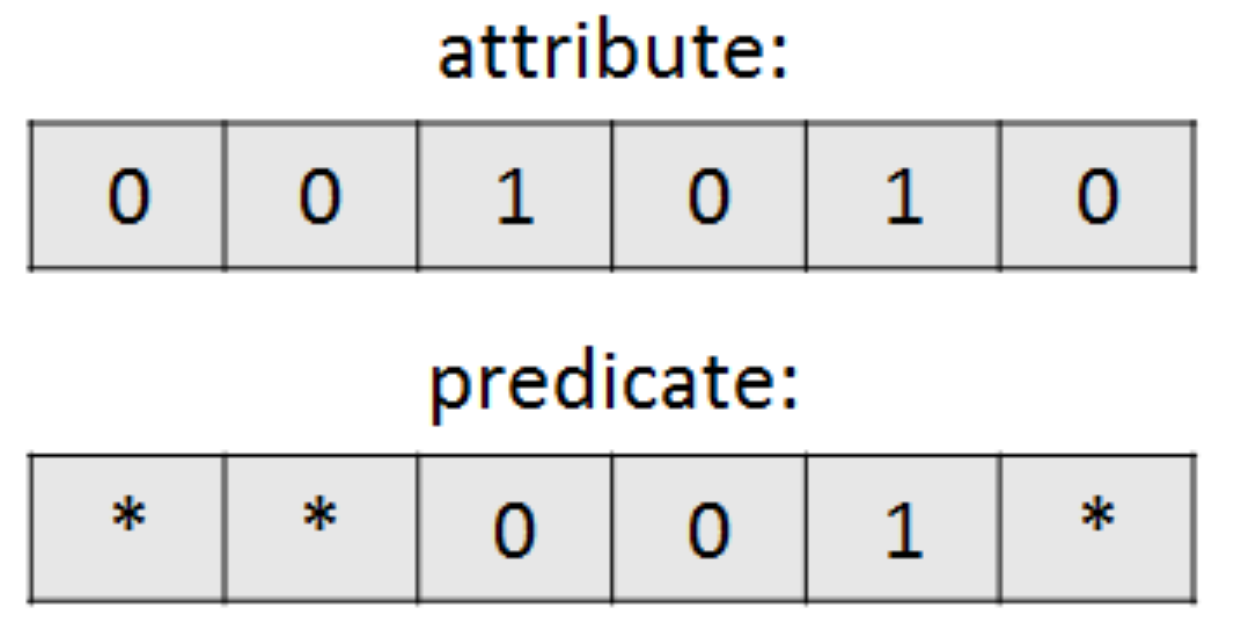}
                \label{fig:hvefail}
	}
	\figspb
        \caption{Predicate evaluation on ciphertexts with HVE}
	\label{fig:hveexamples}
	\figspb
	\figspc
\end{figure}

{\em Hidden Vector Encryption (HVE)} \cite{Boneh07} is a searchable encryption system that supports predicates in the form of conjunctive equality, range and subset queries. Compared to earlier solutions~\cite{Boneh03,Boneh06}, HVE yields ciphertexts with considerably smaller length. Search on ciphertexts can be performed with respect to a number of {\em index attributes}. HVE represents an attribute as a bit vector (each element has value $0$ or $1$), and the search predicate as a {\em pattern} vector where each element can be $0$, $1$ or '*' that signifies a wildcard (or ``don't care'') value. Let $l$ denote the HVE {\em width}, which is the bit length of the attribute, and consequently that of the search predicate. A predicate evaluates to $True$ for a ciphertext $C$ if the attribute vector $I$ used to encrypt $C$ has the same values as the pattern vector of the predicate in all positions that are not '*' in the latter. Figure~\ref{fig:hveexamples} illustrates the two cases of {\em Match} and {\em Non-Match} for HVE, whereas Algorithm~\ref{algo:HVE} provides the matching pseudocode.
We provide additional mathematical background on HVE encryption and its operations in Appendix~\ref{sec:app}.

\begin{algorithm}
\caption{HVE\_Match}
\label{algo:HVE}
\SetKwInOut{Input}{Input}\SetKwInOut{Output}{Output}
\Input{HVE index $I = \left[I_1 : I_l\right]$}
\Input{HVE token $T = \left[T_1 : T_l\right]$}
\Output{$True$ if $I$ matches $T$, $False$ otherwise}
\BlankLine
\eIf{$\forall i \in \left[1 : l\right]$, $I_{i} = T_{i}$ or $T_{i} = *$}{return $True$}
{return $False$}
\end{algorithm}

\figspc
\section{Proposed Spatial HVE Approaches}
\label{sec:contribution}

In Section~\ref{sec:baseline} we outline a naive {\em baseline} technique which applies HVE in a straightforward manner to determine privately which users fall within one or more alert zones. The baseline leads to prohibitive costs, as shown by experiments in Section~\ref{sec:exps}. To bring down the overhead of HVE, we propose in Section~\ref{sec:hier} a {\em hierarchical encoding} technique, which reduces the amount of cryptographic primitives (specifically, bilinear pairings) required during search.  Next, in Section~\ref{sec:gray} we further refine hierarchical encoding and devise the {\em Gray encoding}, which achieves superior computation savings.

\subsection{Baseline Encoding}
\label{sec:baseline}

Recall that the data space is partitioned by a two-dimensional $d \times d$ regular grid. When a user reports its position, it sends to the server an encryption of the grid cell it is enclosed by. Similarly, the TA defines the alert zones as a set of grid cells. Each grid cell can be uniquely identified by a {\em cell identifier}, with values between $1$ and $d^2$. Thus, the straightforward way to support secure location-based alerts is to use an HVE index with width $l=d^2$. The data and query encoding are performed as follows:
\begin{itemize}
\item 
When user $u$ enclosed by grid cell $i$ reports its location, it uses a bitmap index $I$ of width $d^2$ where all the bits are set to '0' except bit $i$ which is set to '1'.
\item 
The $TA$ creates a single token for search, which captures all the alert zones. The token is a bitmap with $d^2$ bits where all bits corresponding to cell identifiers that are included in an alert zone are set to '*'. All other bits are set to '0'.
\item 
At the server (i.e., at query time), according to the rules for HVE query evaluation from Section~\ref{sec:hve}, a user will be determined as a $Match$ if and only if the '1' bit in the encrypted location will correspond to a '*' entry in the token. 
\end{itemize}

\begin{figure}[t]
\centering
\includegraphics[scale=0.22]{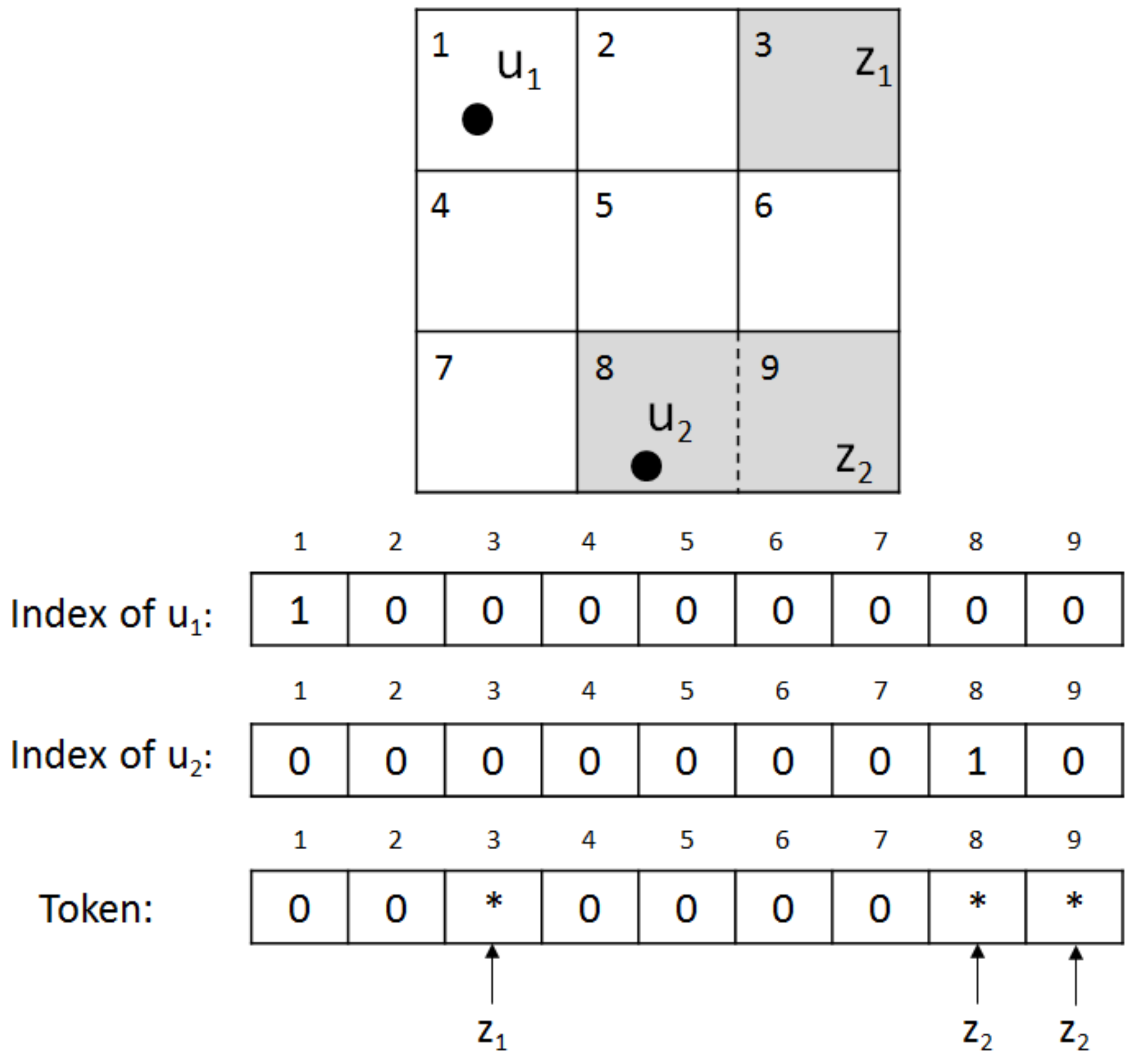}
\figspb
\caption{Naive Baseline Encoding}\label{fig:baseline}
\figspb
\end{figure}

Consider the example in Figure~\ref{fig:baseline}, where $d=3$. We have nine grid cells, so the width of the HVE is $l=9$. There are two alert zones: $z_1$ which consists of a single grid cell ($3$), and $z_2$ which spans two grid cells ($8$ and $9$). Two users report their locations: $u_1$ enclosed by cell $1$, and $u_2$ enclosed by cell $8$. The index vectors of the two users are shown in the diagram. A single token is used to represent both alert zones, and a '*' is placed in the positions corresponding to the cells enclosed by the zones, namely $3$, $8$ and $9$. The predicate evaluation for \(u_2\) will return $Match$, as the position marked by '1' in the index of $u_2$ corresponds to a '*' in the token. Conversely, a {\em Non-Match} is returned for $u_1$, as the bit '1' in position $1$ corresponds to a '0' in the token. Algorithm~\ref{algo:baseline} provides the baseline encoding pseudocode.

\vspace{-20pt}
\begin{algorithm}
\caption{Baseline Encoding}
\label{algo:baseline}
\SetKwInOut{User}{User $u$}
\SetKwInOut{TA}{TA}
\SetKwInOut{Server}{Server}
$p$ = ID of grid cell enclosing user $u$

$A$ = set of grid cell IDs that make up the alert zone

\User{Set $I = \left[I_1 : I_{d^{2}}\right]$, $I_{p} = 1$ and $\forall i \neq p, I_{i} = 0$}
\User{Send encrypted $I$ to Server $S$}
\TA{Set $T = \left[T_1 : T_{d^{2}}\right]$, $T_{i} = * $ if $i \in A$ and $T_{i} = 0$ otherwise}
\TA{Send encrypted $T$ to Server $S$}
\Server{If $HVE\_Match(I, T) = True$, return $Match$}
\Server{Otherwise, return $NotMatch$}
\end{algorithm}

\vspace{-20pt}
As discussed in Appendix~\ref{sec:app}, Eq.~\eqref{eq:query} from the query step executes two pairing operations and multiplies their results for every element in $J$, i.e., for every position that is not '*' in the token. Having a token with one position for each grid cell leads to high cost, so the naive encoding where the HVE width is equal to the number of cells is not practical. Furthermore, the sum of areas of all alert zones is relatively small compared to the entire dataspace, hence the number of '*' entries will be small, and the cardinality of set $J$ will be large, increasing cost. Next, we propose two effective forms of encoding HVEs such that execution cost is reduced.

\subsection{Hierarchical Encoding}
\label{sec:hier}

The main problem of the baseline encoding is that the HVE width grows linearly with the grid cell count. We propose a technique that reduces the HVE width from $d^2$ to $2 \log d$, by using the binary representation of cell identifiers. However, the representation of the search predicates (and thus, that of the tokens) becomes more complicated, since the advantage of the ``bitmap-like'' representation of the baseline is lost. We investigate how to aggregate representations of adjacent cells belonging to the same alert zone, in order to reduce the amount of tokens required. Aggregation is performed according to a hierarchical spatial structure, hence the name of {\em hierarchical encoding}.

\begin{figure}[t]
\centering
\includegraphics[scale=0.22]{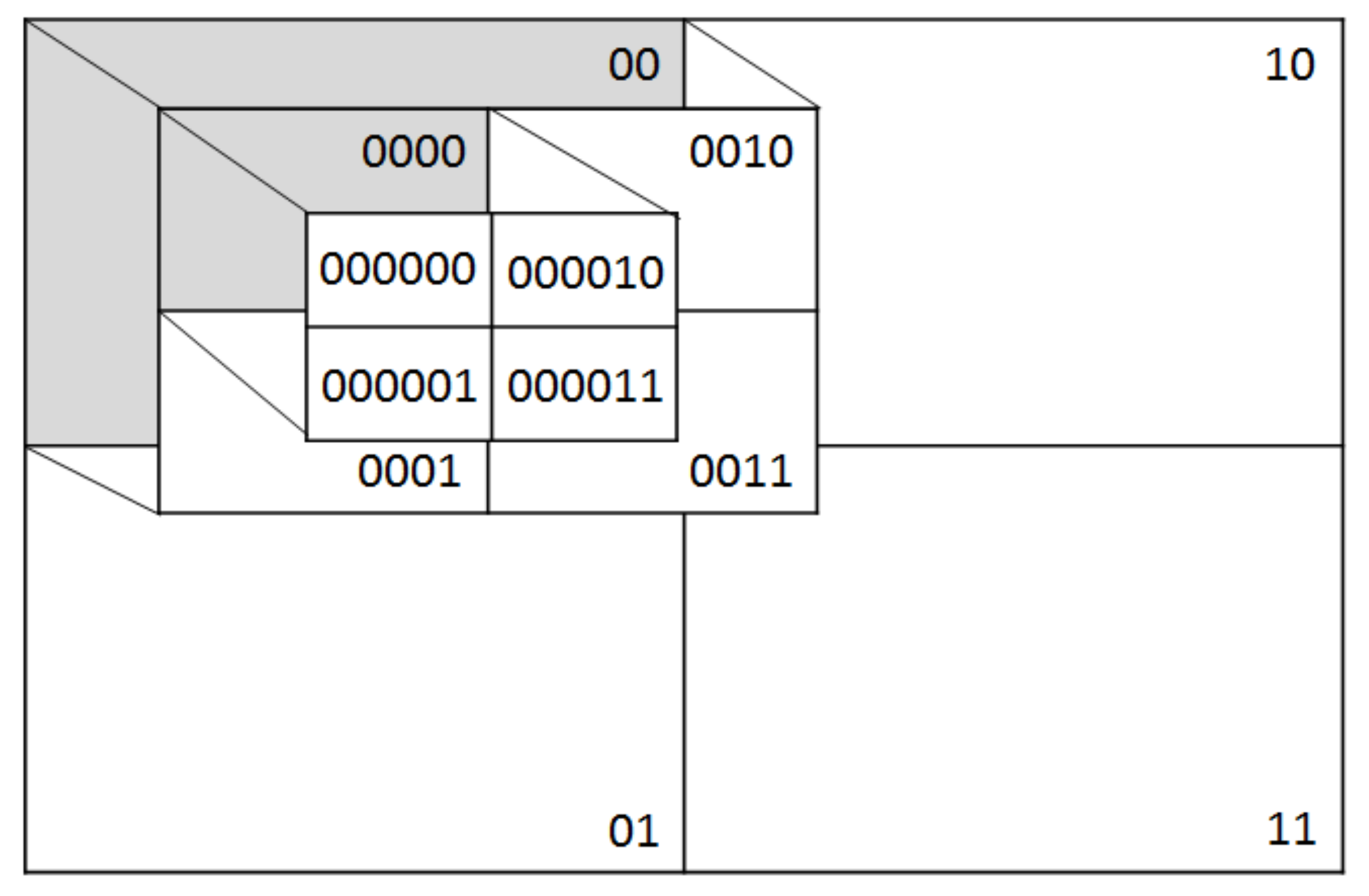}
\figspb
\caption{Hierarchical partitioning on three levels}
\label{fig:hierarchical_partitioning}
\figspa
\end{figure}

We consider a logical organization of the grid cells into a quadtree-like structure\footnote{Note that this is a logical structure, no physical index is required.} \cite{quadtree}. Figure~\ref{fig:hierarchical_partitioning} illustrates the space partitioning into four cells of equal size by using mediators on the $Ox$ and $Oy$ axes. Each of these four cells will have a $2$-bit id: $00$ for top left, $01$ for bottom left, $10$ for top right and $11$ for bottom right. Next, each of these cells is partitioned recursively into four new cells, and the newly obtained $2$-bit identifiers are concatenated as a suffix to the previous step identifiers. For simplicity, in this example we consider that the grid cell count is a power of $4$, but any grid size can be accommodated in this model by using padding. 

The diagram also shows how aggregation of cells from level $j$ is performed into a larger cell at level $j+1$ (i.e., in reverse direction of scoping). Note that, with the binary representation of identifiers, cell aggregation corresponds to binary minimization of a logical 'OR' expression composed of the terms that represent cell identifiers. As a result, instead of using a distinct token (i.e., HVE pattern) for each cell, we can use token aggregation and reduce the number of predicates that need to be tested. If two cells are in the same alert zone and their identifiers differ in just one bit, then a '*' can be used instead of that bit, similar to a wildcard in binary minimization. The newly obtained token is faster to generate and evaluate, because according to the operations described in Appendix~\ref{sec:app}, only the positions in the pattern vector where the value is not '*' need to be considered (i.e., those in set $J$). 
If all of the four partitions belonging to the same quadtree node are in the same alert zone, then they can all be aggregated to the identifier of their parent. In our implementation, in order to generate HVE pattern vectors with aggregation, we use the binary expression minimization tool Espresso \cite{espresso}.

\begin{figure}[t]
\centering
\includegraphics[scale=0.25]{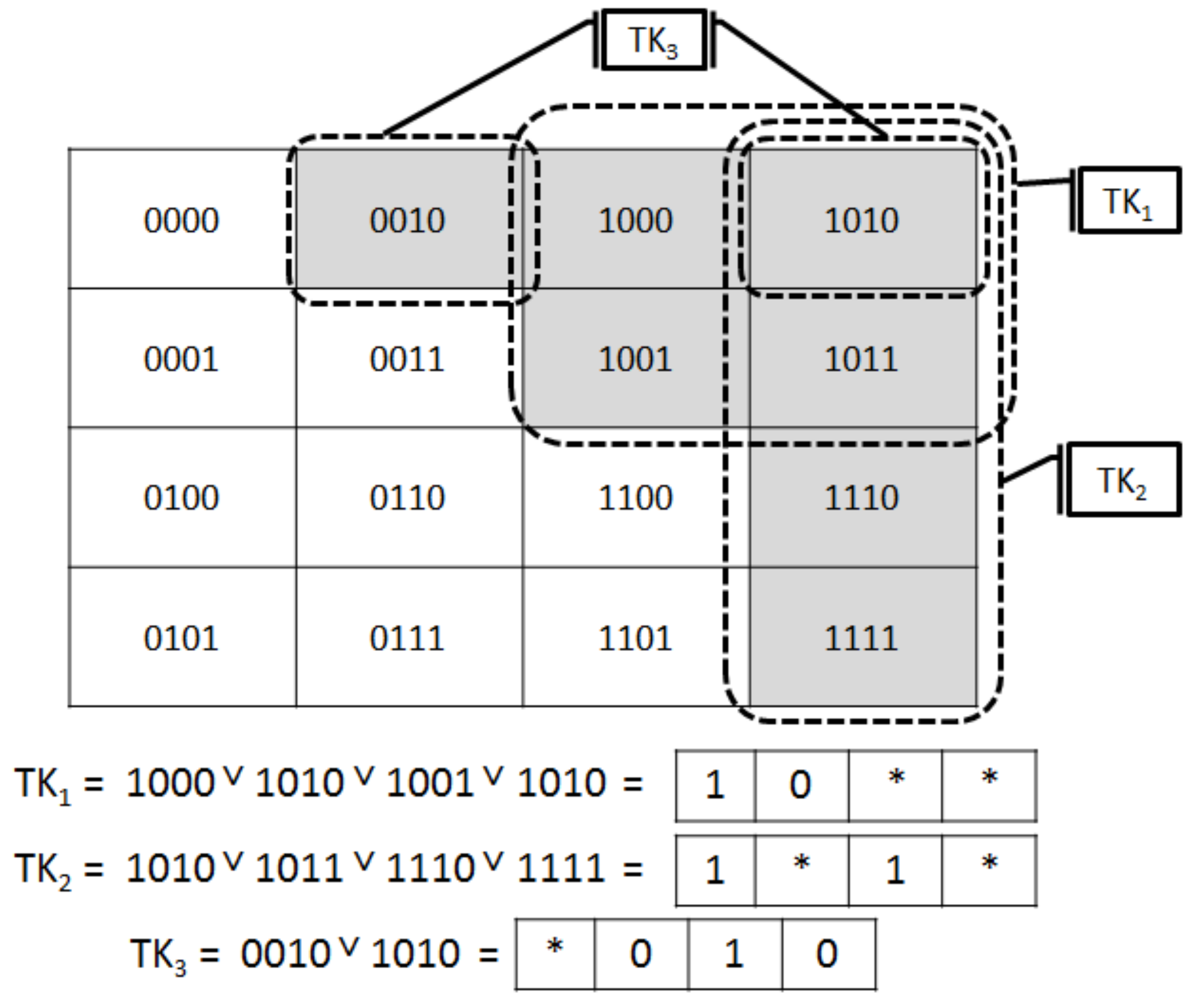}
\figspb
\caption{Hierarchical encoding and token aggregation}
\label{fig:hierarchical_tokens}
\figspa
\end{figure}

Consider for instance the example from Figure~\ref{fig:hierarchical_tokens}, where the alert zone is composed of seven cells. All four cells whose identifiers have prefix $10$ are in the zone, hence they can all be aggregated to \(TK_1=10**\). Also, cells on the last vertical line can be aggregated to \(TK_2=1*1*\). Finally, cells $0010$ and $1010$ can be aggregated to \(TK_3=*010\). Note that, although these tokens overlap, this does not introduce a correctness problem at query (i.e., matching) time at the server. Furthermore, the monitoring server can evaluate them in order from the most general (highest number of '*'s) to the most specific one (lowest number of '*'s). If one token evaluates to a $Match$ on a particular ciphertext, there is no need to evaluate the rest of the tokens, since it is clear that the user is in the alert zone. In addition, creating overlapping tokens helps if these tokens have more '*' symbols in their HVEs, because the cardinality of set $J$ (Eq.~\eqref{eq:query} in Appendix~\ref{sec:app}) decreases, hence query and token generation times decrease as well. 

In summary, the hierarchical scheme works as follows:

{\bf Encryption.} Users determine the binary identifier of the grid cell they are in, and create an HVE index $I$ with that representation, having width $2\log d$, where $d \times d$ is the grid size. The encryption is performed with respect to $I$. Since the grid parameters are public, the user can easily determine its enclosing cell and construct $I$.

{\bf Token Generation.} For each alert zone $z$, the TA creates the set of binary codes of cells within the zone. Next, it computes the minimized binary expression equivalent to the logical 'OR' of all codes in the set. For each resulting term in the minimized expression, the TA creates a token, and the token will have a '*' symbol in each position that was reduced during the minimization. All tokens are sent to the server.

{\bf Query.} For every user and alert zone, the server $S$ performs matching as follows: $S$ evaluates the encrypted user location against every token that represents the zone, in {\em decreasing} order of the number of '*' symbols in the token. In other words, tokens with a higher number of '*'s are considered first. If a $Match$ is obtained, then the remaining tokens for the zone are no longer considered. If a {\em Non-Match} is obtained for all tokens in the zone, then the server concludes that the user is not inside the zone.

Even though the number of tokens increases compared to the baseline, the width of each token is considerably smaller. In addition, the proportion of '*' symbols in a token is much higher for the hierarchical scheme, due to aggregation. Finally, considering tokens with a smaller $J$ set first increases the chances of deciding on a $Match$ without having to consider all tokens of a zone. All these factors make the hierarchical encoding perform much faster than the baseline, as we show in Section~\ref{sec:exps}. Algorithm~\ref{algo:hierarchical2} provides the hierarchical encoding pseudocode.

\vspace{-20pt}
\begin{algorithm}
\caption{Hierarchical Encoding}
\label{algo:hierarchical2}
\SetKwInOut{User}{User $u$}
\SetKwInOut{TA}{TA}
\SetKwInOut{Server}{Server}
$p$ = ID of grid cell enclosing user $u$

$A$ = set of grid cell IDs that make up the alert zone

\User{Set $I = \left[I_1 : I_{2\log_{2}{d}}\right]$, s.t. $\forall i, I_{i} \in \{0, 1\}$ and
$\sum_{i = 1}^{2\log_{2}{d}}{I_{i}\cdot2^{i - 1}} = p$}
\User{Send encrypted $I$ to Server $S$}

\TA{Create initial token set $TK = \left[TK_1 : TK_{2\log_{2}{d}}\right]$, where $\forall a \in A$, $\exists T \in TK$ s.t. $\forall i, T_{i} \in \{0, 1\}$ and 
$\sum_{i = 1}^{2\log_{2}{d}}{T_{i}\cdot2^{i-1}} = a$}
\TA{Reduce $TK$ size by aggregating tokens within Hamming distance of 1}
\TA{Send encrypted $TK$ to Server $S$}
\Server{If $\exists T \in TK$ s.t. $HVE\_Match(I, T) = True$, return $Match$; Else return $NotMatch$}
\end{algorithm}
\vspace{-20pt}
\subsection{Gray Encoding}
\label{sec:gray}

The performance gain of the hierarchical technique comes from the ability to combine adjacent cells into a single search token with many '*' positions. In other words, the performance improves when the binary minimization of the logical 'OR' of cell identifiers is more effective. However, in some cases, no aggregation can be performed between two neighboring cells, as the Hamming distance between their identifiers is more than $1$. As alert zones are composed of groups of neighboring cells, it is desirable to have small Hamming distance between adjacent cell identifiers. To improve the effectiveness of the binary minimization step, hence to increase the number of '*' values in search tokens, we represent cell identifiers using Gray codes \cite{Gray}. This way, cell identifier values are assigned in such a manner that the Hamming distance between two adjacent cells is always $1$, hence binary minimization is facilitated.

A one-dimensional Gray code vector is determined using the following recursive algorithm, where $|$ represents the concatenation operator, and \(G_k\) is the vector of a Gray code instance at step \(k\)).  
\[G_1 = (0,1)\]
\figspb
\figspc
\[G_k = (g_1, g_2, ... g_{2^k})\]
\figspb
\figspc
\[G_{k+1} = (0 |g_1, 0|g_2, ..., 0|g_{2^k}, 1|g_{2^k}, ..., 1|g_2,1|g_1)\]
\figspb

For $k=3$, the following Gray code vectors are obtained:
\[G_1 = (0, 1)\]
\figspb
\figspc
\[G_2 = (00, 01, 11, 10)\]
\figspb
\figspc
\[G_3 = (000, 001, 011, 010, 110, 111, 101, 100)\]
\figspb

Given a $d \times d$ grid, the length of the required Gray code necessary to represent all cells is \(2^{\lceil log_2 d\rceil }\). We employ a Gray code instance independently for each of the two dimensions of the space, thus the identifier of a cell consist of the concatenation between the Gray code value for the $y$ axis and the $x$ axis values. Similar to hierarchical encoding, the scheme assumes a total number of cells that is a power of $4$, but other cases can be handled by padding. 

\begin{figure}[t]
\centering

\subfloat[Gray encoding]{
	\centering
	\includegraphics[scale=0.3]{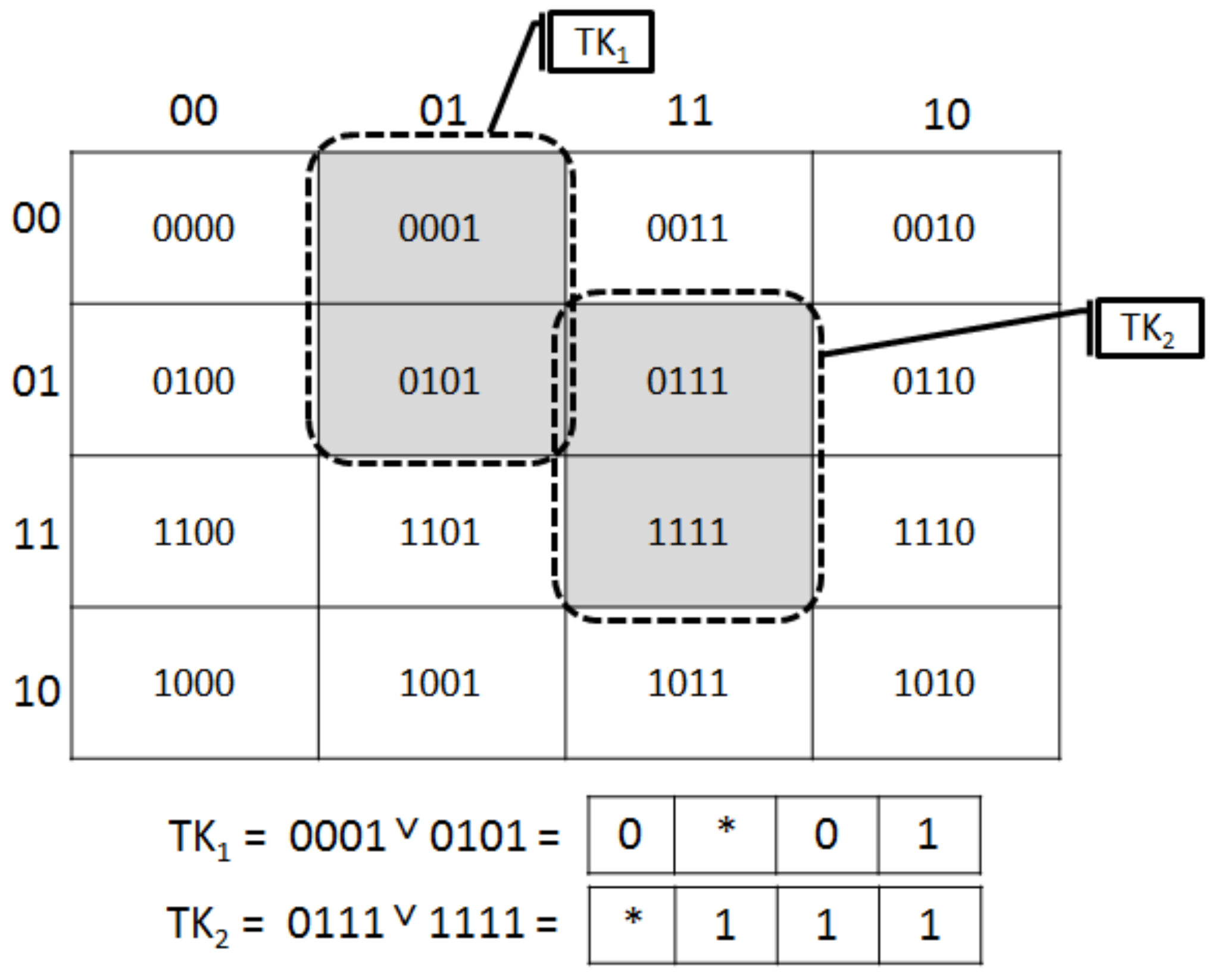}
        \label{fig2:gray-case}
}
\subfloat[Hierarchical encoding]{
	\centering
	\includegraphics[scale=0.3]{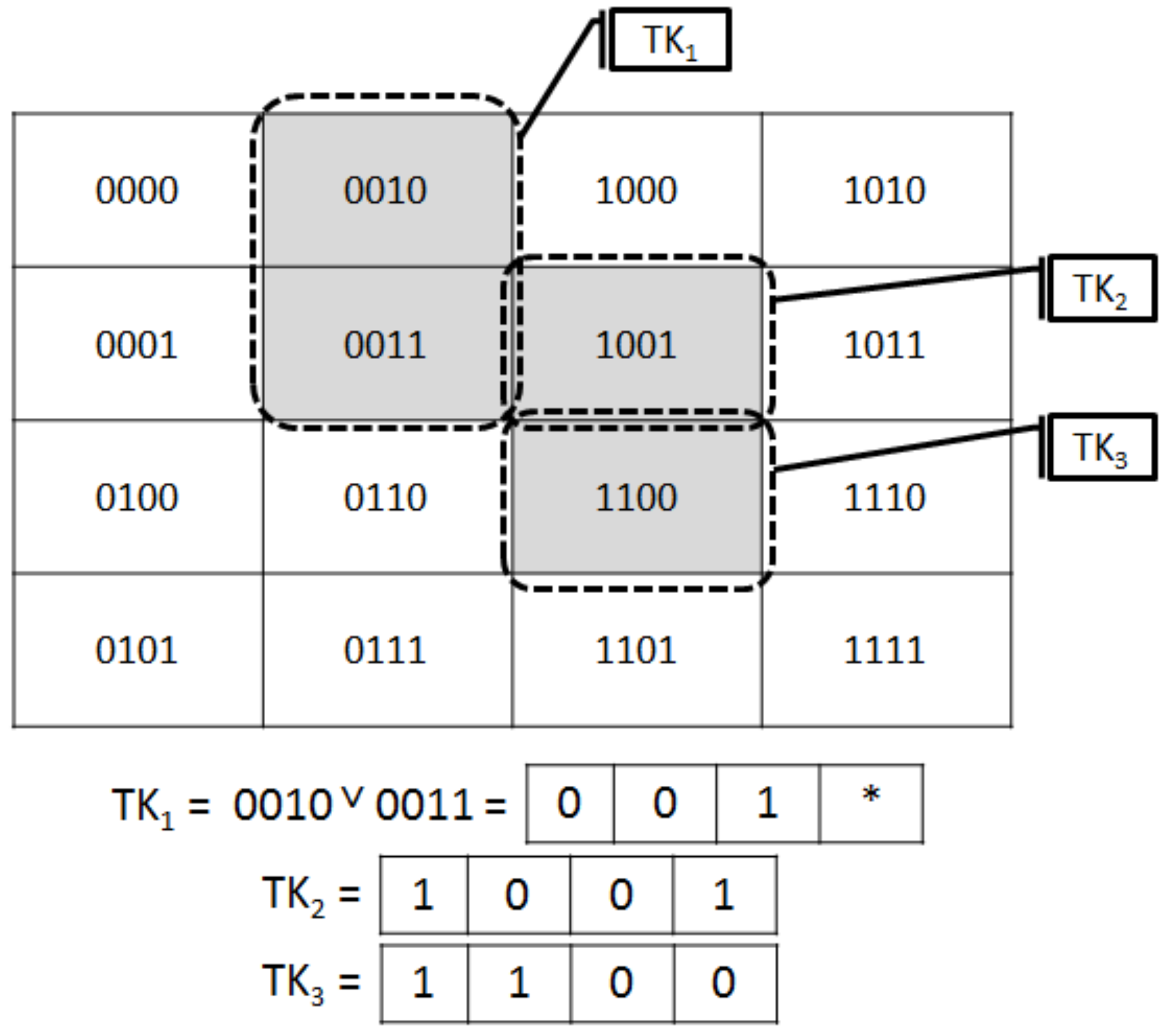}
        \label{fig2:hier-case}
}
\figspb
\caption{Token Aggregation: Gray vs hierarchical encoding}\label{fig:graycode_tokens}
\figspa
\end{figure}

Figure~\ref{fig:graycode_tokens} illustrates the advantage of using Gray encoding instead of hierarchical encoding for a $16$-cell $4 \times 4$ grid. The alert zone consists of four cells. The two digits leading each row in the Gray encoding diagram (Figure~\ref{fig2:gray-case}) mark the two-bit prefix shared by all the cell identifiers in that row. Conversely, the two digits on top of each column mark the two-bit suffix of all the cell identifiers in that column. Using binary minimization, the two tokens shown in the diagram are obtained, each of them having one '*' symbol. Figure~\ref{fig2:hier-case} shows how the hierarchical encoding behaves for the same input. Due to the fact that moving from cell $1001$ to $1100$, or from $0011$ to $1001$ corresponds to a Hamming distance larger than $1$, no aggregation is possible between these cell pairs. As a result, three tokens are necessary to represent this zone. Furthermore, two of these tokens have no '*' symbol, leading to more expensive evaluation. As we will show in Section~\ref{sec:exps}, Gray encoding achieves more effective aggregation, especially in the case of skewed data (e.g., Gaussian distribution of alert zones).

The phases of encryption and query for the Gray encoding method are similar to their counterparts for the hierarchical encoding method of Section~\ref{sec:hier}. The main difference is in the token generation phase, where the binary minimization is performed according to the Gray code cell identifier binary representation. As we show in the experimental evaluation (Section~\ref{sec:exps}), using the Gray code representation can improve performance by achieving more effective binary minimization. This in turn results in either fewer tokens, or tokens with a larger proportion of '*' symbols. 

\section{Performance Optimizations}
\label{sec:optimizations}
\figspc
\subsection{Preprocessing mathematical operations}

As discussed in Appendix~\ref{sec:app}, the HVE mechanism involves a large number of exponentiations with very large integers which incur a significant computational cost. Fortunately, many of these exponentiations are performed on a common base. For example, if we take into consideration the encryption phase, in order to compute \(C^{'}\) and \(C_0\), \(A\) and \(V\) must be raised to the power of \(s\). Even if $s$ is chosen  randomly for each run of this step, $A$ and $V$ depend on the public key, which remains unchanged for long periods of time (in commercial systems, re-keying can be done with frequency of once per year, or even less often). Furthermore, when computing \(C_{i,1}\), because the index attributes consist of a vector of $0$ and $1$ values, the base of the power $s$ can have only two values: \(H_i\) or \(U_i \times H_i\). Because both depend only on the public key, these two exponentiations will always have a constant base. The same logic can be also applied to the exponentiations for token generation. By employing preprocessing on each of these fixed bases, the exponentiations become a lot faster. The preprocessing can be done offline, and the results used during online operation, leading to significant execution time savings. 

When matching a token against an encrypted message, several pairing computations are performed. For a particular token, the values of  \(K_0\), \(K_{i,1}\) and \(K_{i,2}\) remain constant. When applying a pairing, Miller's algorithm is used \cite{Miller}. Typically, for each such operation, it is required to compute several line equations. In \cite{pbc_thesis} it is shown that effective preprocessing can be used as long as the first parameter is constant because the equations of the lines can be calculated and stored ahead of time. At runtime, the coordinates of a given point are substituted into these precomputed expressions. Since HVE requires symmetric elliptic curves, preprocessing can be also done for the second parameter. Preprocessed information is stored with each token and used by the  server to improve the time of each pairing. Preprocessing each token must be done only when the tokens are generated at the trusted authority. 

\subsection{Parallelization}

The server is monitoring a large number of users, and may receive a large number of alert zones. This creates a considerable load on the server. However, we emphasize that the processing of a message from a user can be done independently from messages originating at other users. Furthermore, even for the same user, the matching for different alert zones are completely independent operations. This presents a great potential for parallelization. In fact, the problem is embarrassingly parallel, and significant execution time improvements can be obtained by using several CPUs for matching. Nowadays, even off-the-shelf desktop computers have multiple cores. Commercial cloud services typically have hundreds or thousands of CPUs available for computation. Due to the parallel nature of the problem, the speedup is expected to be close to linear, and the resulting system scales very well as the number of CPUs involved grows.

We consider a message-passing parallel computing paradigm, which is favored by the fact that only a small amount of data needs to be shared among distinct CPUs. 
One master process coordinates all other slave processes. The master process distributes to the slaves the search tokens. Then, as encrypted updates from users arrive, the master receives the requests and dispatches them to slaves. Load balancing can be easily implemented at the master level, which keeps track of the status of all slave processes. No communication is required between slave processes, and the master-slave communication is required only at the start and end of each task. Furthermore, distinct messages originating from the same user can be processed on different CPUs without any loss in correctness or performance (i.e., no state maintenance is required). After processing is done, if the token evaluated successfully, a response action can be taken by the processing CPU, or the event can be sent to a central server responsible only with handling what to do in case of a successful match.

\section{Privacy-Performance Trade-off through Alert Zone Expansion}
\label{sec:queryexpansion}

So far, we considered that alert zones are a fixed input to the system, and we provided data encodings and optimizations to reduce computational overhead under this constraint. Since alert zones were not modified, we maintained the amount of location disclosure to a minimum, i.e., an adversary could only learn whether a specific ciphertext corresponded to a location inside the alert zone or not. In this section, we consider a relaxation of the alert zone extent in order to improve performance. Specifically, given an input alert zone, we investigate whether it is possible to slightly enlarge it such that the resulting set of tokens needed to implement secure notification requires fewer bilinear pairings to evaluate. 

To maintain the level of additional disclosure low, we allow only a relatively small enlargement factor, expressed as a ratio of the alert zone area, and quantified by a bound parameter $\alpha$. Given an enlargement factor $\alpha$, our proposed alert zone expansion heuristic determines an enlarged area with significantly lower processing overhead. In effect, this proposed optimization trades a small amount of additional location disclosure for a significant boost in matching performance. As a salient feature of this optimization, the privacy-performance trade-off can be tuned using a single parameter ($\alpha$).

The optimization is deployed at the TA, which is in charge of generating search tokens. In an actual deployment, since the TA is trusted, it can perform additional steps to check whether the enlarged zone is acceptable from a security standpoint, for instance by comparing it against a set of pre-defined policies. In this paper, we only focus on the performance aspect, and derive effective algorithms that quickly generate enlarged search tokens (the policy aspect is outside our scope). Similar to optimizations from prior sections, the zone expansion is guided by the objective of deriving tokens with fewer non-wildcard elements, which results in less computation.
The expansion technique assumes the same hierarchical data domain representation considered so far, and works in conjunction with either hierarchical or Gray encodings.

We denote by {\em base cell} a cell in the leaf level of the hierarchical domain representation (recall that, the domain is split into $d \times d$ base cells, where $d$ is a power of two). The hierarchy has a number of $1 + \log d$ levels. At level $k$, an {\em aggregate} cell consists of $2^k \times 2^k$ base cells. Specifically, at the leaf level, numbered as $k=0$, each cell is a base cell, whereas at the top of the hierarchy (level $\log d$) there is a single cell with size $d \times d$ (expressed in terms of base cells). We identify a cell at level $k$ by its coordinates within that level: $(x, y)_k$. The {\em binary identifier} of a cell consists of a binary string, which can be immediately derived from its coordinates.

Algorithm~\ref{alg:ExpandQuery} captures the main steps of the proposed heuristic alert zone expansion technique. The input consists of expansion factor $\alpha$ and initial alert zone $A$. The heuristic is given a maximum budget $W = \lfloor\alpha |A|\rfloor$ base cells that it can add to the initial zone, where $|A|$ is the area of the initial zone expressed in terms of base grid cells. The output of Algorithm~\ref{alg:ExpandQuery} is an expanded zone $\hat{A}$ such that $|\hat{A}| \leq |A| + W$ and the number of bilinear pairings required to evaluate $\hat{A}$ is lower than that of $A$.

\begin{algorithm} 
	\caption{ExpandZone} 
	\label{alg:ExpandQuery} 
	\SetKwFunction{SelectPatchesSingleLevel}{SelectPatchesSingleLevel}
	\KwIn{expansion ratio $\alpha$; alert zone $A$ }
	\KwOut{expanded alert zone $\hat{A}$}
	
	$W \leftarrow \lfloor\alpha |A|\rfloor$, $\mathcal{B} \leftarrow \emptyset$
	
	$\hat{A} \leftarrow CrtZone \leftarrow A$
 
	\For{$k \in [0, \dots, \log d] $}{
			
		$\mathit{PSet} \leftarrow $ {\SelectPatchesSingleLevel{$W$, $\frac{d}{2^k}$, $CrtZone$}}   \label{algo:line:callSelectPatches}
		
		\For{$p \in \mathit{PSet}$}{\label{algo:line:startCheckExpansion}
				\For{$(x, y)_k \in p.\mathit{attached\_cells}$}{
								
					$\mathcal{B} \leftarrow \mathcal{B}~\cup $ all base cells belonging to cell $(x, y)_k$
					 
				}
		}\label{algo:line:endCheckExpansion}
		\If{\#pairing of $(\hat{A} \cup \mathcal{B}) \leq$ \#pairings of $\hat{A}$ }{
			$\hat{A} \leftarrow \hat{A} \cup \mathcal{B}$ \label{algo:line:startPrepareNextLevel}
			
			\For{$p \in \mathit{PSet}$}{
				$W \leftarrow W - p.\mathit{cost}$
				
				\For{$(x, y)_k \in p.\mathit{attached\_cells}$}{
					$CrtZone \leftarrow CrtZone \cup (x, y)_k$
				}
			}
		
			$W \leftarrow \lfloor\frac{W}{4}\rfloor$
			
			\If{$W \leq 0$}{
				$\mathbf{break}$
			}

			\For{$(x, y)_k \in CrtZone$}{
				$CrtZone = CrtZone \setminus (x, y)_k \cup (\lfloor\frac{x}{2}\rfloor, \lfloor\frac{y}{2}\rfloor)_{k+1}$
			} 
			\label{algo:line:endPrepareNextLevel}
		}{
		}
	}
	
	\Return $\hat{A}$
\end{algorithm}

The {\tt ExpandZone} routine (Algorithm~\ref{alg:ExpandQuery}) works by considering each level of the data domain hierarchy. 
An essential step of {\tt ExpandQuery} is the {\tt SelectPatchesSingleLevel} routine (detailed in Algorithm~\ref{alg:SelectPatchesSingleLevel}) which finds \textit{patches} to add to the current set of zone cells (line~\ref{algo:line:callSelectPatches}). 
A {\em patch} (formally defined in Section~\ref{sec:constructpatches}) is a set of cells added to the zone in a single iteration.
If the new set of zone cells, denoted as $CrtZone$, does not require more pairings than the current set of cells, 
an expansion is made with the cells in the patch and we continue to the next level.%

In order to prepare for the next level (lines~[\ref{algo:line:startPrepareNextLevel}-\ref{algo:line:endPrepareNextLevel}]), parameters and indices are adjusted. 
Budget $W$ is reduced by a factor of $4$, since in the next level the size of one cell is equal to $4$ cells in the current level. 
Similarly, indices of zone cells are divided by two.
The intuition behind dividing the indices by two is that all $2 \times 2$ areas containing zone cells in this level must be fully covered by the expansion, which means all cells in those areas are in $CrtZone$. 
Algorithm~\ref{alg:ExpandQuery} stops when one of the following conditions is met: {\em (i)} the new set of zone cells increases the number of pairings; {\em (ii)} budget $W$ is exhausted; or {\em (iii)} the zone expands to the entire root level.

\begin{figure}
\figspa
	\centering
	\subfloat[Level $k=0$]{
		\includegraphics[width=0.5 \columnwidth]{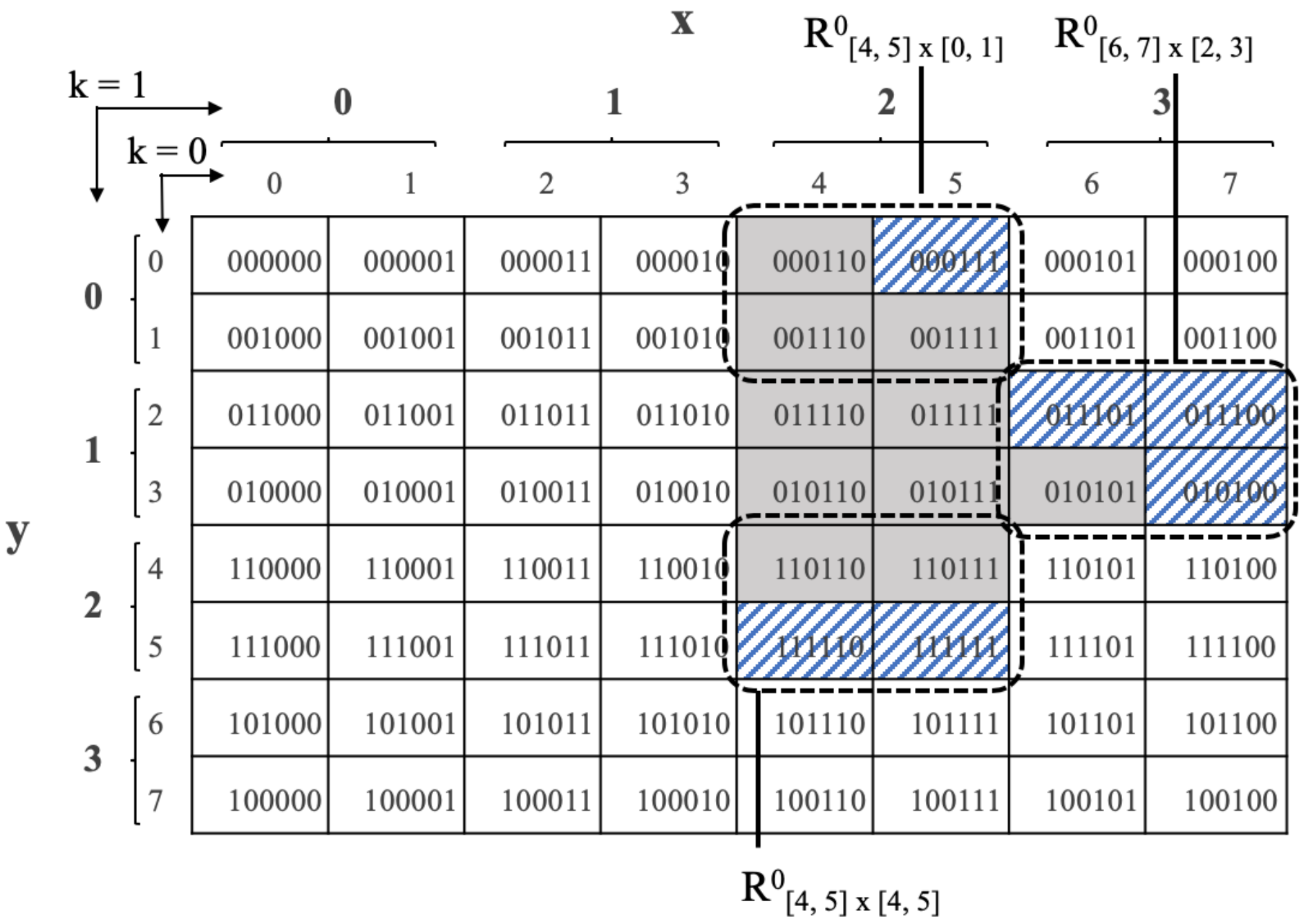}
		\label{fig:expanding1_k0}}
	\subfloat[Level $k=1$]{
		\includegraphics[width=0.5 \columnwidth]{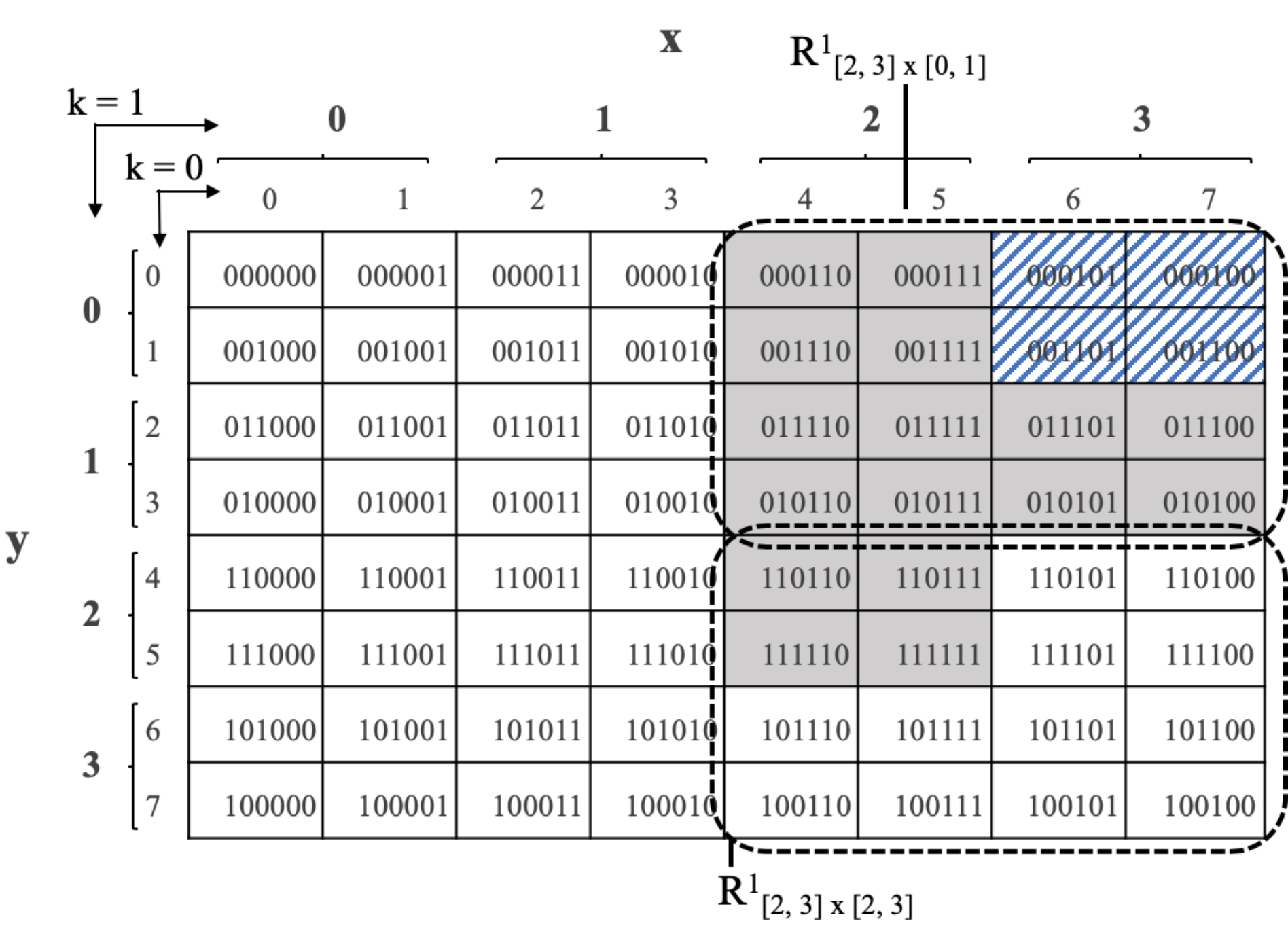}
		\label{fig:expanding1_k1}  }
	\caption{Example of expanding alert zone (in grey) at level $k = 0$~\protect\subref{fig:expanding1_k0} and $k = 1$~\protect\subref{fig:expanding1_k1}. At level $0$, a total of $6$ cells (illustrated with stripes) are added to obtain $3$ fully-covered areas $R^0_{[4, 5] \times [0, 1]}$, $R^0_{[6, 7] \times [2, 3]}$, and $R^0_{[4, 5] \times [4, 5]}$. At level $1$, four additional cells are added to the area $R^1_{[2, 3] \times [0, 1]}$. The final expanded alert zone contains all cells in $R^0_{[4, 7] \times [0, 3]}$ and $R^0_{[4, 5] \times [4, 5]}$ (both grey and diagonally-striped cells in~\protect\subref{fig:expanding1_k1})}
	\label{fig:expanding1}  
\figspa
\end{figure}

To illustrate the zone expansion algorithm, consider the example in Figure~\ref{fig:expanding1}. Zone cells are shown in grey color, and budget is set to $W = 10$. 
An area with $x \in [x_1, x_2], y \in [y_1, y_2]$ at level $k$ is denoted as $R^k_{[x_1, x_2] \times [y_1, y_2]}$.
Starting at level $k = 0$ (Figure~\ref{fig:expanding1_k0}), the $2 \times 2$ areas $R^0_{[4, 5] \times [0, 1]}$, $R^0_{[6, 7] \times [2, 3]}$, and $R^0_{[4, 5] \times [4, 5]}$ are considered for expansion. All six cells with diagonal stripes are added to the current set of zone cells to fill those three areas.
To prepare for expansion at the next level, the coordinate ids of zone cells must be adjusted for each $2 \times 2$ area. For example, for $k=1$, area $R^0_{[4, 5] \times [0, 1]}$ becomes cell $(2, 0)_1$, area $R^0_{[6, 7] \times [2, 3]}$ becomes cell $(3, 1)_1$ and so on.
The budget $W$ is reduced to $\lfloor\frac{10 - 6}{4}\rfloor = 1$.
Next, at level $k = 1$ (Figure~\ref{fig:expanding1_k1}), the areas $R^1_{[2, 3] \times [0, 1]}$ and $R^1_{[2, 3] \times [2, 3]}$ are considered for expansion.%
The cells with diagonal stripes in range $R^1_{[3, 3] \times [0, 0]}$ (which equals $R^0_{[6, 7] \times [0, 1]}$ in the base grid) are added to the zone.

\subsection{Patch Assembly}
\label{sec:constructpatches}
Next, we focus on the process of assembling \textit{patches} at each level $k$ of the data domain hierarchy.
A patch is a set of cells that can be combined with existing zone cells to reduce the number of non-wildcard elements in a search token. 
We denote the cells belonging to a patch as \textit{attached cells}, and the zone cells adjacent to the patch as \textit{attaching cells}.
A patch is associated with a local \textit{cost} and \textit{gain}: the cost measures the increase in alert zone area, whereas the gain quantifies the resulting reduction in bilinear pairing operations when the patch is added to the zone.

We consider as patch candidate each $2 \times 2$ cell\footnote{We emphasize that, as patch candidates are considered at each level of the hierarchy, a patch cell may include many base grid cells.} $R^k_{[x, x + 1] \times [y, y + 1]}$ that satisfies the following conditions:
{\em (i)} has even $x$ and $y$ coordinates, {\em (ii)} contains at least one zone cell, and {\em (iii)} has at least one non-zone cell.
Revisiting the example in Figure~\ref{fig:expanding1_k0}, the area $R^0_{[4, 5] \times [0, 1]}$ composed of $2 \times 2$ base cells is a patch candidate. 
Note that, not all $2 \times 2$ cell areas are valid candidates for patches. For instance, $R^0_{[5, 6] \times [0, 1]}$ has an odd $x$; $R^0_{[6, 7] \times [0, 1]}$ does not contain any zone cell; and $R^0_{[4, 5] \times [2, 3]}$ does not contain any non-zone cell.

\begin{figure}	
	\centering
	\subfloat[]{
		\includegraphics[width=0.25 \columnwidth]{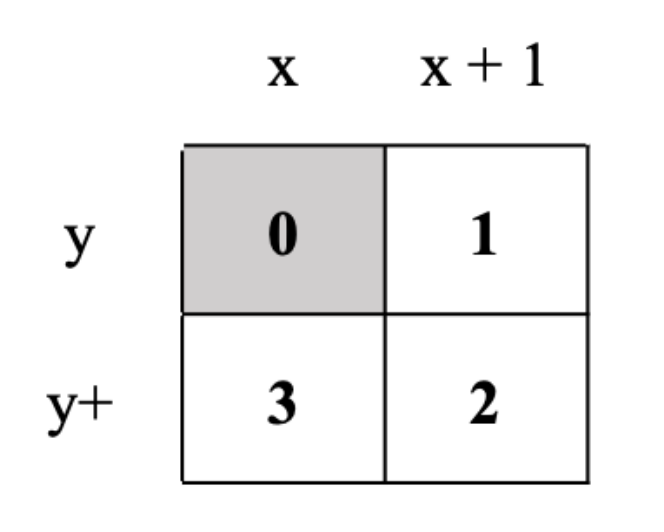}
		\label{fig:2by2_index}}
	\hspace{\fill}
	\subfloat[]{
		\includegraphics[width=0.4 \columnwidth]{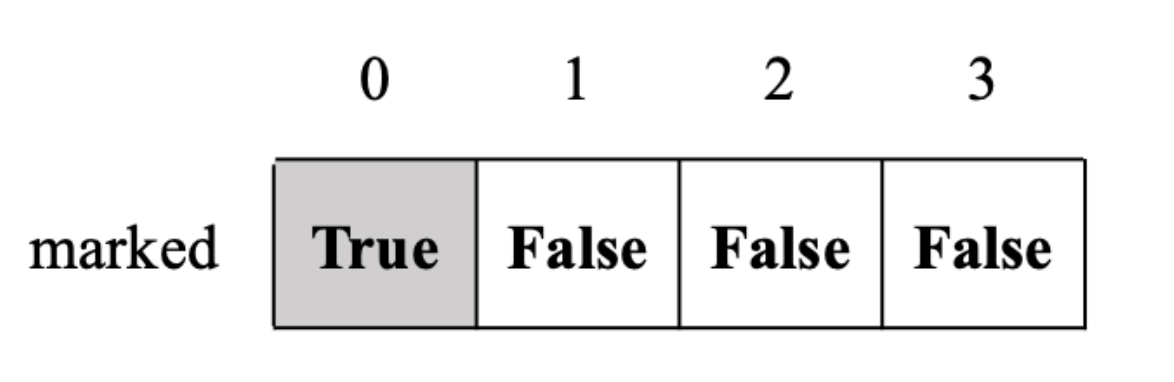}
		\label{fig:marked_eg}  }
	
	\caption{(a) Cell indices within a $2 \times 2$ patch; (b) Array $marked$ for a $2 \times 2$ patch.}
	\label{fig:2by2_marked}
\vspace{-20pt}
\end{figure}

For each valid patch candidate $R^k_{[x, x + 1] \times [y, y + 1]}$, cells are indexed in a spiral order, as shown in Figure~\ref{fig:2by2_index}. We use this indexing order because it simplifies the process of patch assembly, as will be described later in Section~\ref{sec:selectpatches}.
In order to keep track of zone and non-zone cells, a boolean array $marked$ is maintained, such that $\mathit{marked}[i] = \mathit{True}$ if $i^{th}$ cell within $R^k_{[x, x + 1] \times [y, y + 1]}$ is a zone cell, and  $\mathit{marked}[i] = \mathit{False}$, otherwise.
Figure~\ref{fig:marked_eg} shows a $\mathit{marked}$ array for the area in Figure~\ref{fig:2by2_index}.
The $\mathit{marked}$ array is constructed by checking for each cell within the area whether or not it belongs to the alert zone. 
The marking procedure is summarized in Algorithm~\ref{alg:MarkZoneCells}.

\vspace{-20pt}

\begin{figure}
\vspace{-20pt}
	\centering
	\hspace{\fill}
	\subfloat[]{
		\includegraphics[width=0.2 \columnwidth]{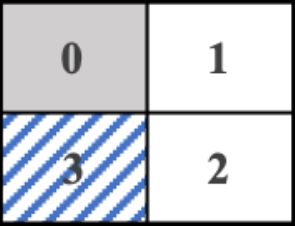}
		\includegraphics[width=0.2 \columnwidth]{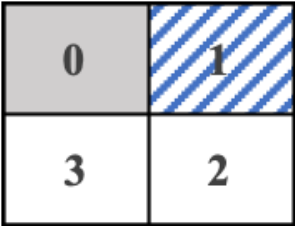}
		\includegraphics[width=0.2 \columnwidth]{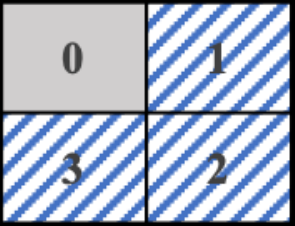}
		\label{fig:patch2by2-1}}
	\hspace{\fill}
	
	\hspace{\fill} 
	\subfloat[]{ 
		\includegraphics[width=0.2 \columnwidth]{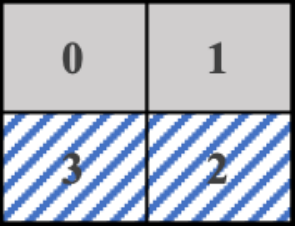}
		\label{fig:patch2by2-2}  }
	\hspace{\fill}
	\subfloat[]{
		\includegraphics[width=0.2 \columnwidth]{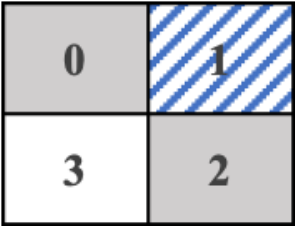}
		\includegraphics[width=0.2 \columnwidth]{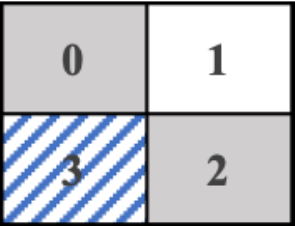}
		\label{fig:patch2by2-3}  }
	\hspace{\fill}
	\subfloat[]{
		\includegraphics[width=0.2 \columnwidth]{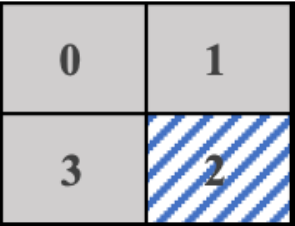}
		\label{fig:patch2by2-4}  }
	\hspace{\fill}
	
	\caption{Example of candidate patches for an area of size $2 \times 2$
	\protect\subref{fig:patch2by2-1} Three candidate patches for an area with one zone cell;
	\protect\subref{fig:patch2by2-2} One candidate patch for an area with two zone cells;
	\protect\subref{fig:patch2by2-3} Two candidate patches for an area with two zone cells; 
	\protect\subref{fig:patch2by2-4} One candidate patch for an area with three zone cells.}
	\label{fig:patch2by2}
	\vspace{-20pt}
\end{figure}

Using the $marked$ array, patch candidates are constructed such that one or more non-zone cells can be attached to zone cells to reduce the number of pairings. 
Figure~\ref{fig:patch2by2} shows several examples of patches for an area with $2 \times 2$ cells containing one, two, or three zone cells.
In each example, the non-zone cell (striped fill) is attached to the zone cell (grey fill) to form a patch. Note that in Figure~\ref{fig:patch2by2-3}, a striped cell can be attached to either grey cell.

\begin{algorithm} 
	\caption{MarkZoneCells} 
	\label{alg:MarkZoneCells} 

	\KwIn{zone cells $\mathit{A}$; start values $x$ and $y$ of area $R^k_{[x, x + 1] \times [y, y + 1]}$}
	\KwOut{boolean array indicating which cells in $2 \times 2$ block are zone cells}
	
	Initiate $\mathit{marked}$ array with $\mathit{marked}[i] = \mathit{False}, \forall i \in [0, 3]$ 
	
	\ForEach{$i \in [0, 3]$}{
		\If{ $i^{th}$ cell of $R^k_{[x, x + 1] \times [y, y + 1]} \in A$ }{
			$\mathit{marked}[i] \leftarrow True $
		}
	}
	
	\Return $\mathit{marked}$
\end{algorithm}

However, when the area contains only one zone cell, although there are three potential patches, only one of these is selected (Figure~\ref{fig:patch2by2-1}). 
The reason is that if two patches, each having a single zone cell, are selected, the number of pairings is not reduced; on the other hand, if the patch with three zone cells is selected, there is no need to select other patches with a single zone cell. 
Therefore, for each area, we construct \textit{patch groups} that include all potential patches such that no more than one patch can be selected from that group.
For example, in Figure~\ref{fig:patch2by2-1}, there is only one patch group containing all three patches; in Figure~\ref{fig:patch2by2-2} and \ref{fig:patch2by2-4}, there is only one patch group containing one patch; in Figure~\ref{fig:patch2by2-3}, there are two patch groups, each containing one patch.

The {\tt GetPatchGroupsInsideArea} routine (Algorithm~\ref{alg:GetPatchGroupsInsideArea}) shows the details of constructing patches and patch groups.
The algorithm handles separately each case based on the number of zone cells in the area. 
For a single zone cell (line~\ref{algo:line:getPatchGroupCase1}), similar to the example in Figure~\ref{fig:patch2by2-1}, one patch group is constructed which includes two patches: one with one non-zone cell and another with all three non-zone cells.
If there are two zone cells (line~\ref{algo:line:getPatchGroupCase2}), the algorithm further considers if those two zone cells are adjacent or opposite (similar to Figure~\ref{fig:patch2by2-2} and \ref{fig:patch2by2-3}, respectively) and either one or two patch groups are created, corresponding to the two situations.
Finally, when there are three zone cells (line~\ref{algo:line:getPatchGroupCase3}), a single patch group is created.
\vspace{-20pt}

\begin{algorithm} 
	\caption{GetPatchGroupsInsideArea} 
	\label{alg:GetPatchGroupsInsideArea} 

	\KwIn{array $\mathit{marked}$ for $2 \times 2$ cell block}
	\KwOut{Patch groups with cell ids 0, 1, 2, or 3}
	
	$\mathcal{G}_{\mathit{inside}} \leftarrow \emptyset$
	
	$\mathit{num\_zone\_cells} \leftarrow$ number of 1's in $\mathit{marked}$

	\If{$\mathit{num\_zone\_cells} = 1$}{\label{algo:line:getPatchGroupCase1}
		
		$p_1 \leftarrow$ new patch with an non-zone cell adjacent to the zone cell as \textit{attached\_cells} and the zone cell as \textit{attaching\_cells}
		
		$p_2 \leftarrow$ new patch with all three non-zone cells as \textit{attached\_cells}  and the zone cell as \textit{attaching\_cells}

		$\mathcal{G}_{\mathit{inside}} \leftarrow \{p_1, p_2\}$
		
	}
	\ElseIf{$\mathit{num\_zone\_cells} = 2$}{ \label{algo:line:getPatchGroupCase2}
		\If{two zone cells are adjacent}{
			$p \leftarrow$ new patch with all non-zone cells as \textit{attached\_cells}  and the zone cells as \textit{attaching\_cells}

			$\mathcal{G}_{\mathit{inside}} \leftarrow \{p\}$
		}
		\Else{
			$p_1 \leftarrow$ new patch with an non-zone cell adjacent to the first zone cell as \textit{attached\_cells} 
			
			$p_2 \leftarrow$ new patch with the other non-zone cell adjacent to the second zone cell as \textit{attached\_cells} 
			
			$G_1 \leftarrow \{p_1\}$
			
			$G_2 \leftarrow \{p_2\}$
			
			$\mathcal{G}_{\mathit{inside}} \leftarrow \{G_1, G_2\}$
		}
	}
	\ElseIf{$\mathit{num\_zone\_cells} = 3$}{ \label{algo:line:getPatchGroupCase3}
		$p \leftarrow$ new patch with all non-zone cells as \textit{attached\_cells}  and the zone cells as \textit{attaching\_cells}

		$\mathcal{G}_{\mathit{inside}} \leftarrow \{p\}$
	}
	
	\Return $\mathcal{G}_{\mathit{inside}}$
\end{algorithm}

At the end of Algorithm~\ref{alg:GetPatchGroupsInsideArea}, each patch has its cells numbered from set $\{0, 1, 2, 3\}$. 
In order to recover the original cell ids (i.e., the coordinates in current level $k$ of hierarchy), we use Algorithm~\ref{alg:RecoverOriginalCoordId},
which takes as inputs a cell id $i \in [0, 3]$ and the $x$, $y$ values of the area $R^k_{[x, x + 1] \times [y, y + 1]}$, and utilizes the 
spiral index to recover the original values.

\begin{algorithm} 
	\caption{RecoverOriginalCoordId} 
	\label{alg:RecoverOriginalCoordId} 
	
	\KwIn{cell id $i \in [0, 3]$; $x$, $y$ values of the area $R^k_{[x, x + 1] \times [y, y + 1]}$} 
	\KwOut{Original cell coordinates}
	
	\If{$i = 1$ $\mathbf{or}$  $i = 2$}{
		$x \leftarrow x + 1$
	}
	
	\If{$i = 2$ $\mathbf{or}$  $i = 3$}{
		$y \leftarrow y + 1$
	}
	
	\Return $(x, y)$

\end{algorithm}

Next, we need to evaluate which patches are more desirable to use in the enlarged zone, by computing the local {\em cost} and {\em gain} for each patch.
Algorithm~\ref{alg:CalCostGain} takes as inputs a candidate patch and the grid dimension $d_k$ at current level $k$. It outputs as cost the number of attached cells (i.e., non-zone cells) of the patch (line~\ref{algo:line:costgaincost}). Effectively, the cost measures the amount of enlargement of the expanded alert zone caused by this patch. The gain measures the amount of saved computation: specifically, the number of search token non-wildcards that are eliminated when we combine the attached cells with the attaching cells for the current patch.
There are two cases to consider when determining the gain of the patch: {\em (i)} when only one cell is attached to form a $1 \times 2$ patch (line~\ref{algo:line:costgain2}), we can remove one non-wildcard element (line~\ref{algo:line:costgain2Gain1}); {\em (ii)} when the entire $2\times 2$ area is filled (line~\ref{algo:line:costgain4}), the number of zone cells inside the area (i.e., $n_1$) is further considered to determine the gain. Specifically, when $n_1 = 3$, the gain is larger ($2 \times k$) since we can remove a token in its entirety.

\vspace{-20pt}

\begin{algorithm} 
	\caption{CalculateCostGain} 
	\label{alg:CalCostGain} 

	\KwIn{patch $p$; grid length at current level $k$;}
	\KwOut{updated patch with cost and gain calculated}
	
	$n_1 \leftarrow$ number of zone cells of $p$
	
	$n_2 \leftarrow$ number of non-zone cells of $p$
	
	$p.\mathit{cost} \leftarrow n_2$ \label{algo:line:costgaincost}
	
	\If{$n_1 + n_2 = 2$}{ \label{algo:line:costgain2}
		$p.\mathit{gain} \leftarrow 1$ \label{algo:line:costgain2Gain1}
	}
	\ElseIf{$n_1 + n_2 = 4$}{ \label{algo:line:costgain4}
		\If{$n_1 = 1$}{
			$p.\mathit{gain} \leftarrow 2$
		}
		\ElseIf{$n_1 = 2$}{
			$p.\mathit{gain} \leftarrow 1$
		}
		\Else{ \label{algo:line:costgainN1_3}
			$p.\mathit{gain} \leftarrow 2 \times k$ \label{algo:line:costgainLogL}
		}
	}
	
	\Return $p$
	
\end{algorithm}

In the previous example from Figure~\ref{fig:expanding1_k0}, there are three patch groups corresponding to three areas: $G_1$ for area $R^0_{[4, 5] \times [0, 1]}$, $G_2$ for area $R^0_{[6, 7] \times [2, 3]}$, and $G_3$ for area $R^0_{[4, 5] \times [4, 5]}$.
The patches in each patch group along with their cost, gain, attaching cells, and attached cells are shown in Table~\ref{tbl:patchgroupsexample}.
For instance, to express area $R^0_{[4, 5] \times [0, 1]}$ one can look at patches $p_1$ of $G_1$, and use two tokens ``00*110" and ``00111*", with a total of $10$ non-wildcard elements.
By adding one cell, only a single token ``00*11*" is needed to represent the area.
Thus, the number of non-wildcard elements is reduced from $10$ to $4$, or an improvement of $6$. 
The high gain when applying patch $p_1$ results not only from the number of non-wildcards reduced in one token, but also from the reduction in the number of tokens (as one of the initial tokens is completely eliminated).

\begin{table}[]
	\centering
	\begin{tabular}{|l||l|r|r|r|r|}
		\hline
		\textbf{Patch Group} & \textbf{Patch} & $\mathit{cost}$ & $\mathit{gain}$ & $\mathit{attached\_cells}$ & $\mathit{attaching\_cells}$ \\ \hline
		\hline
		$G_1$                   & $p_1$             & $1$             & $6$             & $(5, 0)$                  & $(4, 0), (4, 1), (5, 1)$   \\ \hline
		\multirow{3}{*}{$G_2$}  & $p_2$             & $1$             & $1$             & $(6, 2)$                  & $(6, 3)$                   \\ \cline{2-6} 
		& $p_3$             & $1 $            & $1$             & $(7, 3)$                  & $(6, 3) $                 \\ \cline{2-6} 
		& $p_4$             & $3$             & $2$            & $(7, 2), (7, 3), (6, 2)$  & $(6, 3)$                   \\ \hline
		$G_3$                   & $p_5$             & $2$             & $1$             & $(4, 5), (5, 5) $         & $(4, 4), (5, 4)$           \\ \hline
	\end{tabular}
	\caption{Example of candidate patch groups for expanding the alert zone at level $k = 0$. Patch groups $G_1, G_3$ contain one patch, while patch group $G_2$ contains 3 patches. No more than one patch is selected per patch group.}
	\label{tbl:patchgroupsexample}
\vspace{-20pt}
\end{table}

\figspa
\subsection{Patch Selection}
\label{sec:selectpatches}

Once we have the set of patches and patch groups, as well as their respective costs and gains, we need a method to select the actual patches to expand the current alert zone. Algorithm~\ref{alg:SelectPatchesSingleLevel} outlines the patch selection process, which takes as inputs budget $W$, the grid dimension at current level $d_k=\frac{d}{2^k}$, and current alert zone $A_k$. It outputs a set of patches $\mathit{PSet}$ that has total cost at most $W$ and maximizes the gain compared to other candidate patches.

The selection algorithm works within an expanding search boundary determined by the call to routine {\tt FindExpandingBoundary} in line~\ref{algo:line:findexpandingboundary} ({\tt FindExpandingBoundary} is summarized in Algorithm~\ref{alg:FindQueryBoundary}: the boundaries consist of the maximum and minimum coordinate values of zone cells, and they always have even values). 
Then, for each $2 \times 2$ area starting with even values (lines [\ref{algo:line:2x2areastartx}-\ref{algo:line:2x2areastarty}] in Algorithm~\ref{alg:SelectPatchesSingleLevel}), 
if the current $2 \times 2$ area is a valid area to expand (line~\ref{algo:line:hasemptycells}), the patches and patch groups within this area are constructed (according to the procedure detailed in Section~\ref{sec:constructpatches}).
First, the cells that already belong to the current zone are marked by calling Algorithm~\ref{alg:MarkZoneCells} (line~\ref{algo:line:callMarkQueryCell}).
Then, using the marking information, the set $\mathcal{G}_{\mathit{inside}}$ of \textit{patch groups} within that area is constructed by calling Algorithm~\ref{alg:GetPatchGroupsInsideArea} (line~\ref{algo:line:callGetPatchGroups}).
Next, for each patch in the patch groups of $\mathcal{G}_{\mathit{inside}}$, the original coordinate ids of cells in the attaching and attached set of that patch are recovered by calling Algorithm~\ref{alg:RecoverOriginalCoordId} (line~\ref{algo:line:RecoverCellValue}), and the local \textit{cost} and \textit{gain} are calculated by calling Algorithm~\ref{alg:CalCostGain} (line~\ref{algo:line:callCalCostGain}).
Finally, a set of patches $\mathit{PSet}$ is selected for expansion by calling Algorithm~\ref{alg:knapsack} (line~\ref{algo:line:callKnapsack}).

\begin{algorithm} 
	\caption{SelectPatchesSingleLevel} 
	\label{alg:SelectPatchesSingleLevel} 
	\SetKwFunction{FindExpandingBoundary}{FindExpandingBoundary}
	\SetKwFunction{MarkQueryCells}{MarkQueryCells}
	\SetKwFunction{GetPatchGroupsInsideArea}{GetPatchGroupsInsideArea}
	\SetKwFunction{RecoverOriginalCoordId}{RecoverOriginalCoordId}
	\SetKwFunction{CalCostGain}{CalculateCostGain}
	\SetKwFunction{KnapsackForGroups}{KnapsackForGroups}	
	\KwIn{budget $W$; grid dimension $d_k$ at level $k$; zone cells $A_k$ }
	\KwOut{patches with total cost $\leq W$ and positive gain}
	
	$\mathit{min\_x}, \mathit{max\_x}, \mathit{min\_y}, \mathit{max\_y} \leftarrow$ {\FindExpandingBoundary{$d_k, A_k$}}  \label{algo:line:findexpandingboundary}
	
	$\mathcal{G} \leftarrow \emptyset $
	
	$x \leftarrow \mathit{min\_x}$ \label{algo:line:2x2areastartx}
	
	\While{$x < \mathit{max\_x}$}{
		
		$y \leftarrow \mathit{min\_y}$
		
		\While{$y < \mathit{max\_y}$}{\label{algo:line:2x2areastarty}
			
			$c \leftarrow$ number of zone cells in this area $R^k_{[x, x + 1] \times [y, y + 1]}$
			
			\If{$0 < c < 4$} {\label{algo:line:hasemptycells}
				
				$\mathit{marked} \leftarrow$ {\MarkQueryCells{$A_k, x, y$}} \label{algo:line:callMarkQueryCell}
				
				$\mathcal{G}_{\mathit{inside}} \leftarrow$ {\GetPatchGroupsInsideArea{$\mathit{marked}$}} \label{algo:line:callGetPatchGroups}
				
				\For{$G \in \mathcal{G}_{\mathit{inside}}$}{
					\For{$p \in G$}{
						Recover original coordinate ids of each cell $i$ in $\mathit{attaching\_cells}$ and $\mathit{attached\_cells}$ using {\RecoverOriginalCoordId{$i, x, y$}} \label{algo:line:RecoverCellValue}

						Update $p \leftarrow$ {\CalCostGain{$p, d_k$}} \label{algo:line:callCalCostGain}
					}
				
					$\mathcal{G} \leftarrow \mathcal{G} \cup G $
					
				}
			}
		}
	}

	$\mathit{PSet} \leftarrow $ {\KnapsackForGroups{$W, \mathcal{G}$}} \label{algo:line:callKnapsack}
	
	\Return $\mathit{PSet}$
		
\end{algorithm}

\begin{algorithm} 
	\caption{FindExpandingBoundary} 
	\label{alg:FindQueryBoundary} 

	\KwIn{Grid length $d_k$; zone cells $A_k$ }
	\KwOut{the boundary for expansion}

	($\mathit{min\_x}, \mathit{max\_x}, \mathit{min\_y}, \mathit{max\_y}$) = minimum bounding rectangle of $A_k$ expanded to even coordinates
			
	\Return $\mathit{min\_x}, \mathit{max\_x}, \mathit{min\_y}, \mathit{max\_y}$
\end{algorithm}

The patches are selected such that the total cost is no more than $W$, the total gain is maximized, and there is no more than one patch selected per group.
This can be modeled as a variant of a multiple-choice knapsack problem (MCKP) where a class in MCKP is represented by a patch group, and we may choose a single item from a class, instead of being required to choose at least one item.
The reduction is as follows: Given an instance of MCKP with capacity $W$, $m$ classes, and each item $j$ in class $j$ having cost $c_{i,j}$ and gain $g_{i, j}$, for each class $i$, a new item $j'$ (or patch in our setting) is added with cost $c_{i,j'} = 0$ and gain $g_{i, j'} = 0$.
However, our patch selection problem is not NP-hard, because $W$ is restricted to a fraction of the alert zone, which in turn is restricted to a fraction of the entire grid.

In our setting, each patch group contain either one or two patches. As a result, a dynamic programming approach for traditional binary knapsack problem can be used.
Algorithm~\ref{alg:knapsack} shows the dynamic programming solution that returns the selected patches for expansion.
In the example summarized in Figure~\ref{fig:expanding1} and Table~\ref{tbl:patchgroupsexample}, patches $p_1, p_4, p_5$ are selected for expansion at level $k=0$.

\begin{algorithm} 
	\caption{KnapsackForGroups} 
	\label{alg:knapsack} 

	\KwIn{capacity $W$; groups of $N_g$ patches $\mathcal{G} = \{G_1, G_2, \dots, G_{N_g}\}$; }
	\KwOut{patches with total cost $\leq W$, total gain is maximized, and each patch belongs to a different group}
	
	$K$ $\leftarrow$ matrix of size $(W+1) \times (N_g + 1)$ 
	
	\For{$i \in  [0, N_g]$}{
		\For{$w \in [0, W]$}{
			\If{$i = 0$ $\mathbf{or}$  $w = 0$}{
				$K[i][w] \leftarrow 0$
			}
			\Else{
				$K[i][w] \leftarrow K[i - 1][w]$
				
				\For{$p \in G_i$}{
					\If{$\mathit{p.cost} \leq w$}{
						$K[i][w] \leftarrow \max(K[i][w], \mathit{p.gain} + K[i-1][w - \mathit{p.cost}])$
					}
				}
			}
		}
	}
	
	$\mathit{PSet}$ $\leftarrow \emptyset$
	
	$i \leftarrow N_g$
	
	$w \leftarrow W$
	
	\While{$0 < K[i][w]$} {
		\For{$p \in G_i$}{
			\If{$\mathit{p.cost} \leq w$ $\mathbf{and}$  $K[i][w] = \mathit{p.gain} +  K[i-1][w - \mathit{p.cost}]$}{
				
				$\mathit{PSet}$ $\leftarrow \mathit{PSet} \cup \{p\}$
				
				$w \leftarrow w - \mathit{p.cost}$
			}
		}
	}
	
	\Return $\mathit{PSet}$
		
\end{algorithm}

\subsection{Complexity Analysis}
The complexity of the alert zone expansion (Algorithm~\ref{alg:ExpandQuery}) depends on the complexity of the binary minimization step (line~\ref{algo:line:endCheckExpansion}) in which the algorithm decides whether or not to continue expansion. 
In the worst case, Algorithm~\ref{alg:ExpandQuery} needs to expand through all $\log d$ levels of the hierarchy, and in each level its invokes Algorithm~\ref{alg:SelectPatchesSingleLevel} and the binary minimization procedure (in our implementation, we use the Espresso tool~\cite{espresso}).

Since Algorithm~\ref{alg:SelectPatchesSingleLevel} finds the patch groups within the boundary of the query, and the size of the alert zone is often much smaller than the size of the data domain, we formulate the complexity of Algorithm~\ref{alg:SelectPatchesSingleLevel} based on the alert zone size. 
Let $P_k = |A_k|$ be the number of cells of the alert zone at level $k$.
After finding patch groups, Algorithm~\ref{alg:SelectPatchesSingleLevel} invokes the dynamic programming solution in Algorithm~\ref{alg:knapsack} to select patches. 
In the worst case, the number of patch groups $N_g$ at level $k$ equals the number of cells $P_k$ of the alert zone. 
In our setting, there are only one or two patches in each patch group. Hence, the complexity of Algorithm~\ref{alg:knapsack} becomes $\mathcal{O}(N_g W) = \mathcal{O}(\alpha P_k^2)$ since $W = \alpha P_k$. Thus, the complexity of Algorithm~\ref{alg:SelectPatchesSingleLevel}  is $\mathcal{O}(P_k + \alpha P_k^2)$.

However, since the value of $P_k$ is divided by a factor of $4$ each time $k$ increases, the complexity of the alert zone expansion (Algorithm~\ref{alg:ExpandQuery}) becomes $\mathcal{O}(P_0 + \alpha P_0^2 +  T_{\mathit{Es}}((1 + \alpha) P_0) \log d)$ where $P_0$ is the size of the zone at the base level (i.e., original grid) and $T_{\mathit{Es}}(t)$ is the time to run the binary minimization procedure for $t$ inputs.

\begin{figure*}[tbh]
\vspace{-20pt}
\subfloat[Token generation]{
	\centering
	\includegraphics[width=0.33\textwidth]{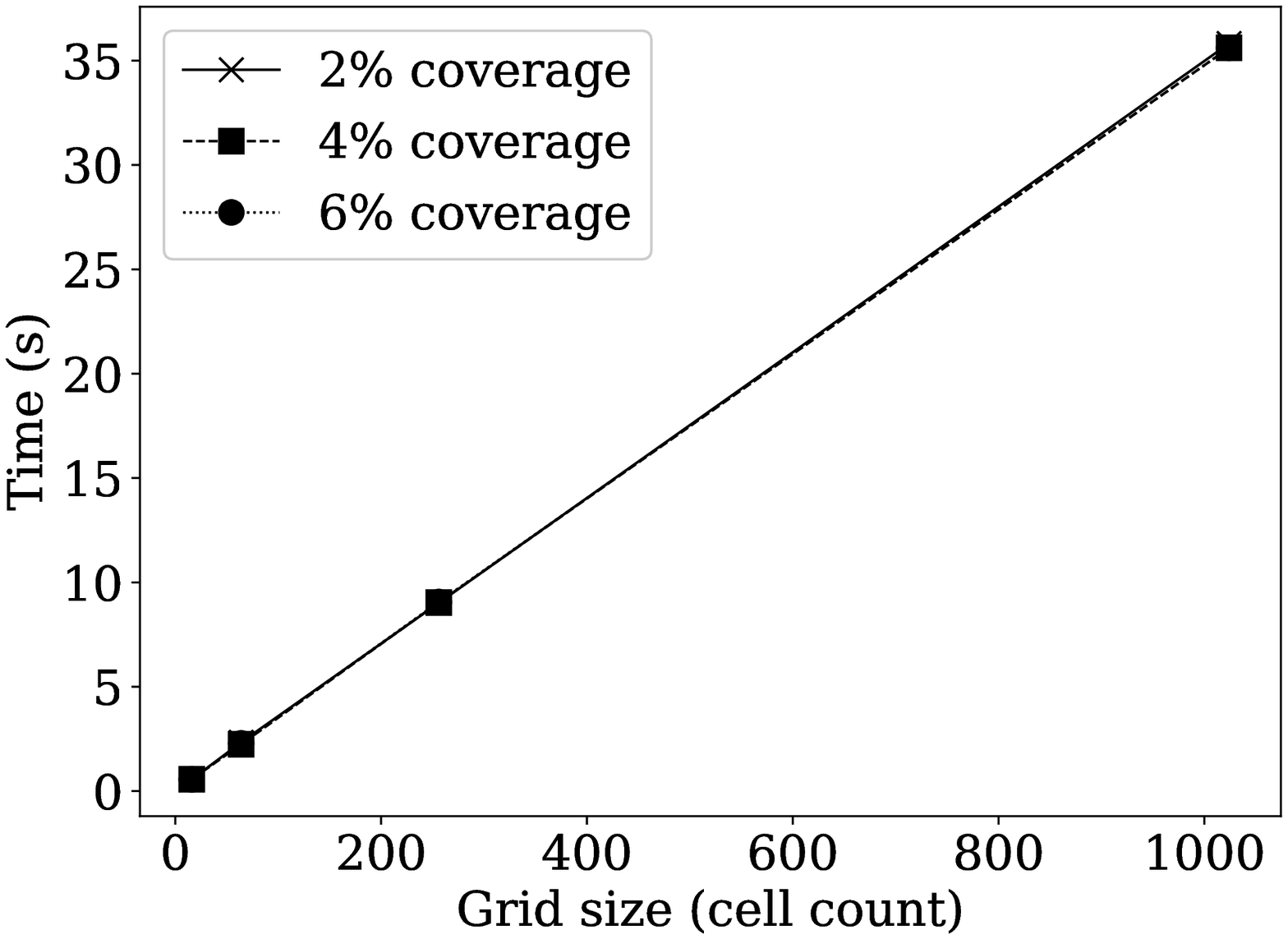}
        \label{fig:baseline_token}
}
\subfloat[Message encryption]{
	\centering
	\includegraphics[width=0.33\textwidth]{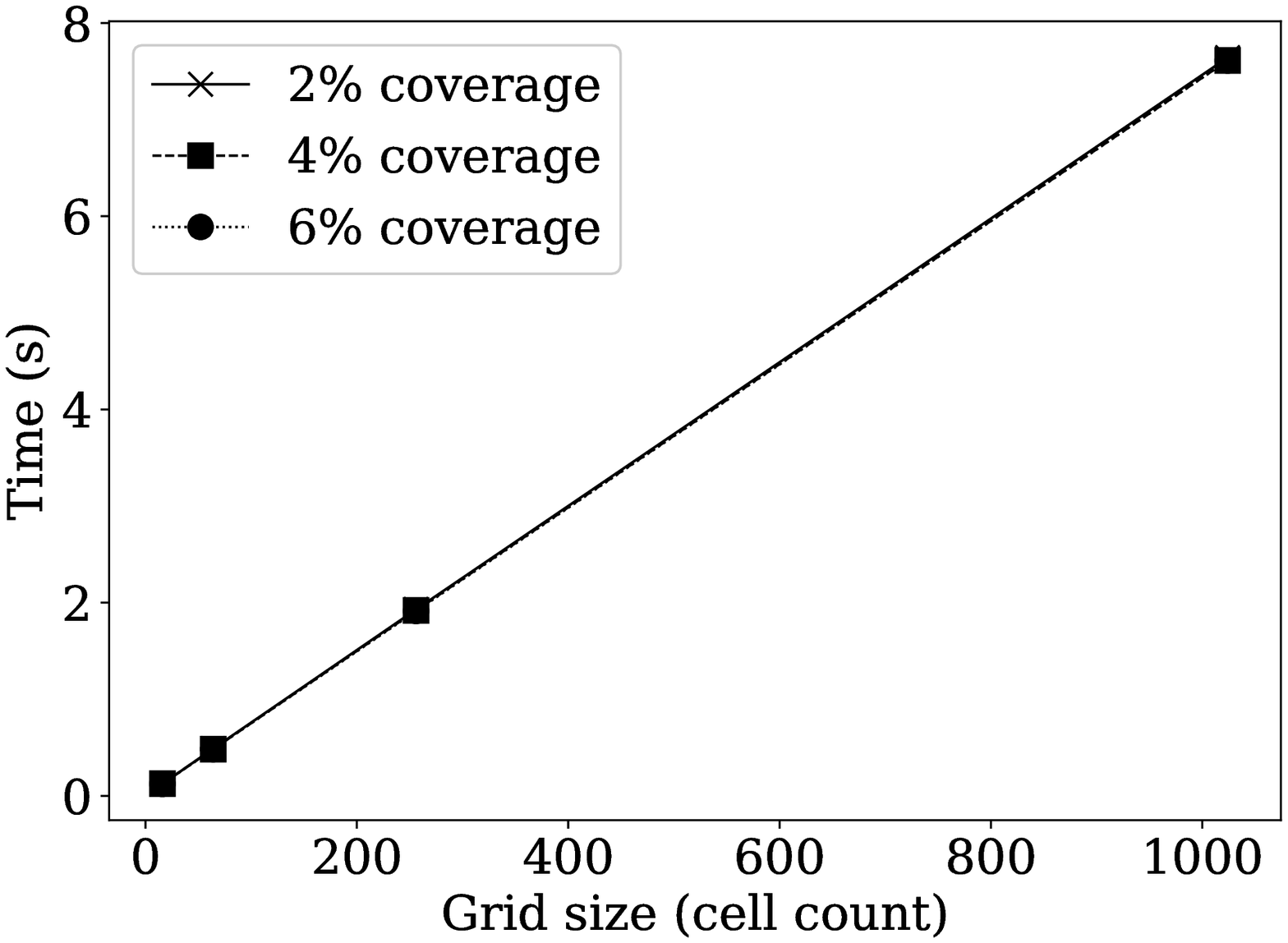}
        \label{fig:baseline_encrypt}
}
\subfloat[Query]{
	\centering
	\includegraphics[width=0.33\textwidth]{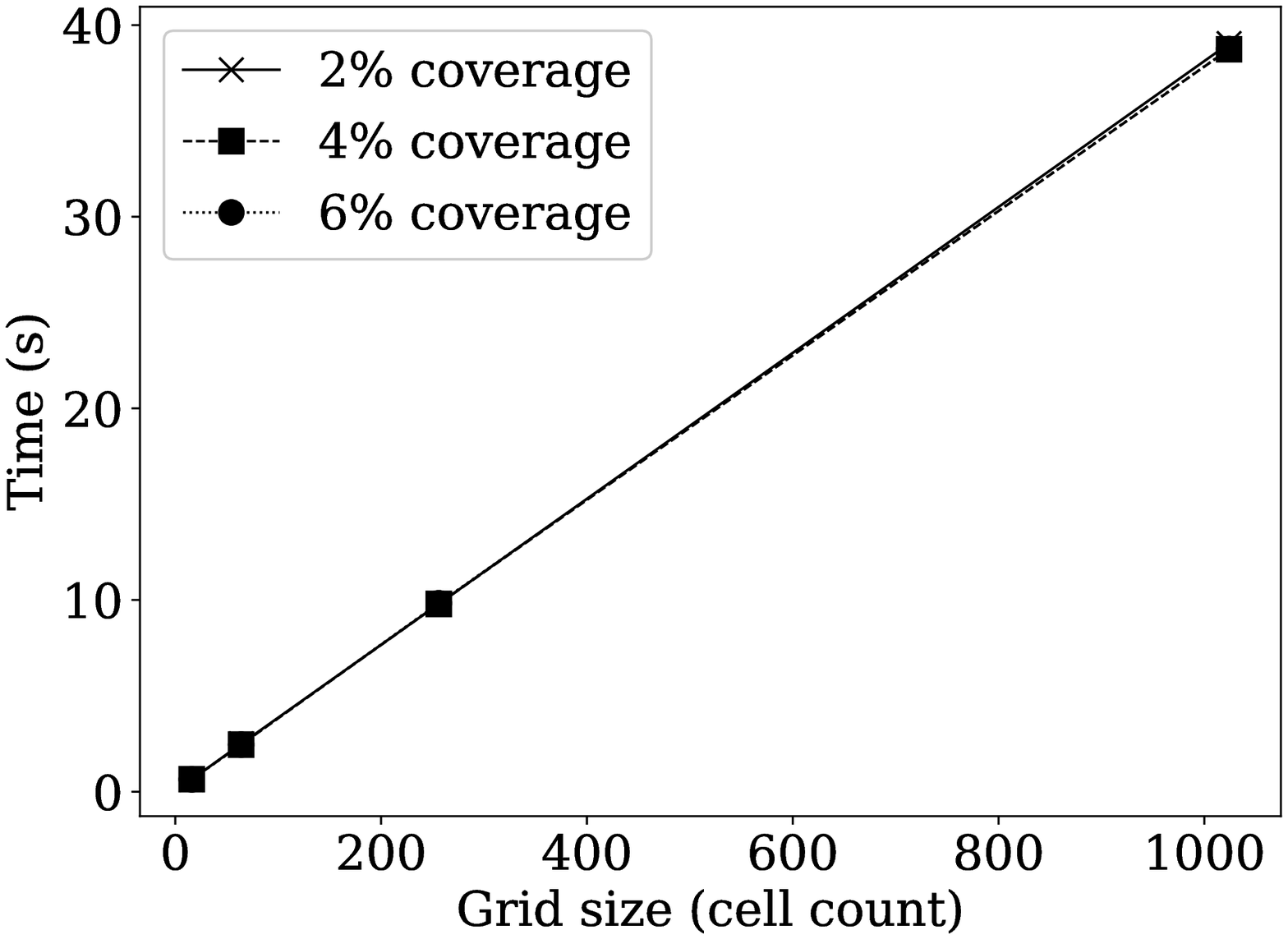}
        \label{fig:baseline_query}
}
\figspb
\caption{Baseline encoding results}
\figspb
\label{exp:baseline}
\subfloat[Token generation]{
	\centering
	\includegraphics[width=0.33\textwidth]{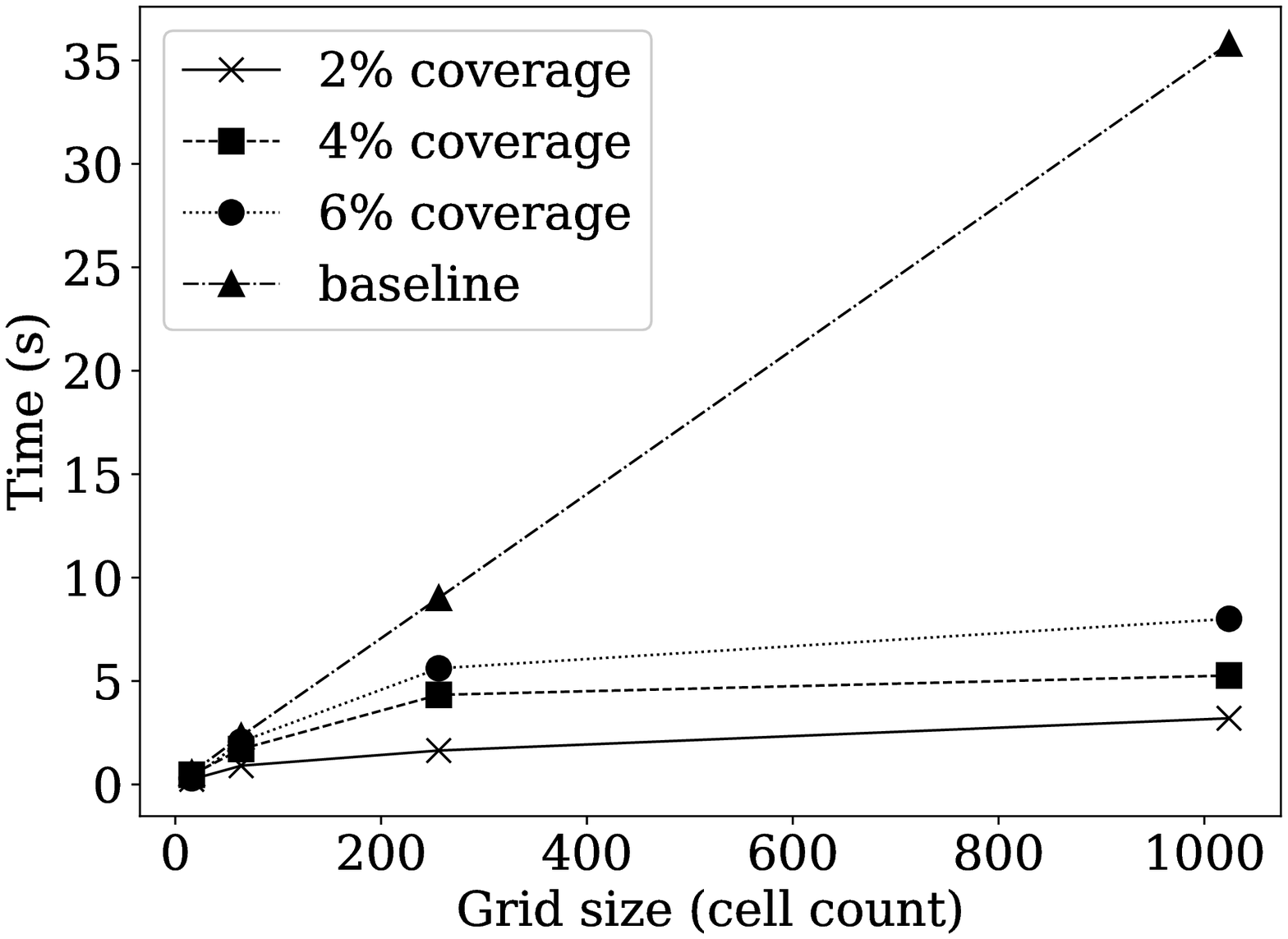}
        \label{fig:uniform_hierarchical_token}
}
\subfloat[Message encryption]{
	\centering
	\includegraphics[width=0.33\textwidth]{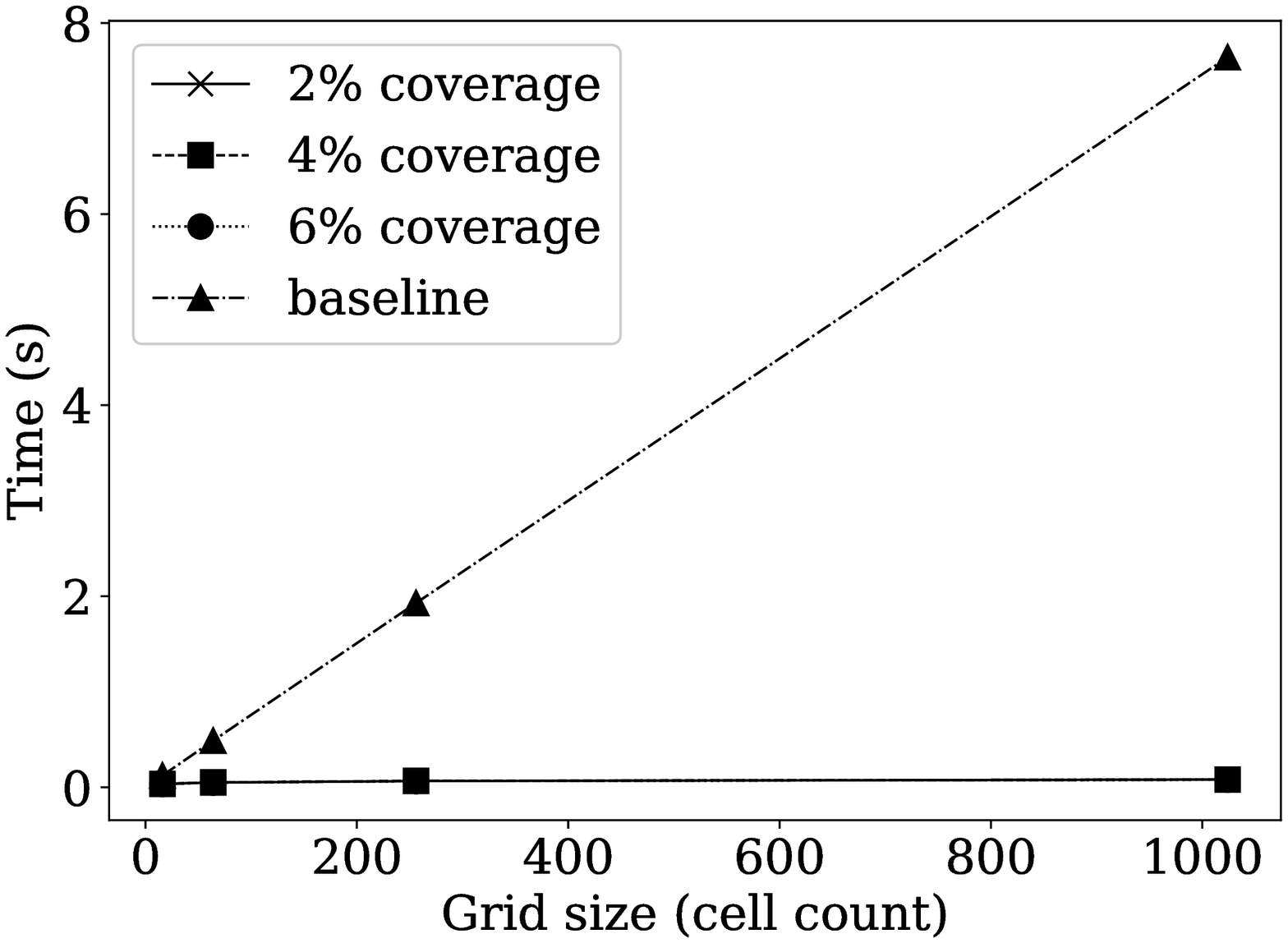}
        \label{fig:uniform_hierarchical_encrypt}
}
\subfloat[Query]{
	\centering
	\includegraphics[width=0.33\textwidth]{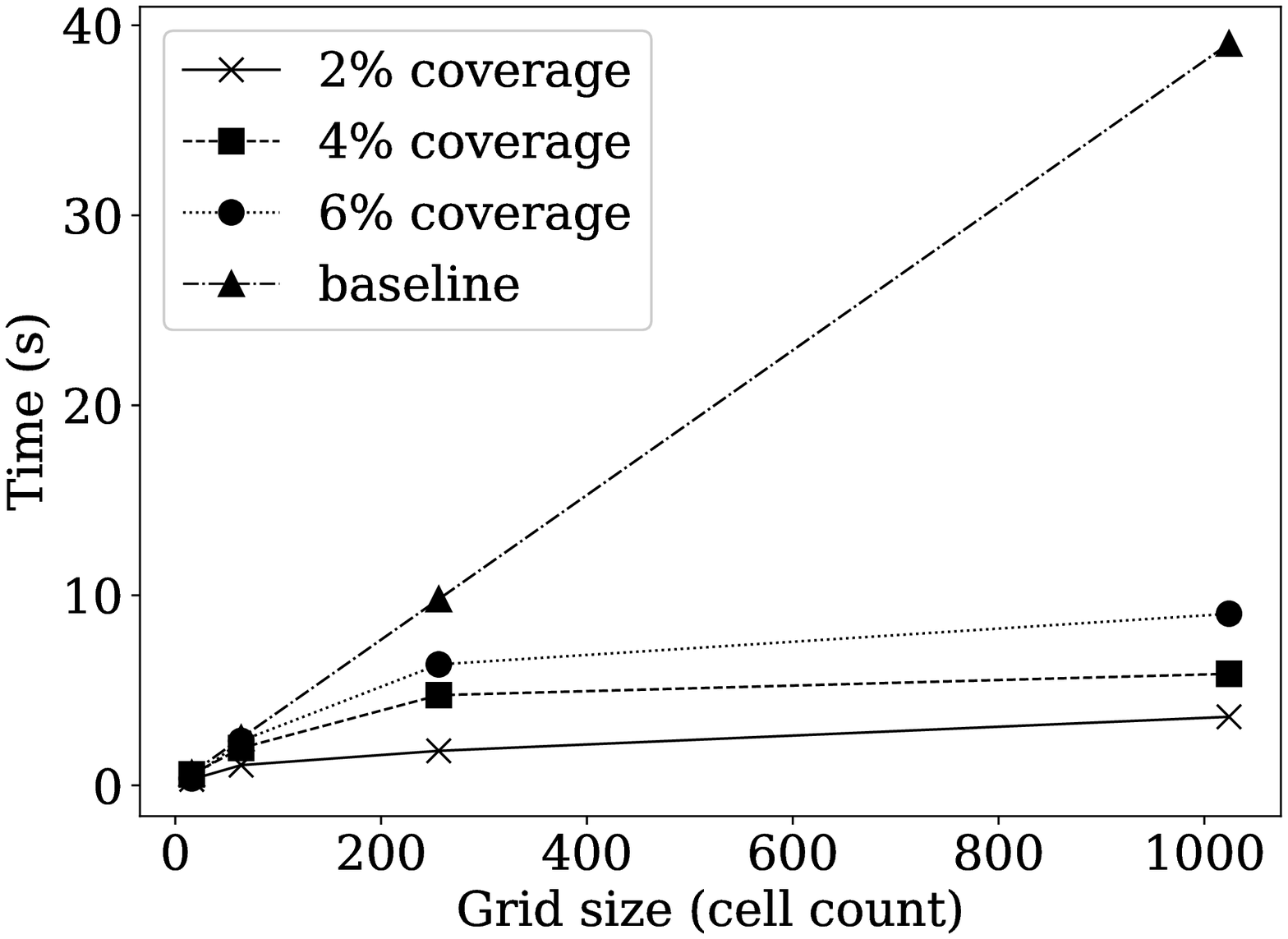}
        \label{fig:uniform_hierarchical_query}
}
\figspb
\caption{Hierarchical encoding results on uniform data}
\figspa
\label{exp:hier-unif}
\end{figure*}

\figspb
\section{Experimental Evaluation}
\label{sec:exps}

We implemented a Python prototype of the proposed HVE-based location-based alert system and performance optimizations. We have used as dataset the city of Oldenburg, and generated user movements using Brinkhoff's {\em IAPG Network-based Generator of Moving Objects\footnote{http://iapg.jade-hs.de/personen/brinkhoff/generator/}} \cite{Brinkhoff}. We generated alert zones within the boundaries of the dataset domain according to two distributions: uniform and Gaussian. We vary the percentage of space covered by alert zones compared to the entire dataspace extent from $1\%$ to $10\%$, and we denote this parameter as {\em coverage}. We consider a regular grid partitioning the two-dimensional space with size ranging from $16$ to $1024$. The HVE cryptographic functions were implemented using the Gnu MP v6.1.2 library and the Pairing-Based Cryptography v0.5.14 library\footnote{Available online at http://gmplib.org/ and http://crypto.stanford.edu/pbc/}. 
We use key lengths of 768, 1024 (default value), 1280 and 1536 bits. 
The experimental testbed consisted of a Intel(R) Core(TM) i9-9980XE CPU (3.00GHz) with $18$ cores and 128GB of RAM, running Ubuntu 18.04 LTS. All code was written in Python 3.6.9.

\vspace{-10pt}
\subsection{Baseline Evaluation}
\label{sec:exp:baseline}

\vspace{-5pt}
Figure~\ref{exp:baseline} shows the execution time results obtained for token generation, encryption and query. The times presented are for a single operation, and present the average value obtained for a particular grid size and percentage of the area covered by alert zones (each percentage value has a different line in the graphs). First, we note that the coverage does not have a significant effect on the execution time, because the width of the HVE obtained is so large that the associated overhead overshadows the influence of the additional '*' symbols obtained as the area of alert zones grows. Second, it can be observed that the values obtained are very large, and clearly not acceptable in practice. 

Token generation can take up to $35$ seconds. Although expensive, it can be argued that the TA does not execute this phase very often (only when a new alert zone occurs), hence its performance is not critical. However, encryption is very frequent, and it is executed at the resource-constrained mobile users. 
It can take up to $8$ seconds to generate a single encrypted update on a high-end CPU (in practice, this would be executed on a mobile phone). Furthermore, the time required at the server to process a single user update (i.e., perform matching against all alert zones) can reach $40$ seconds.

\begin{figure}[t]
\centering
\subfloat[Token generation]{
	\centering
	\includegraphics[width=0.4\textwidth]{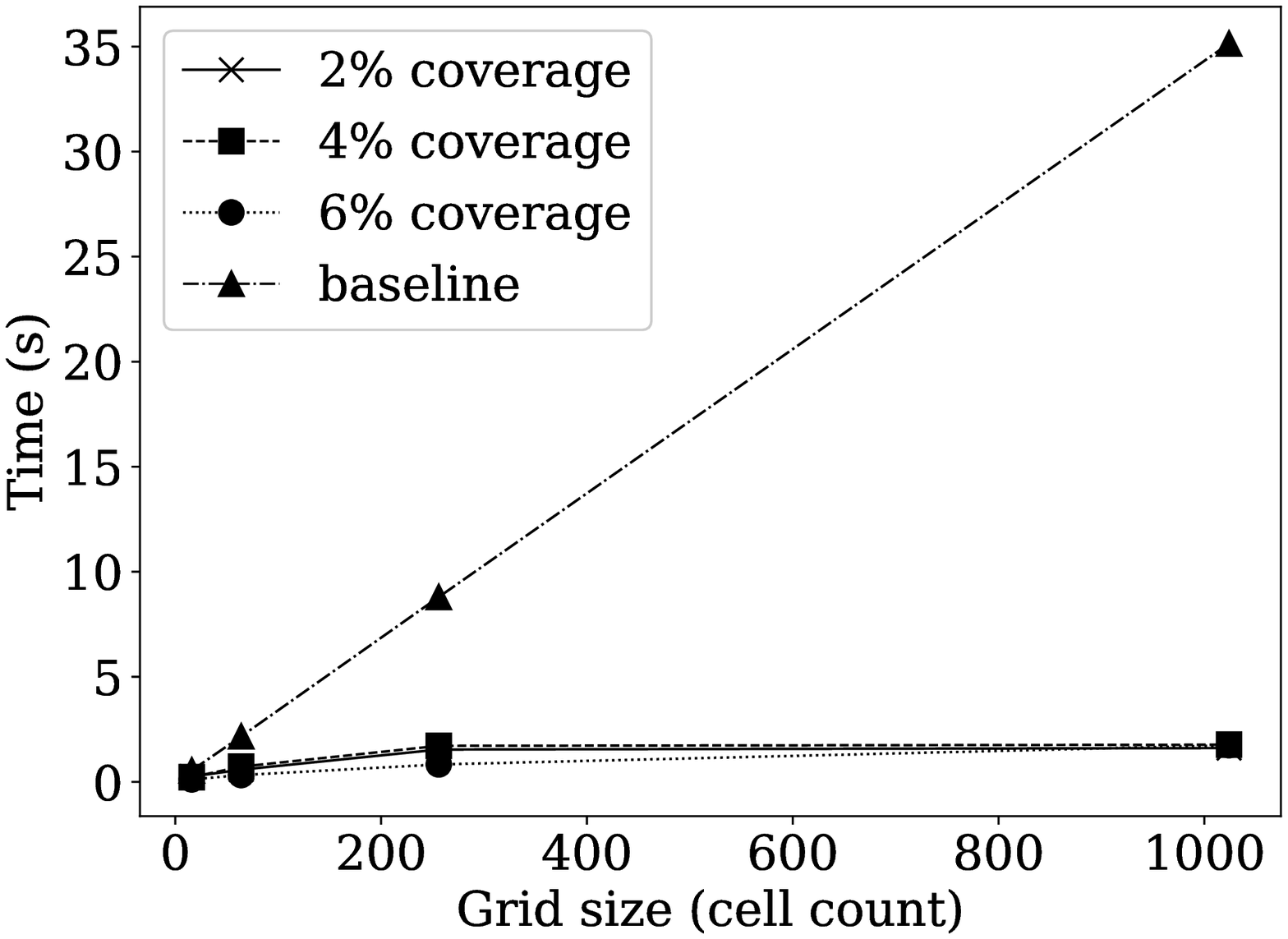}
        \label{fig:normal_hierarchical_token}
}
\subfloat[Query]{
\figspa
	\centering
	\includegraphics[width=0.4\textwidth]{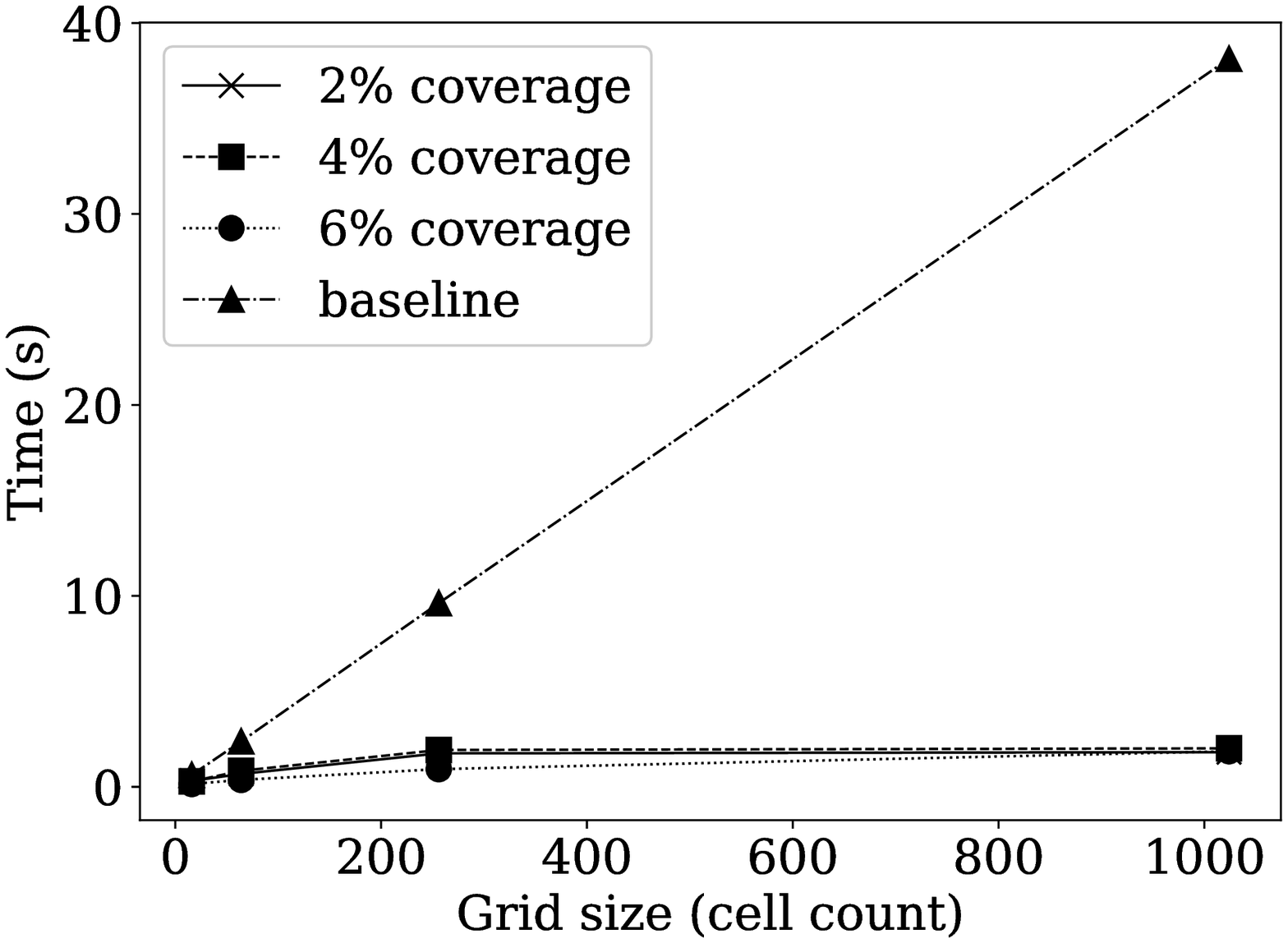}
        \label{fig:normal_hierarchical_query}
}
\figspb
\caption{Hierarchical encoding results on Gaussian data}
\figspa
\label{exp:hier-normal}
\end{figure}

\begin{figure}[t]
\centering
\subfloat[Token generation]{
	\centering
	\includegraphics[width=0.4\textwidth]{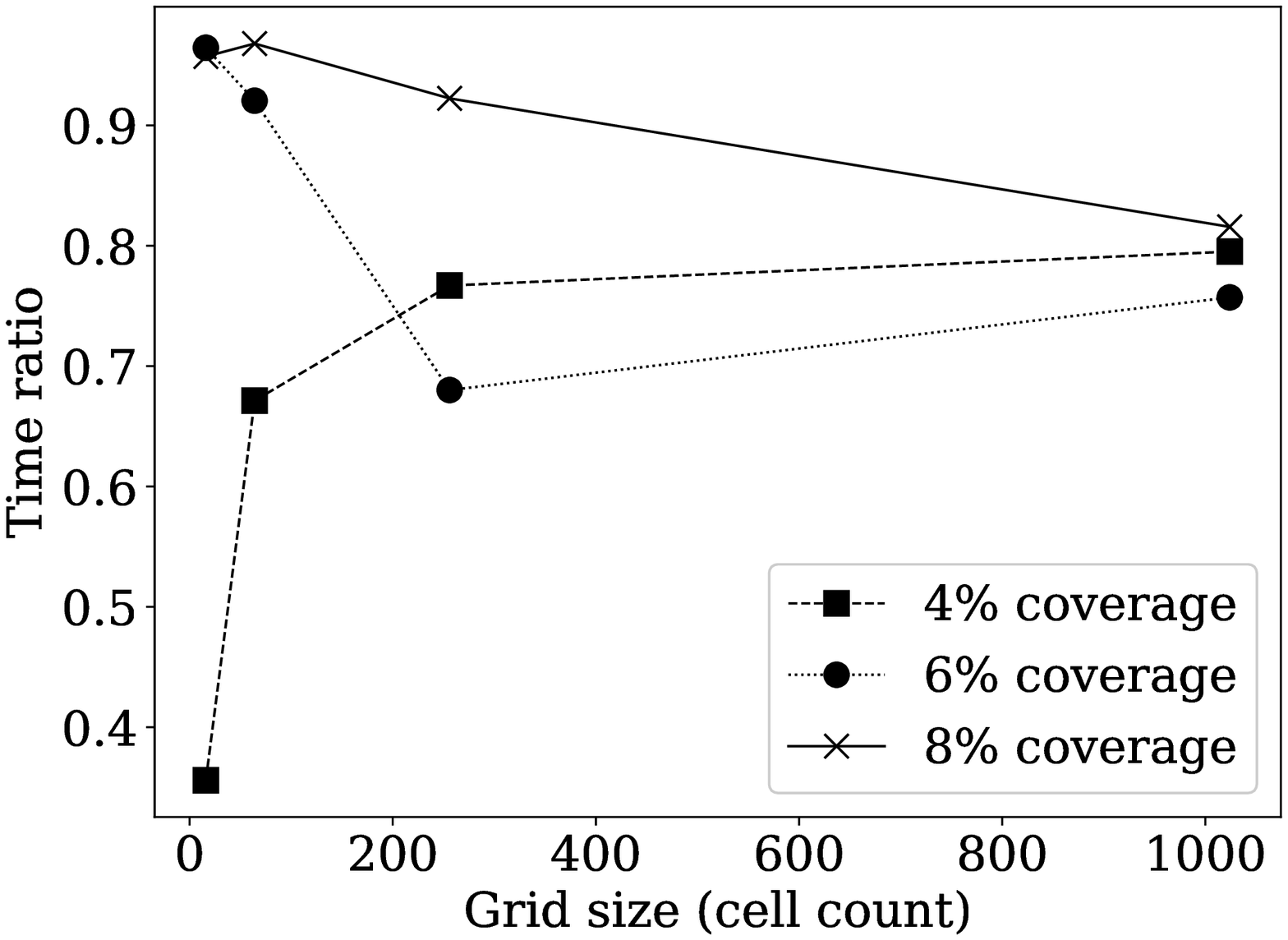}
        \label{fig:normal_graycode_token}
}\vspace{-0pt}
\subfloat[Query]{
	\centering
	\includegraphics[width=0.4\textwidth]{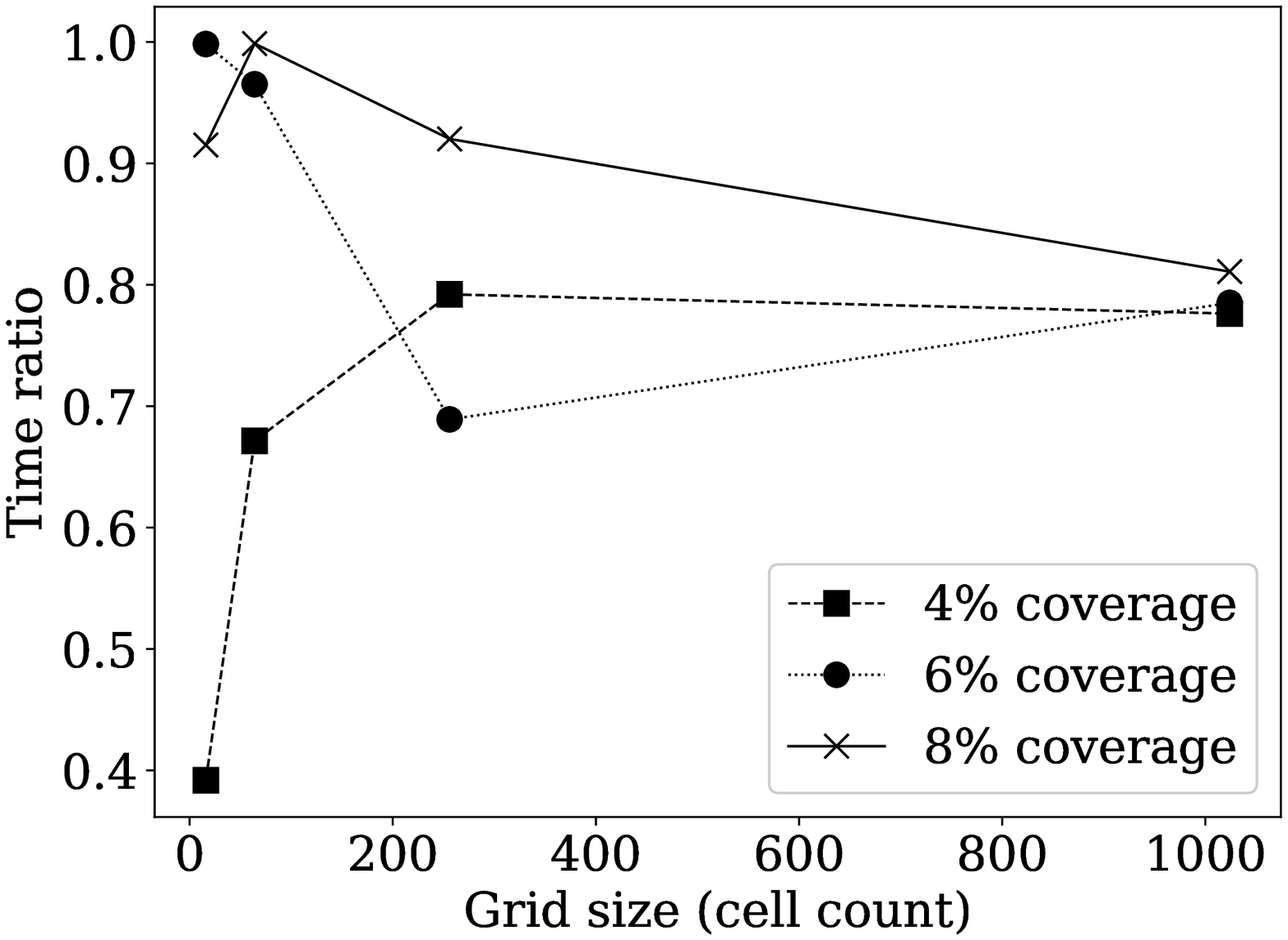}
        \label{fig:normal_graycode_query}
}\vspace{-0pt}
\figspb
\caption{Gray vs. hierarchical encoding on Gaussian data}
\figspa
\label{fig:normal_graycode_results}
\end{figure}
\vspace{-10pt}
\subsection{Hierarchical and Gray Encoding}
\label{sec:exp:hier}

Figures~\ref{exp:hier-unif} and~\ref{exp:hier-normal} show the comparison results for uniform and Gaussian alert zone distributions, respectively. Hierarchical encoding clearly outperforms the baseline, especially in terms of encryption time. The maximum time required for encryption is less than $0.2$ seconds, in contrast with $8$ seconds for the baseline (Figure~\ref{fig:baseline_encrypt}). Recall that alert zones do not influence encryption, so the hierarchical encoding lines present in Figure~\ref{fig:uniform_hierarchical_encrypt} overlap. Encryption is also independent of alert zone distribution, so we do not show encryption in Figure~\ref{exp:hier-normal}.

In terms of token generation and query time, the gain in performance is higher for the Gaussian distribution, since there is more potential for token aggregation. The reason is that minimization of binary expressions of cell identifiers is more effective when zones are clustered, which is likely to be the case in practice.

As expected, execution time is higher for finer-grained grids. However, as opposed to the baseline, in the case of hierarchical encoding the coverage has a significant effect on token generation and query performance, as more alert zone cells translate into a larger number of tokens. Still, the variation with coverage is sublinear, due to the good effectiveness of the aggregation strategy employed (note how when coverage doubles from $2\%$ to $4\%$ for uniform data and largest grid size, the query time increases only by $25\%$). Although the absolute execution times are still high, hierarchical encoding significantly outperforms the baseline. Later in Section~\ref{sec:exp:opt} we show how optimizations can be used to further cut down the performance overhead. For the rest of the experimental evaluation, we will omit the baseline results.  

Next, we evaluate the effect of using Gray encoding on performance. Recall from Section~\ref{sec:gray} that using Gray codes provides better potential for aggregation, thus reducing the number of required tokens and/or increasing the proportion of '*' symbols in a token. For uniform data (graph omitted due to space considerations), both encodings perform similarly, without a clear winner, due to the fact that the aggregation potential is equal in the two cases. On the other hand, for Gaussian data (Figure~\ref{fig:normal_graycode_results}) where alert zones are clustered, Gray encoding favors aggregation of cells. For clarity, to keep the number of lines in the graph low, we present the ratio between the execution time of Gray divided by that of hierarchical encoding. Lower values of the ratio correspond to higher gains for the Gray encoding.
In practice, as alert zones are likely to be clustered, Gray can bring significant performance benefits, of up to $60\%$.

\begin{figure}[t]
\centering
\subfloat[Token generation]{
	\centering
	\includegraphics[width=0.35\textwidth]{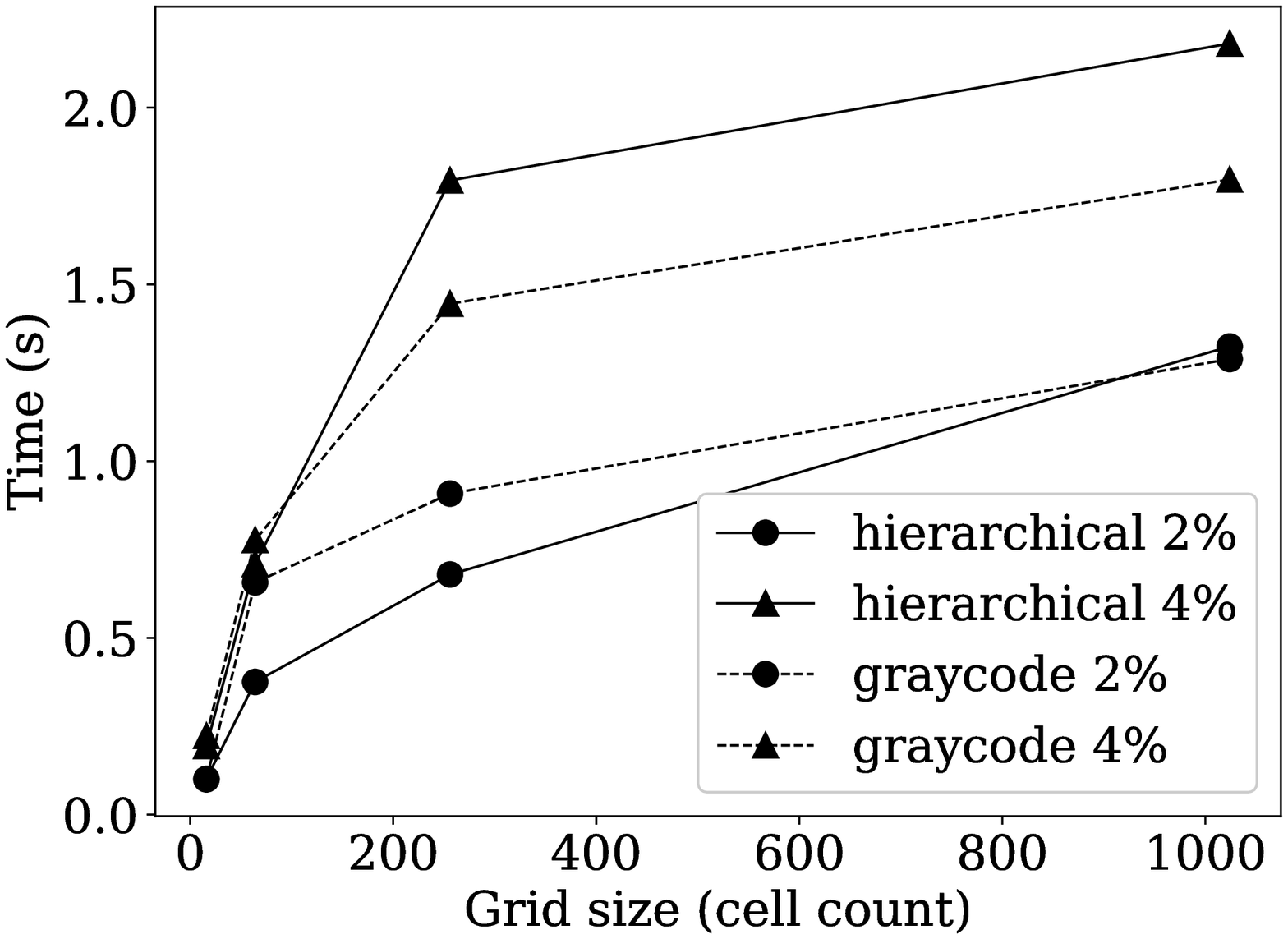}
        \label{fig:uniform_preproc_token}
}
\subfloat[Query]{
	\centering
	\includegraphics[width=0.35\textwidth]{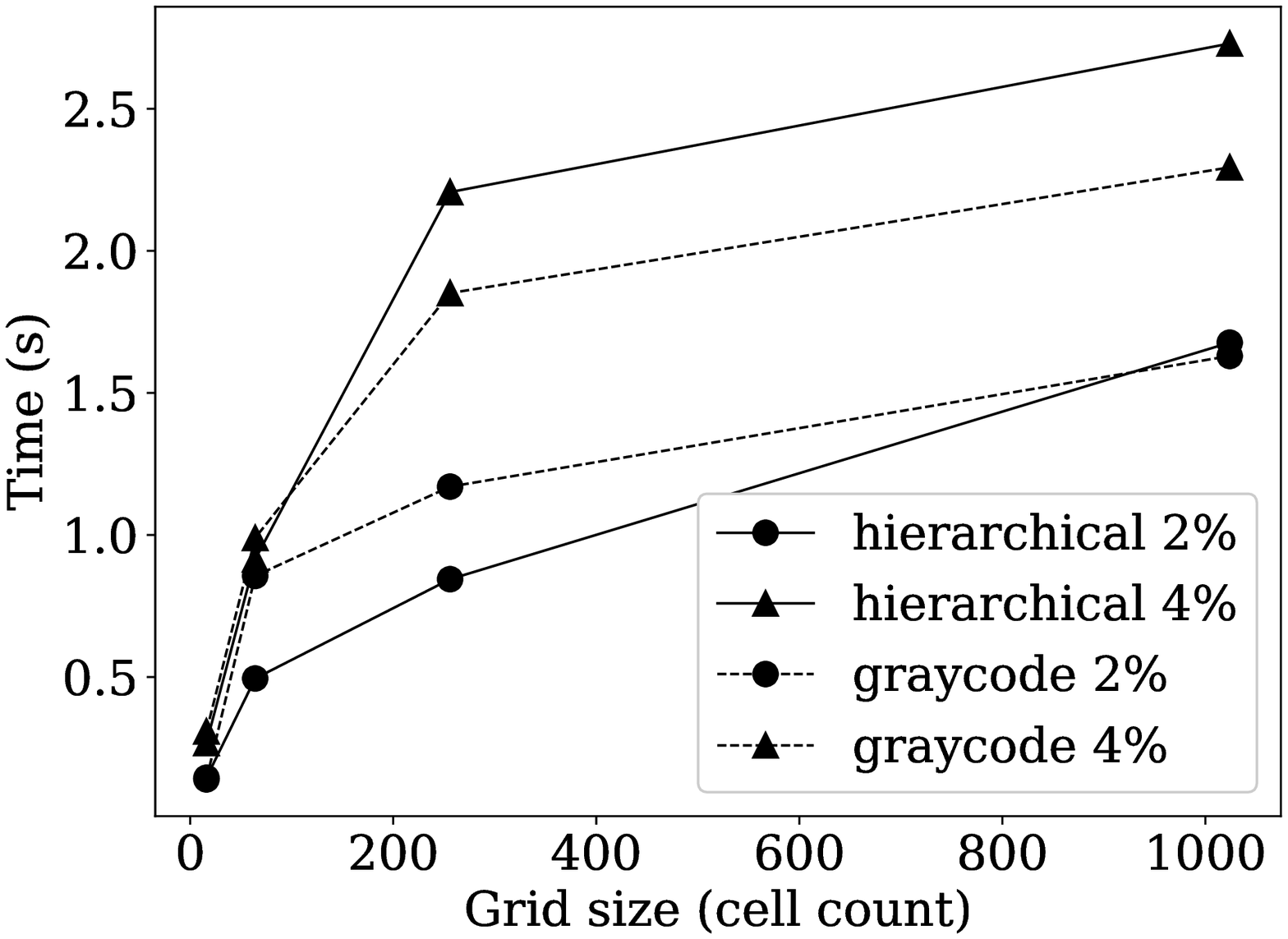}
        \label{fig:uniform_preproc_query}
}
\figspb
\caption{Effect of preprocessing on uniform data}
\figspa
\label{fig:uniform_preproc_results}
\end{figure}

\subsection{Optimization Effect}
\label{sec:exp:opt}

We evaluate the performance of the proposed techniques when incorporating the performance optimizations discussed in Section~\ref{sec:optimizations}. 
First, we show the effect of incorporating preprocessing to pre-compute and re-use some of the results to expensive mathematical operations, such as exponentiations with large numbers.
Figure~\ref{fig:uniform_preproc_results} present the absolute token generation and query times for both proposed encoding techniques on uniform data and two values of the alert zone coverage, namely $2\%$ and $4\%$ (Gaussian data results show similar trends, so we omit them for brevity). 
Token generation computation requirements are improved by roughly $25-30\%$. 
As the coverage increases, more tokens are required to represent a zone, so the generation time increases. We believe that such times are reasonable in practice, especially since creation of alert zones is not a frequent event in the system operation. 
In terms of querying, the execution times are approximately cut in half compared to the non-optimized case (Figures~\ref{exp:hier-unif} and~\ref{exp:hier-normal}). 

\begin{figure*}[bth]
\vspace{-20pt}
\centering
\subfloat[Token generation]{
	\centering
	\includegraphics[width=0.33\textwidth]{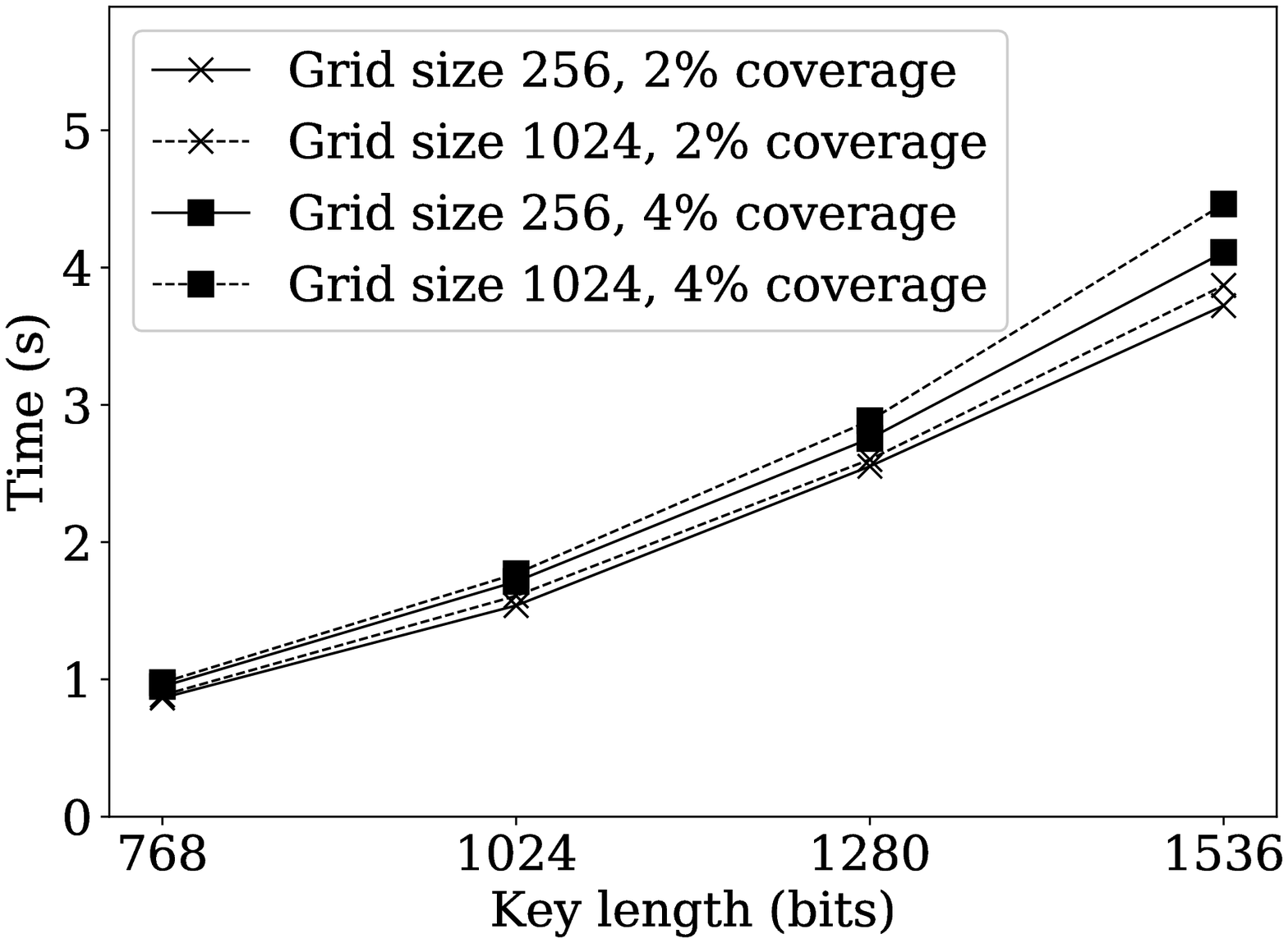}
        \label{fig:key_hierarchical_token}
}
\subfloat[Message encryption]{
	\centering
	\includegraphics[width=0.33\textwidth]{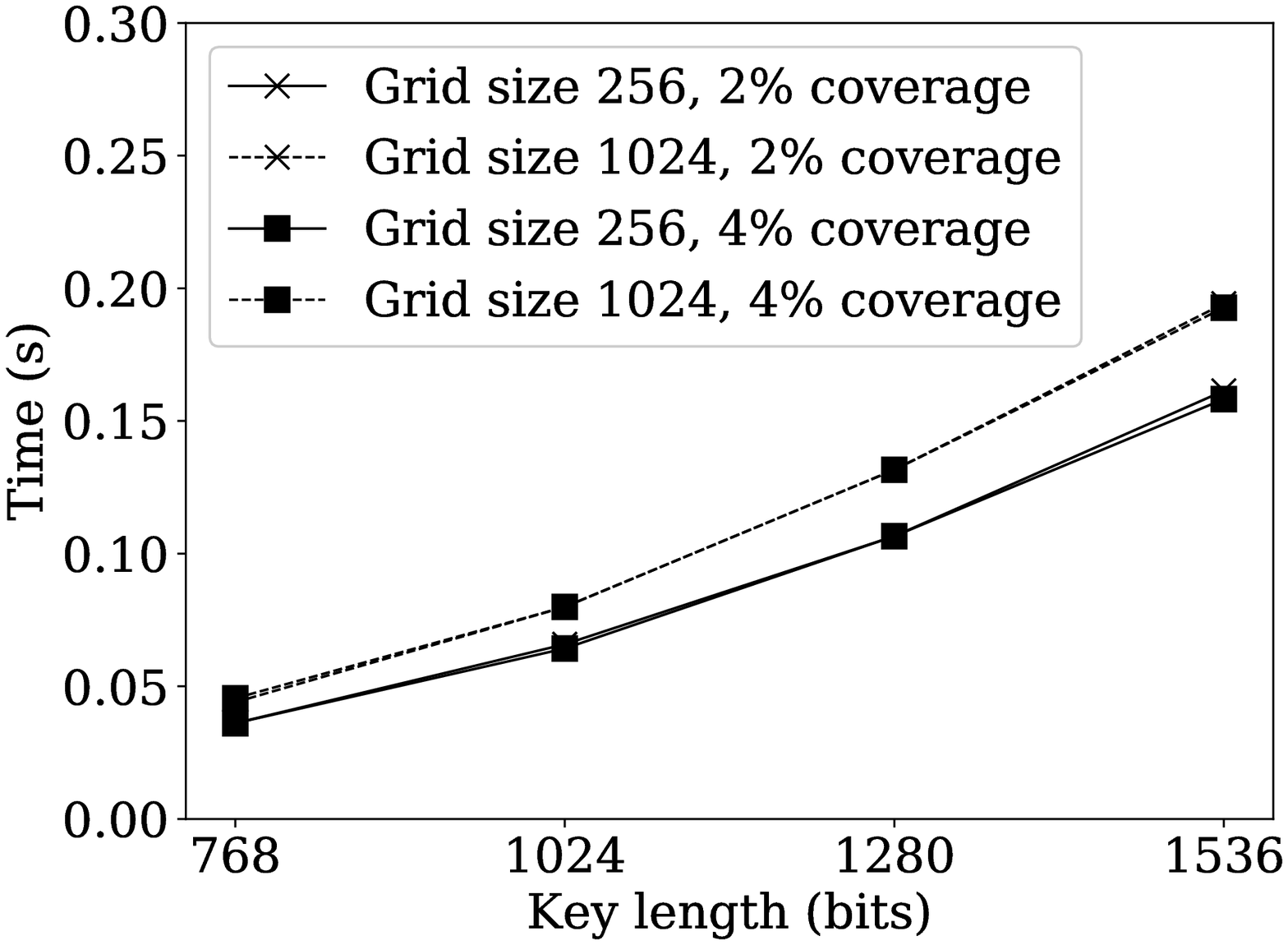}
        \label{fig:key_hierarchical_encrypt}
}
\subfloat[Query]{
	\centering
	\includegraphics[width=0.33\textwidth]{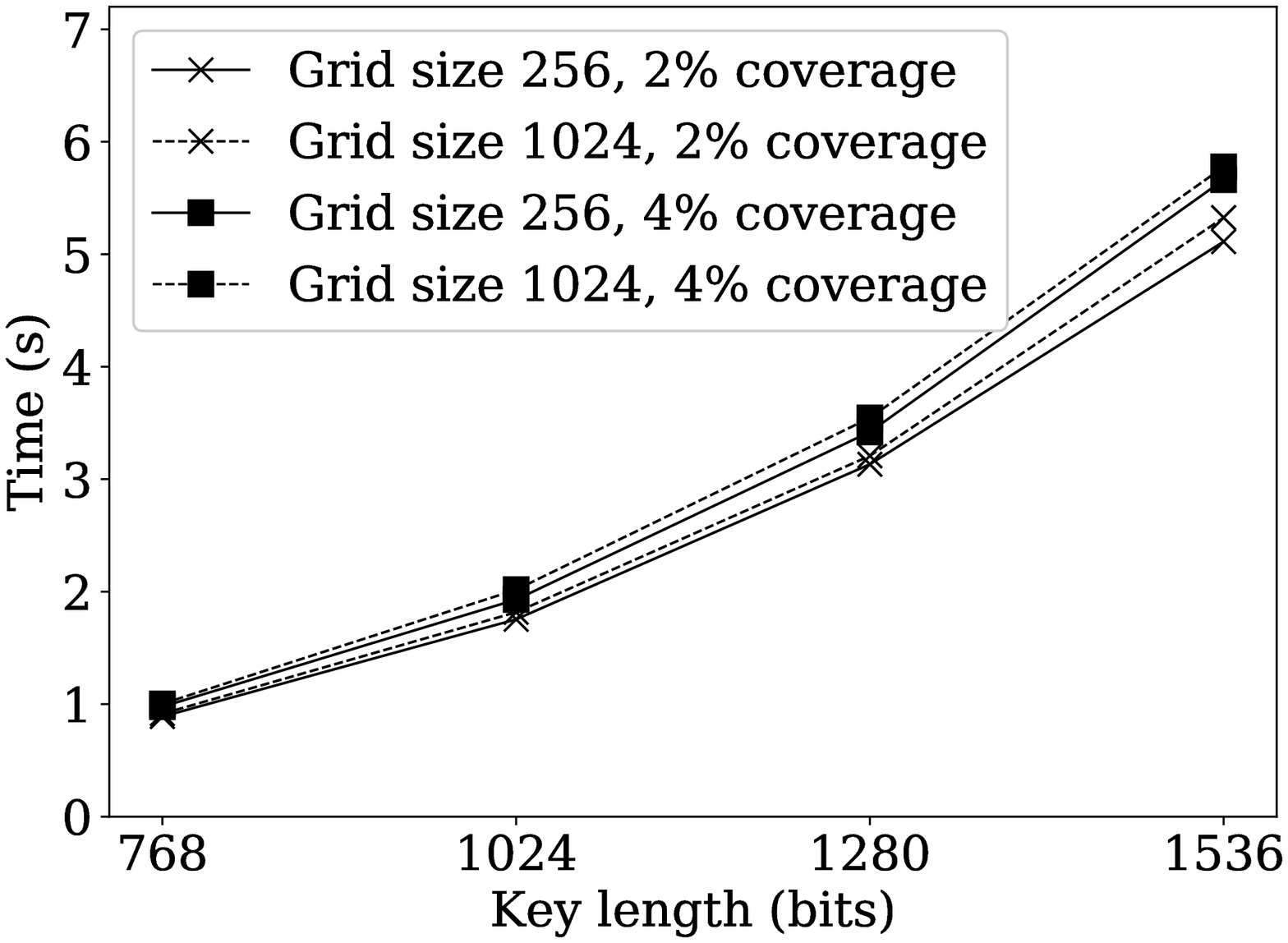}
        \label{fig:key_hierarchical_query}
}
\figspb
\caption{Hierarchical encoding results for Variable Key Length, Gaussian data}
\vspace{-20pt}
\label{exp:hier-key}
\end{figure*}

Figure~\ref{exp:hier-key} presents the behavior of hierarchical encoding with preprocessing when varying encryption key length. We show results for two different grid granularities and coverage values, with Gaussian zone distribution. As expected, the performance decreases when key length increases. However, the 1024-bit setting, which according to industry standards is sufficient for securing individuals' information, does not incur a steep increase in performance overhead. Gray encoding results exhibit similar behavior.

\begin{figure}[t]
\centering
\subfloat[Fixed area percentage]{
	\centering
	\includegraphics[width=0.35\textwidth]{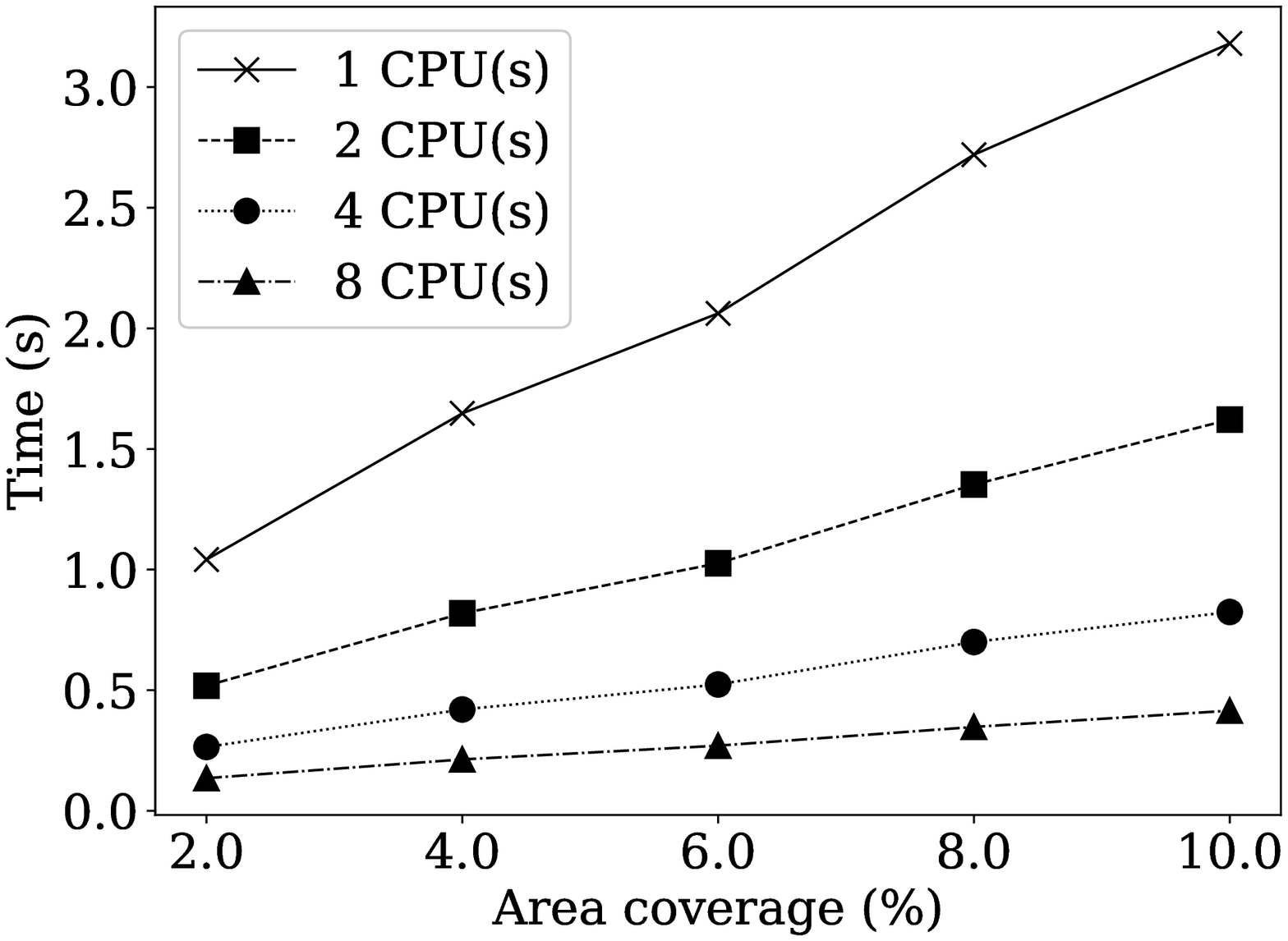}
        \label{fig:uniform_parralel_fixedpercent}
}
\subfloat[Fixed grid size]{
	\centering
	\includegraphics[width=0.35\textwidth]{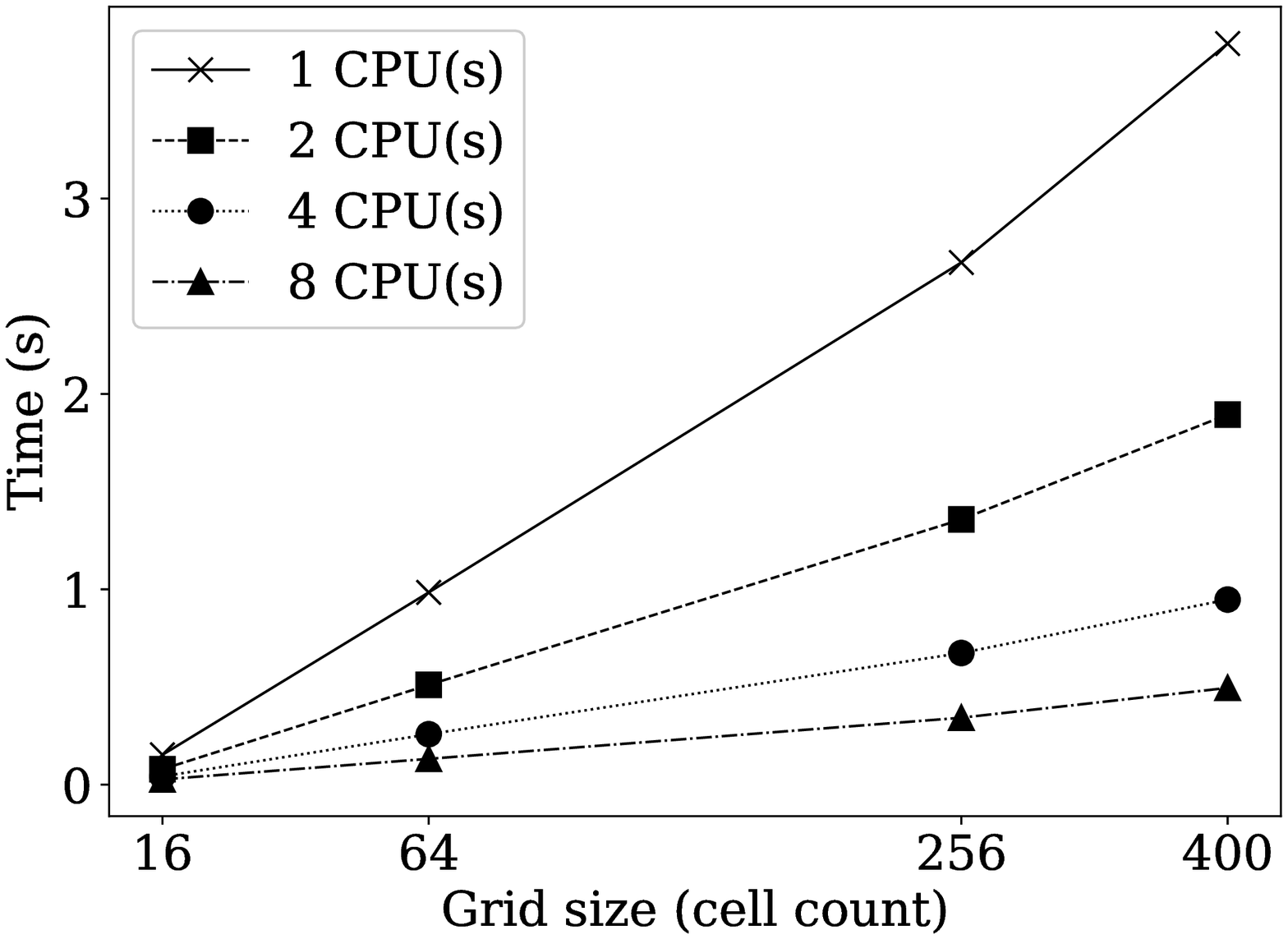}
        \label{fig:uniform_parralel_fixedsize}
}
\figspb
\caption{Parralel results on uniform data}
\vspace{-20pt}
\label{fig:uniform_parralel_results}
\end{figure}

\begin{figure}[t]
\centering
\subfloat[Fixed area percentage]{
	\centering
	\includegraphics[width=0.35\textwidth]{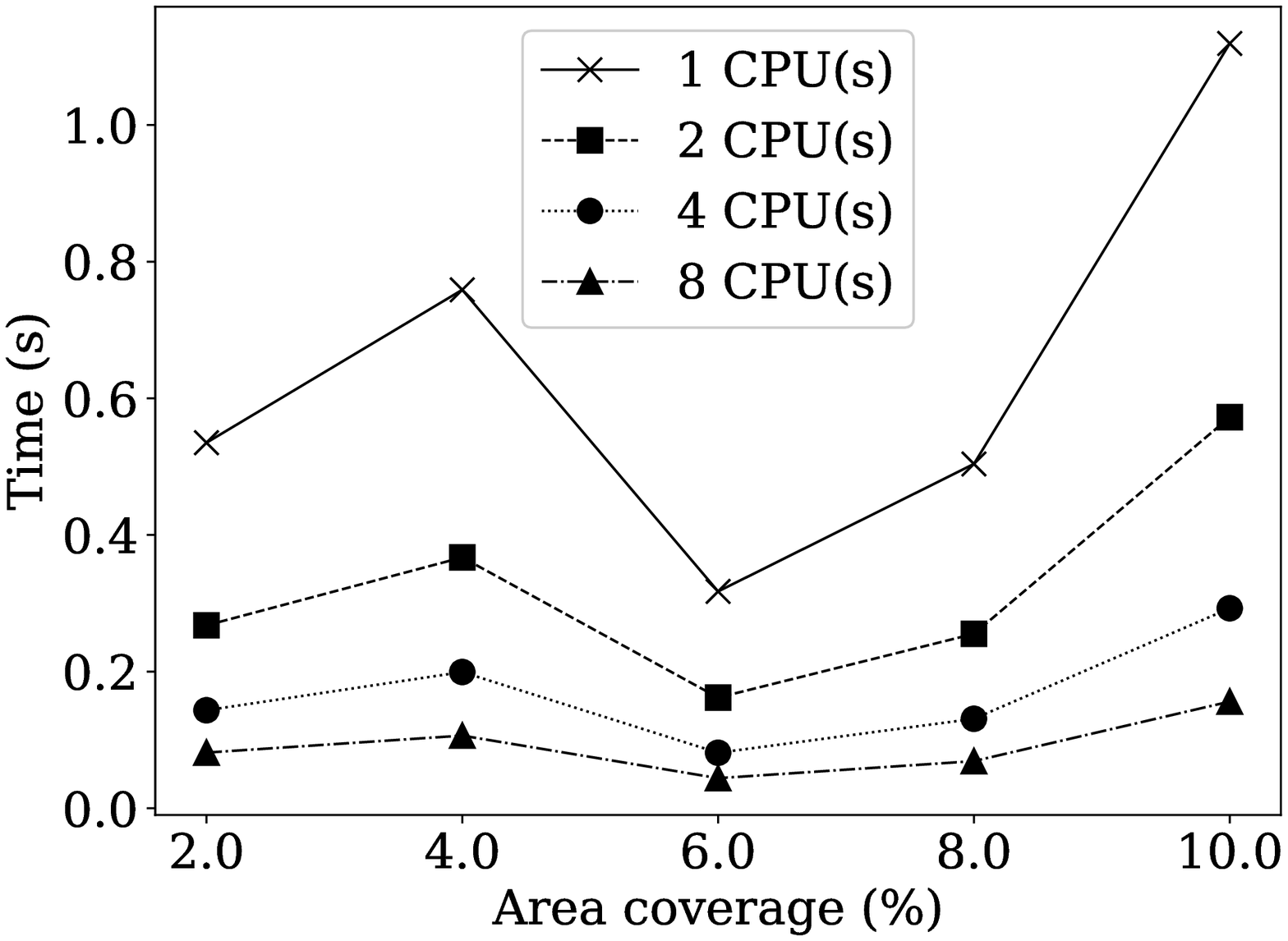}
        \label{fig:normal_parralel_fixedpercent}
}
\subfloat[Fixed grid size]{
	\centering
	\includegraphics[width=0.35\textwidth]{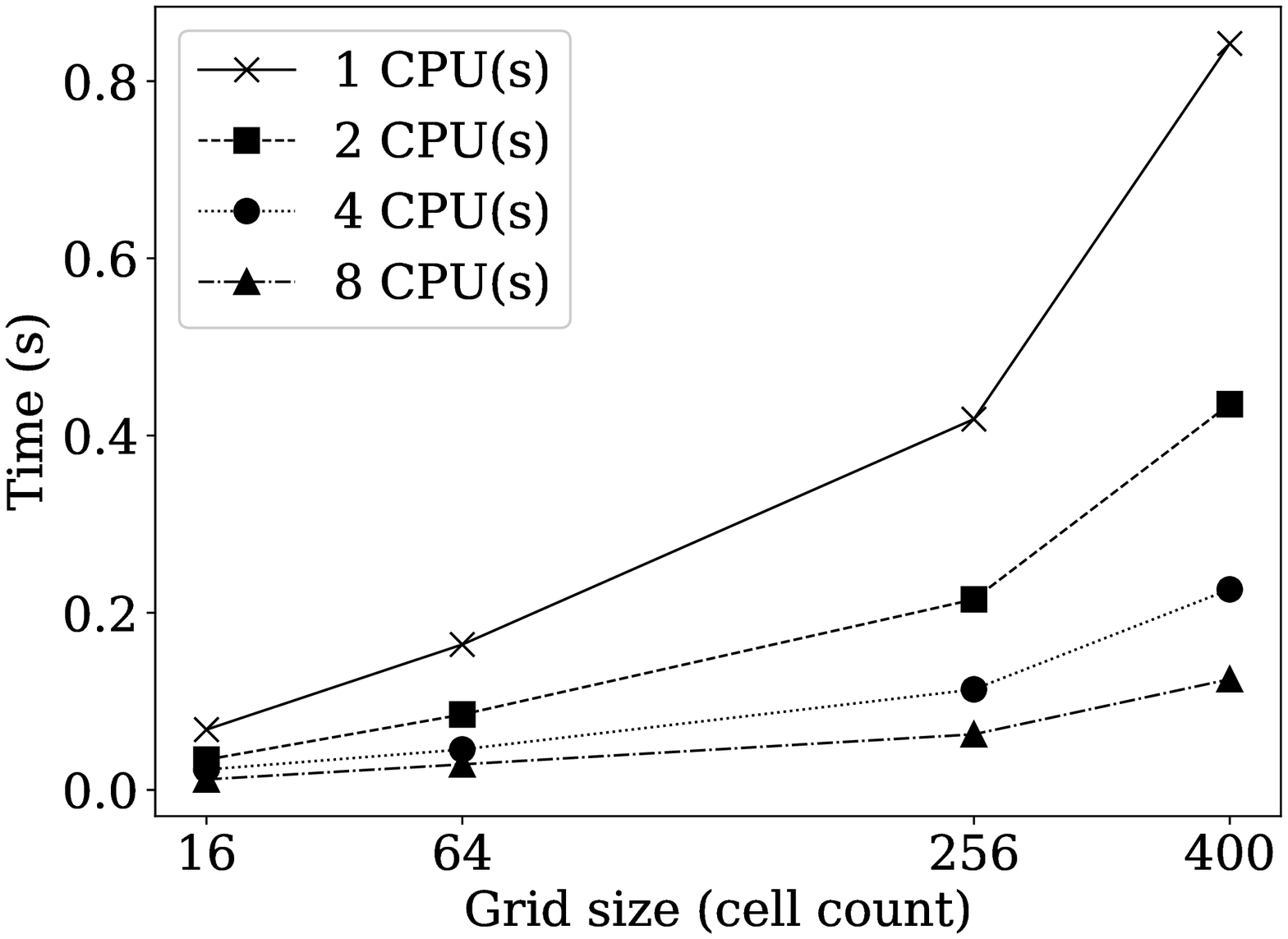}
        \label{fig:normal_parralel_fixedsize}
}
\figspb
\caption{Parralel results on Gaussian data}
\vspace{-20pt}
\label{fig:normal_parralel_results}
\end{figure}

Figures~\ref{fig:uniform_parralel_results} and~\ref{fig:normal_parralel_results} present the results when employing the parallel processing optimization. We used $2$, $4$ and $8$ CPUs for computation. We considered both variable grid size for a fixed coverage of $6\%$, as well as variable coverage of alert zones for a fixed grid size of $256$. 
The results show that a close-to-linear speedup can be obtained. For $8$ CPUs for instance, the speedup is $7.2$. This is a very encouraging outcome, and the query time is this way reduced to less than one second in the worst case. For median-scale settings of the grid size and coverage, we obtain absolute execution times of under $0.1$ seconds per query. We emphasize that, although we only had available $8$ CPUs for testing, the problem studied is embarrassingly parallel in nature, so the availability of a larger number of CPUs is likely to lead to close-to-linear speedup values as well.

\vspace{-10pt}

\subsection{Alert Zone Expansion Evaluation}
\label{sec:exp:expansion}
In this section, we evaluate the performance gain obtained by the alert zone expansion heuristic introduced in Section~\ref{sec:queryexpansion}. 
We use the same settings as in the previous experiments, except that we allow a finer-grained partitioning of the space, to better evaluate the impact of the expansion heuristic. Specifically, we consider grids of granularity $d \times d$, where $d \in \{ 64, 128, 256 \}$, resulting to a total number of grid cells of $4096$, $16384$ and $65536$, respectively. We keep the same alert zone size ranging from $1\%$ to $10\%$ of data space size, but we consider three distinct shapes: square, rectangular (with a skew ratio of $2.5$), and circular. The latter case is used to capture scenarios where there is an event epicenter, and individuals are notified if they are situated within a certain Euclidean distance of it. The resulting circular zone is mapped to the grid. This is representative for cases when mobile users are alerted to stay away from a dangerous location (e.g., a toxic gas spill).
The alert zone expansion ratio $\alpha$ is varied within  set $\{ 0.02, 0.04, 0.06, 0.08, \mathbf{0.10}\}$ (recall that a larger value results in a more significant privacy leakage, but is also likely to yield a higher performance gain).

\begin{figure}[t]
	\centering
	\subfloat[Square Zone]{
		\centering
		\includegraphics[width=0.32\textwidth]{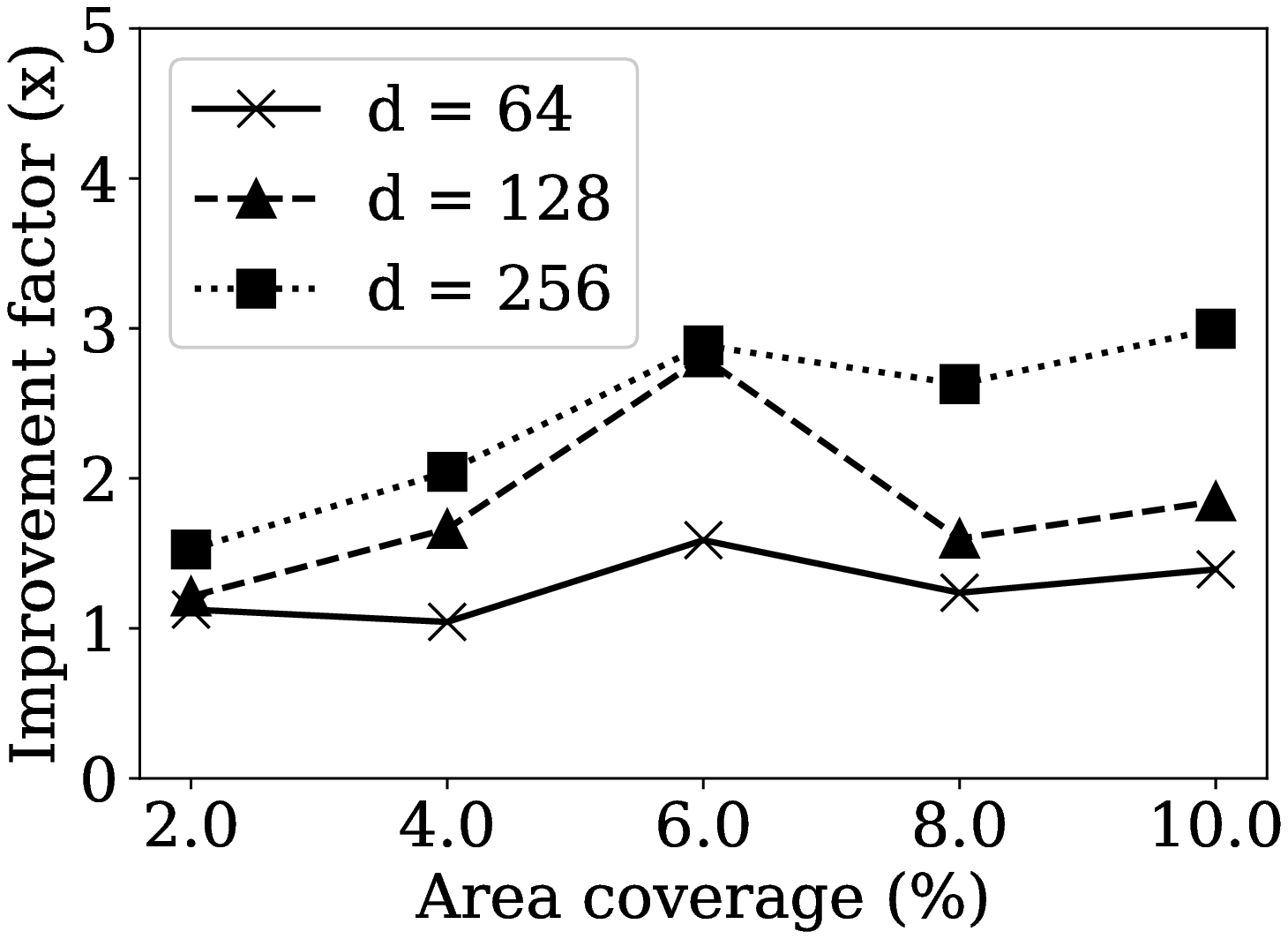}
		\label{fig:expansion_time_paring:query_ratio:square}
	}
	\subfloat[Rectangular Zone]{
		\centering
		\includegraphics[width=0.32\textwidth]{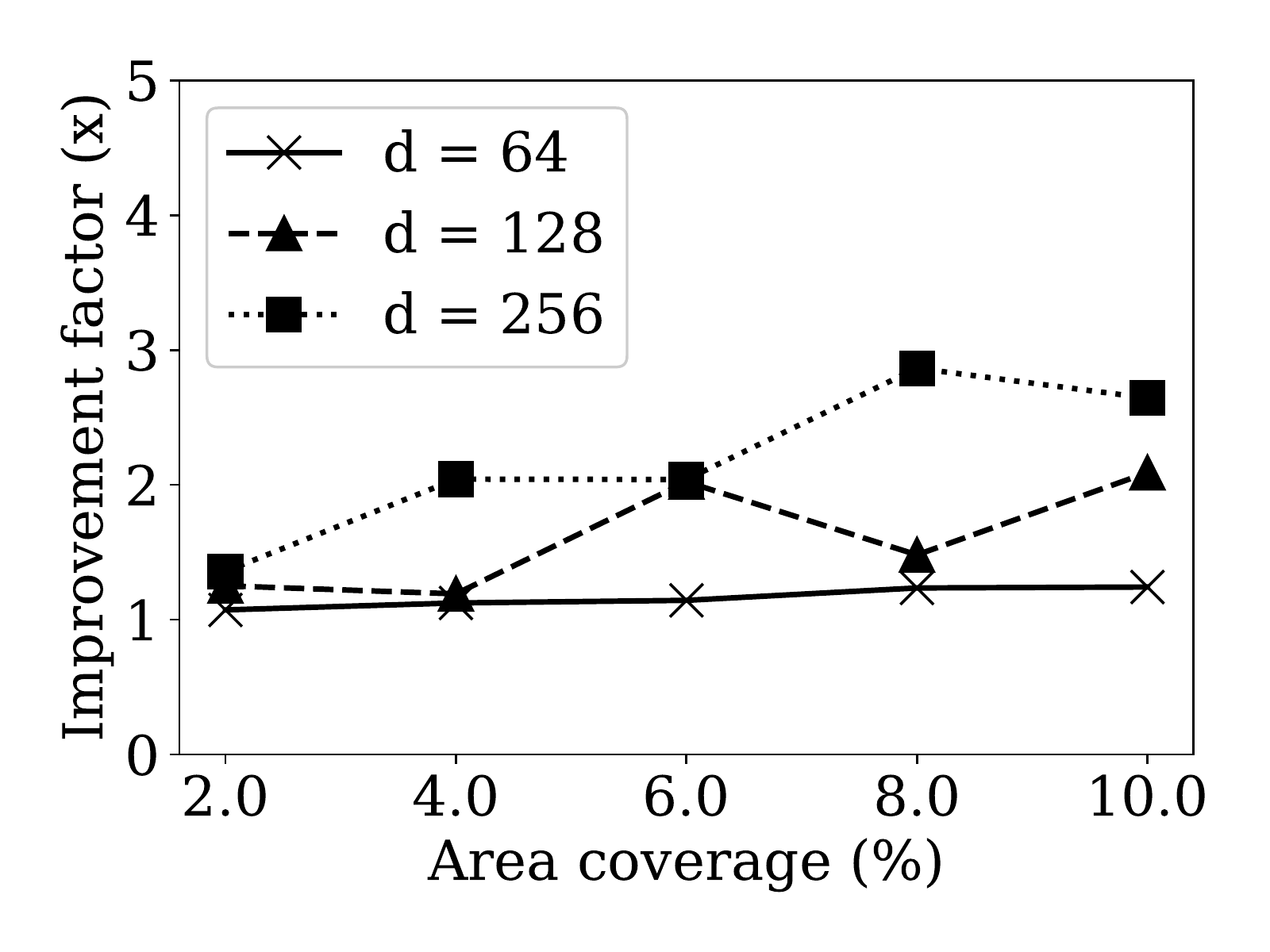}
		\label{fig:expansion_time_paring:query_ratio:rect}
	}
	\subfloat[Circular Zone]{
		\centering
		\includegraphics[width=0.32\textwidth]{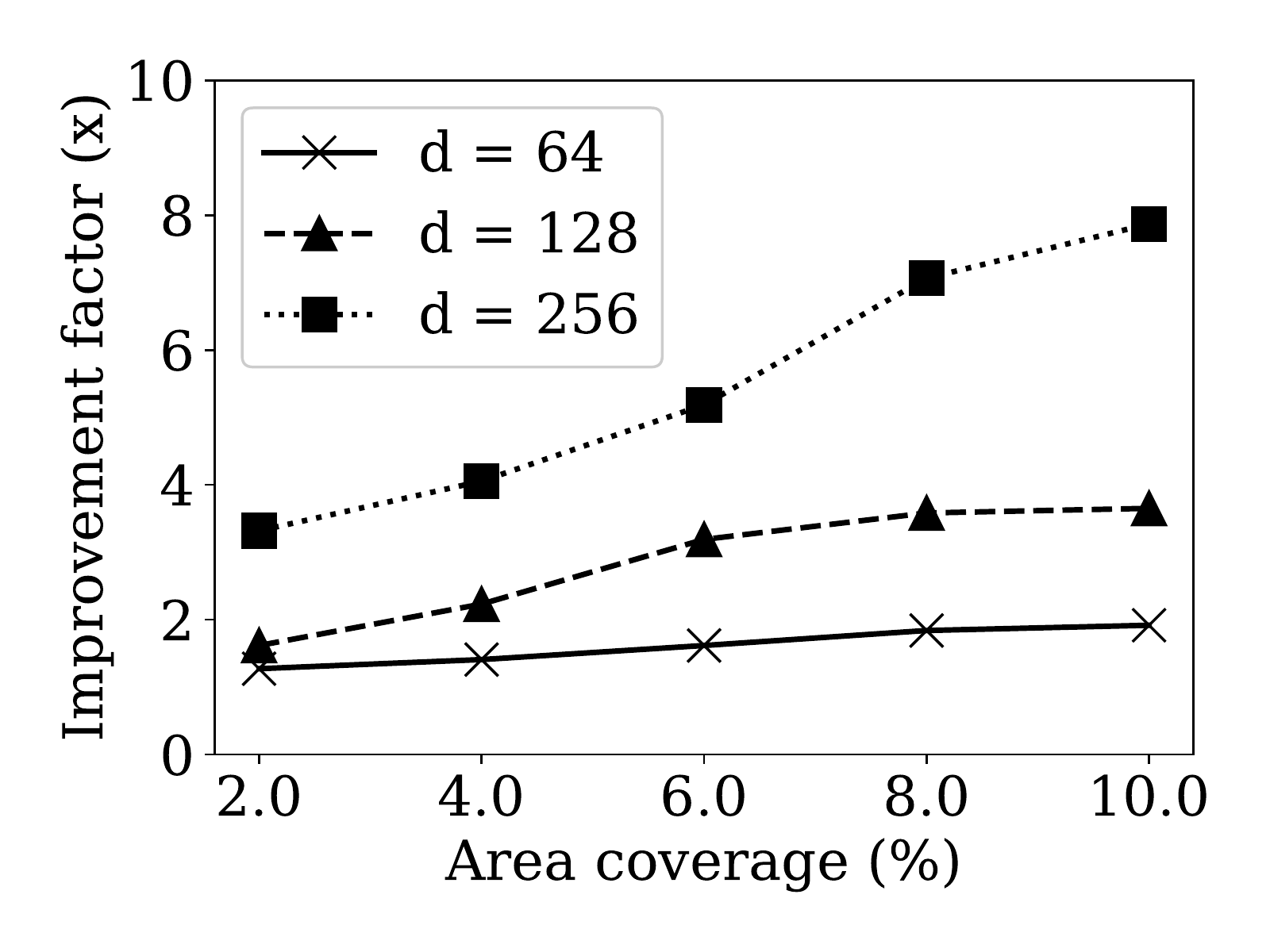}
		\label{fig:expansion_time_paring:query_ratio:circle}
	}
		\figspb
	
	\caption{Zone expansion improvement vs alert zone coverage}
	\vspace{-20pt}
	\label{fig:varzone}
\end{figure}

\begin{figure}[t]
	\subfloat[Square Zone]{
		\centering
		\includegraphics[width=0.32\textwidth]{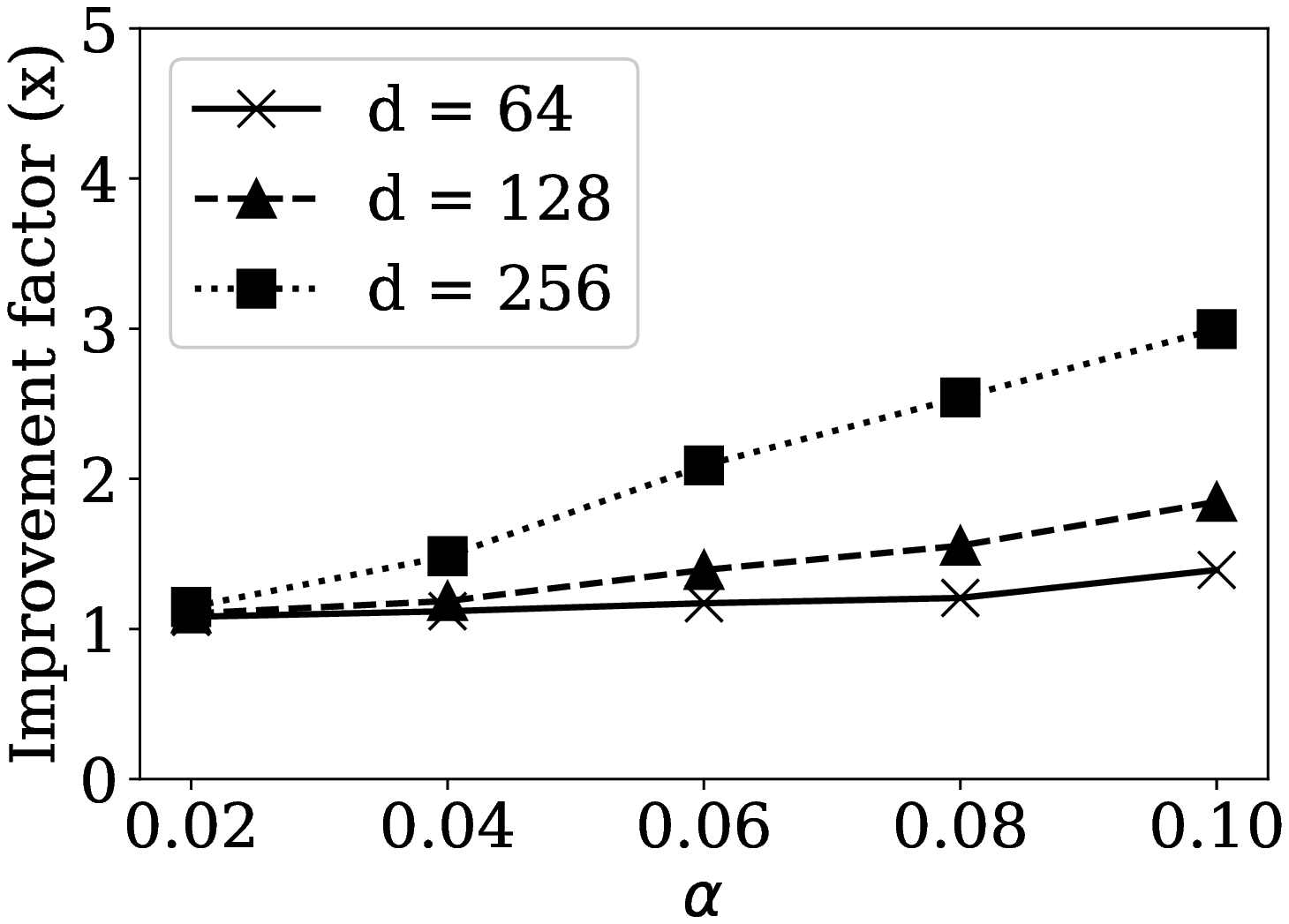}
		\label{fig:expansion_time_paring:expansionratio:square}
	}
	\subfloat[Rectangular Zone]{
		\centering
		\includegraphics[width=0.32\textwidth]{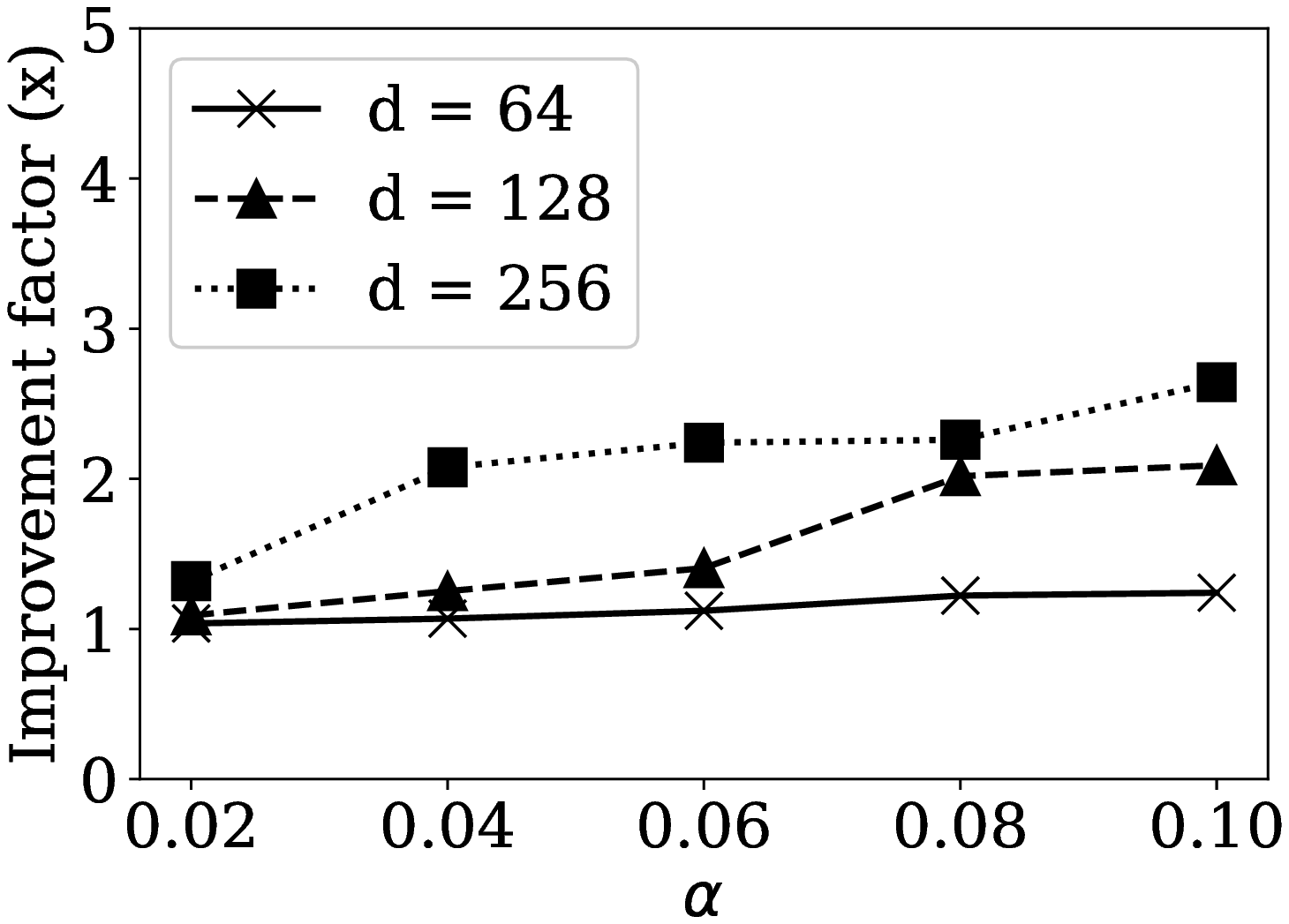}
		\label{fig:expansion_time_paring:expansionratio:rect}
	}
	\subfloat[Circular Zone]{
		\centering
		\includegraphics[width=0.32\textwidth]{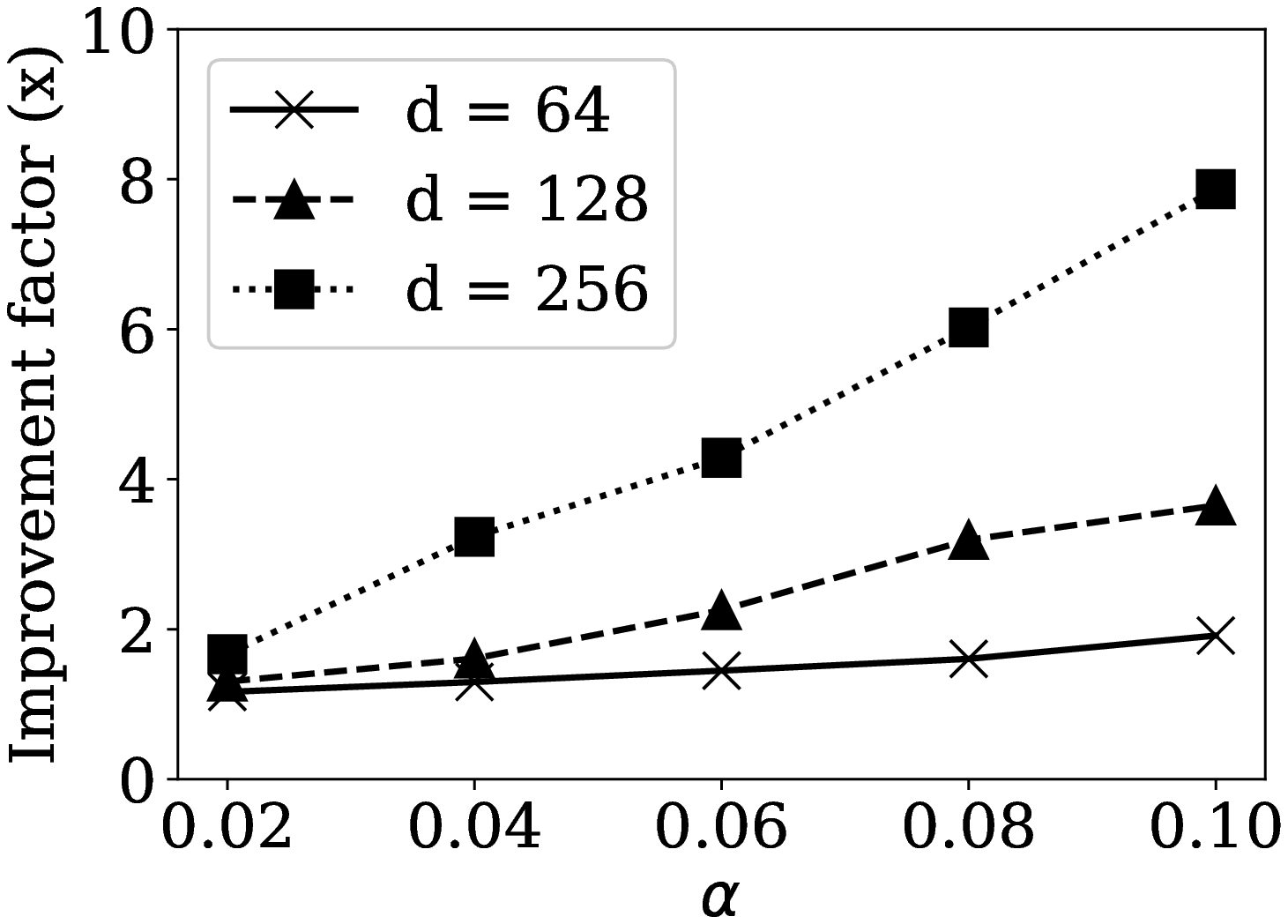}
		\label{fig:expansion_time_paring:expansionratio:circle}
	}
		\figspb
	
	\caption{Zone expansion improvement vs expansion factor $\alpha$}
	\vspace{-20pt}
	\label{fig:varalpha}
\end{figure}

Fig.~\ref{fig:varzone} shows the performance gain of expansion when varying the size of the initial alert zone. For ease of presentation, the gain is expressed as improvement factor ($\times$), i.e., the ratio between the matching time when there is no enlargement over the matching time when enlargement is used (a higher value represents a better performance gain). Each line in the graph corresponds to a different grid granularity $d$. First, we note that enlargement always results in improvement (the value is always greater than $1$). Second, the improvement factor shows a general increasing trend with alert zone size (except for some random outcomes). This is explained by the fact that the enlargement factor $\alpha$ is expressed as a percent of initial alert zone size. When the initial alert zone is larger, the heuristic can select patches from more candidate cells. Finally, the improvement factor is larger for finer granularity cases (i.e., larger $d$). This is also due to the fact that the heuristic has more candidate patches to choose from. A finer granularity also allows the search boundary to advance slightly more. When cells are larger, including an extra cell may cause the $\alpha$ threshold to be exceeded, so the heuristic will not consider that cell for enlargement. We also note that the shape of the zone impacts significantly the gain. Specifically, a circular zone is better for expansion, since the heuristic does not favor any particular expansion direction. When the initial zone is circular (or to be precise, a circle aligned to the grid), the heuristic can bring into the zone cells from all directions, and therefore the amount of possible choices is increased. The zone is also likely to grow uniformly in all directions, leading to more compact tokens due to the binary representations of cells. Conversely, the rectangular case, which leads to the most skewed zones in terms of shape, performs the worst.

Fig.~\ref{fig:varalpha} shows the improvement factor when varying the enlargement factor $\alpha$. As expected, there is a clear increasing trend in execution time improvement. Since more cells are available as patch candidates, the heuristic is able to either completely eliminate some tokens, or significantly increase the number of wildcards in the remaining tokens through binary minimization. %
As in the previous experiment, we note that an increase in granularity $d$ results in a higher improvement factor. Also, when the initial alert zone is circular, the highest improvement is obtained, with values of up to $9$ times. The gain is less pronounced for rectangular alert zones, but the heuristic is still providing significant gains, with an improvement factor of up to $3$ times.

\begin{figure}[t]
	\centering
	\subfloat[Square Zone]{
		\centering
		\includegraphics[width=0.32\textwidth]{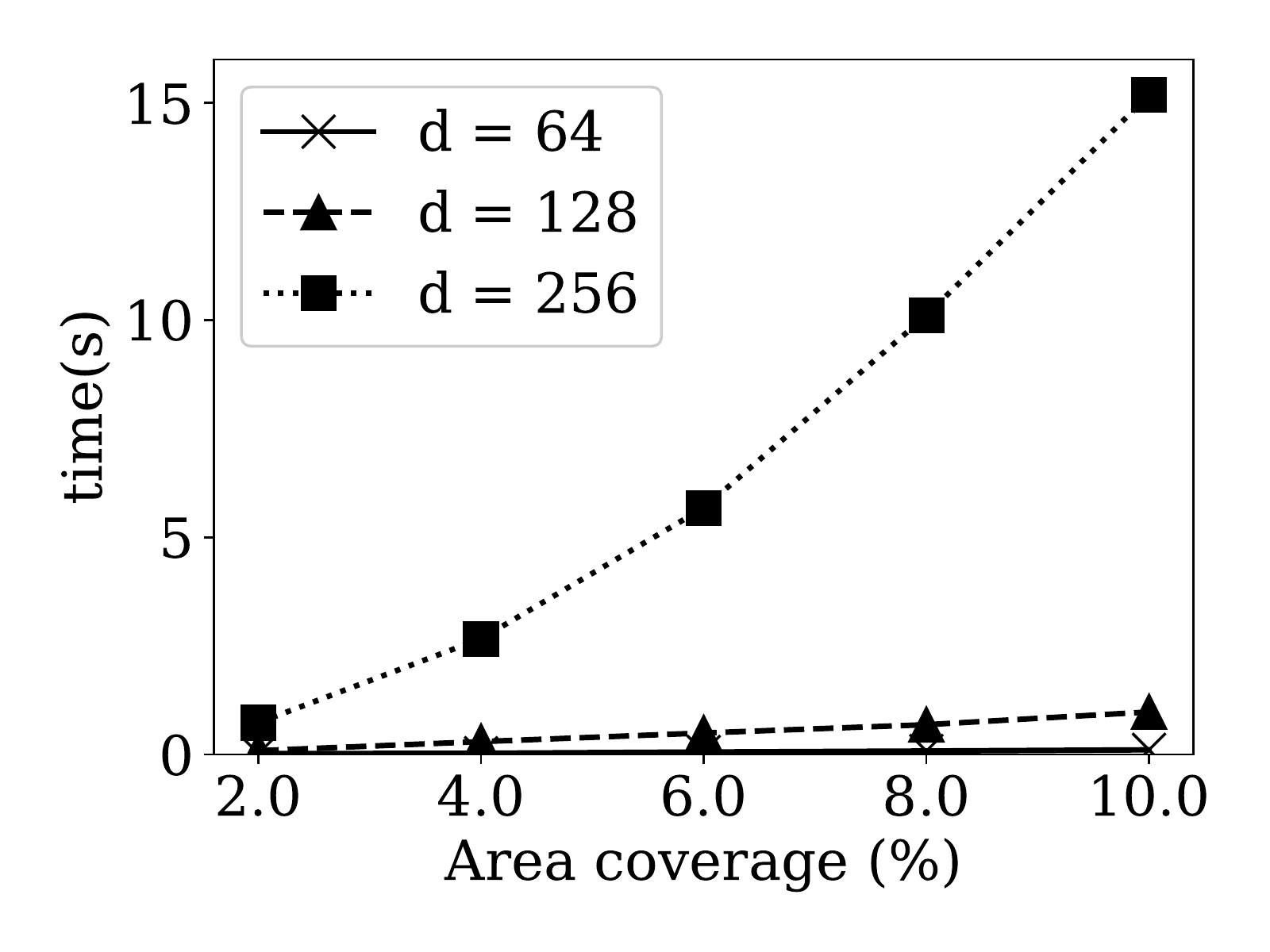}
		\label{fig:expansion_expanding_time:query_ratio:square}
	}
	\subfloat[Rectangular Zone]{
		\centering
		\includegraphics[width=0.32\textwidth]{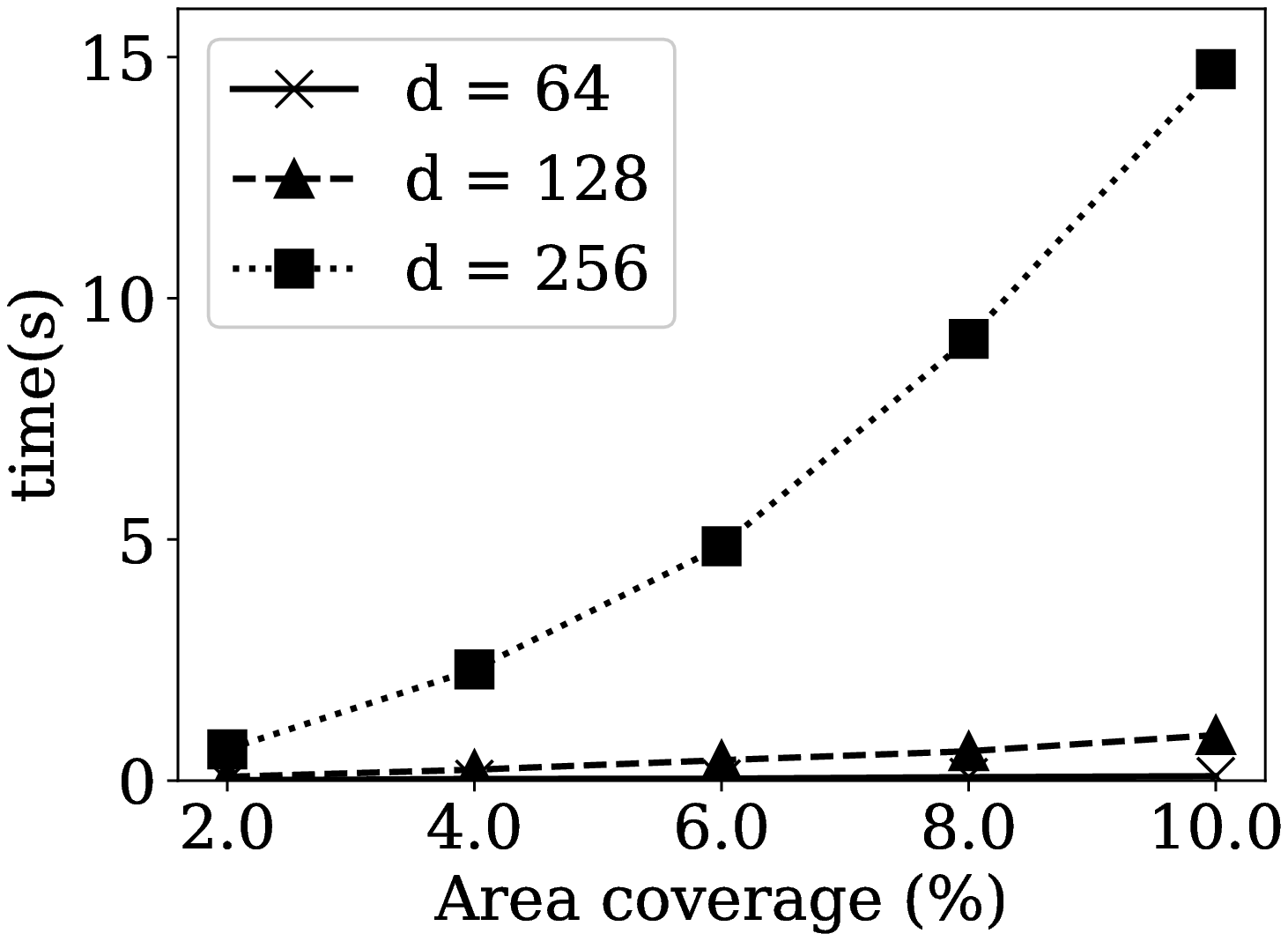}
		\label{fig:expansion_expanding_time:query_ratio:rect}
	}
	\subfloat[Circular Zone]{
		\centering
		\includegraphics[width=0.32\textwidth]{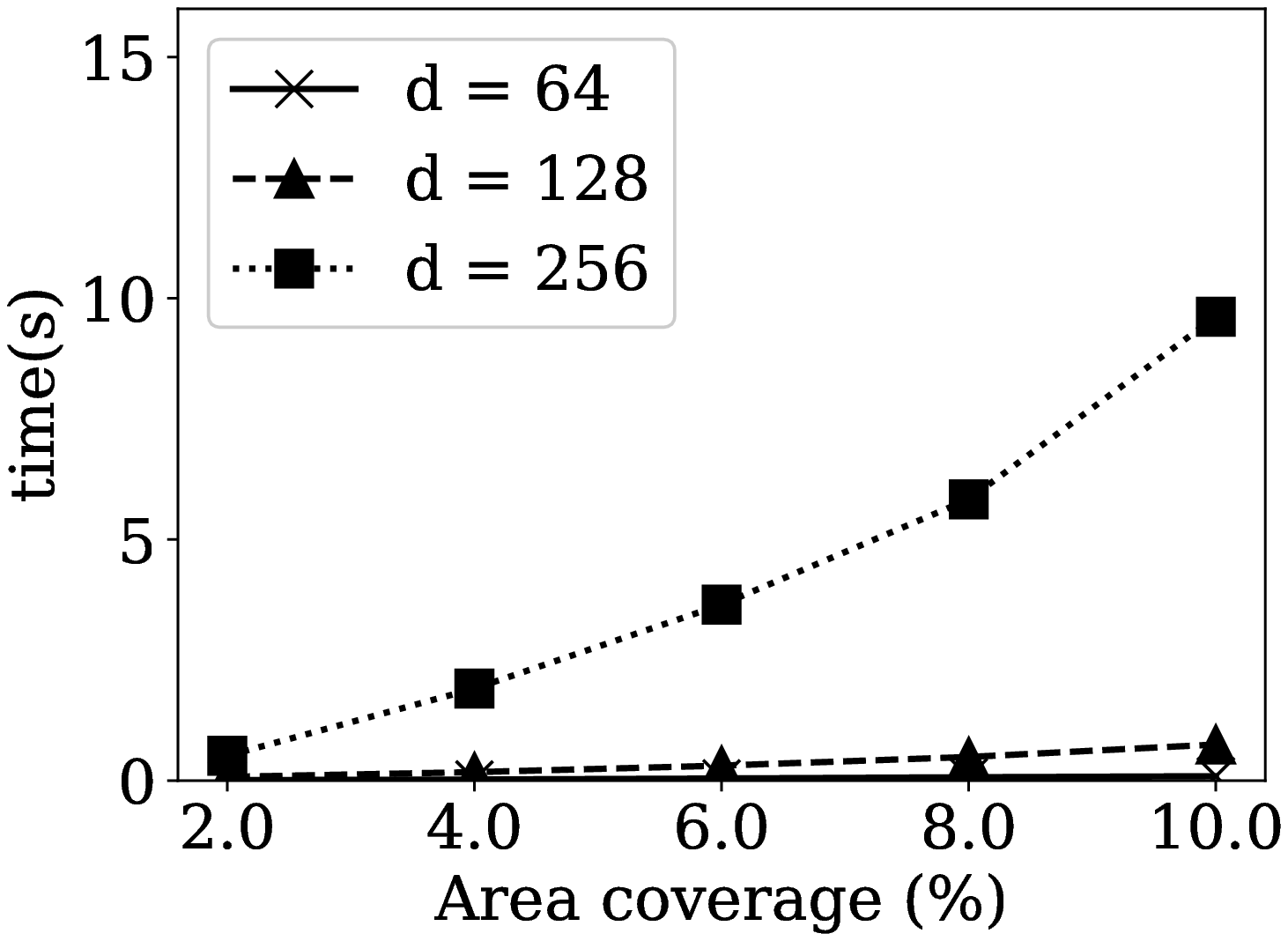}
		\label{fig:expansion_expanding_time:query_ratio:circle}
	}
	\figspb
	\caption{Zone expansion time vs alert zone coverage}
	\vspace{-20pt}
	\label{fig:timevarzone}
\end{figure}

\begin{figure}[t]
	\subfloat[Square Zone]{
		\centering
		\includegraphics[width=0.32\textwidth]{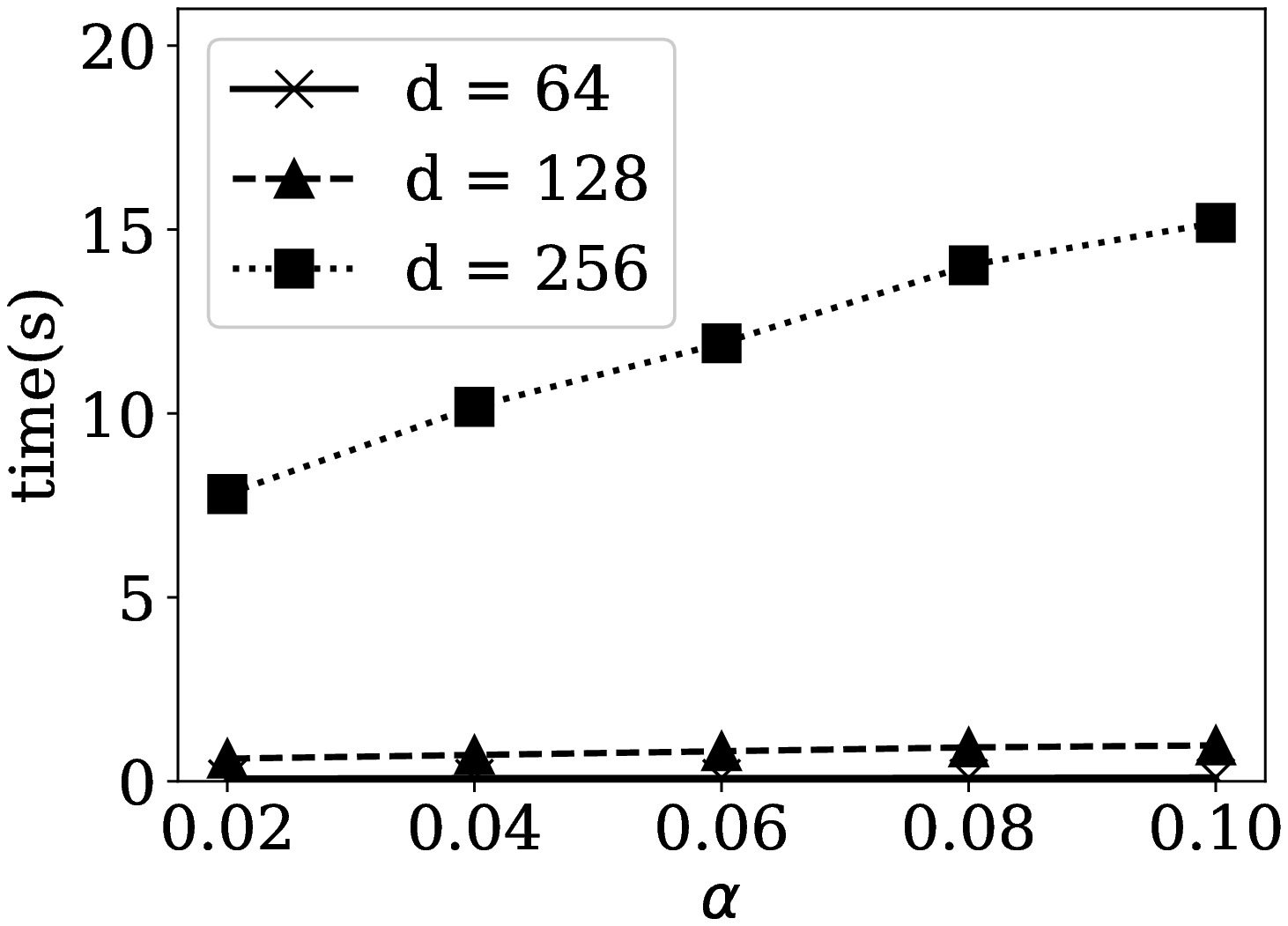}
		\label{fig:expansion_expanding_time:expansionratio:square}
	}
	\subfloat[Rectangular Zone]{
		\centering
		\includegraphics[width=0.32\textwidth]{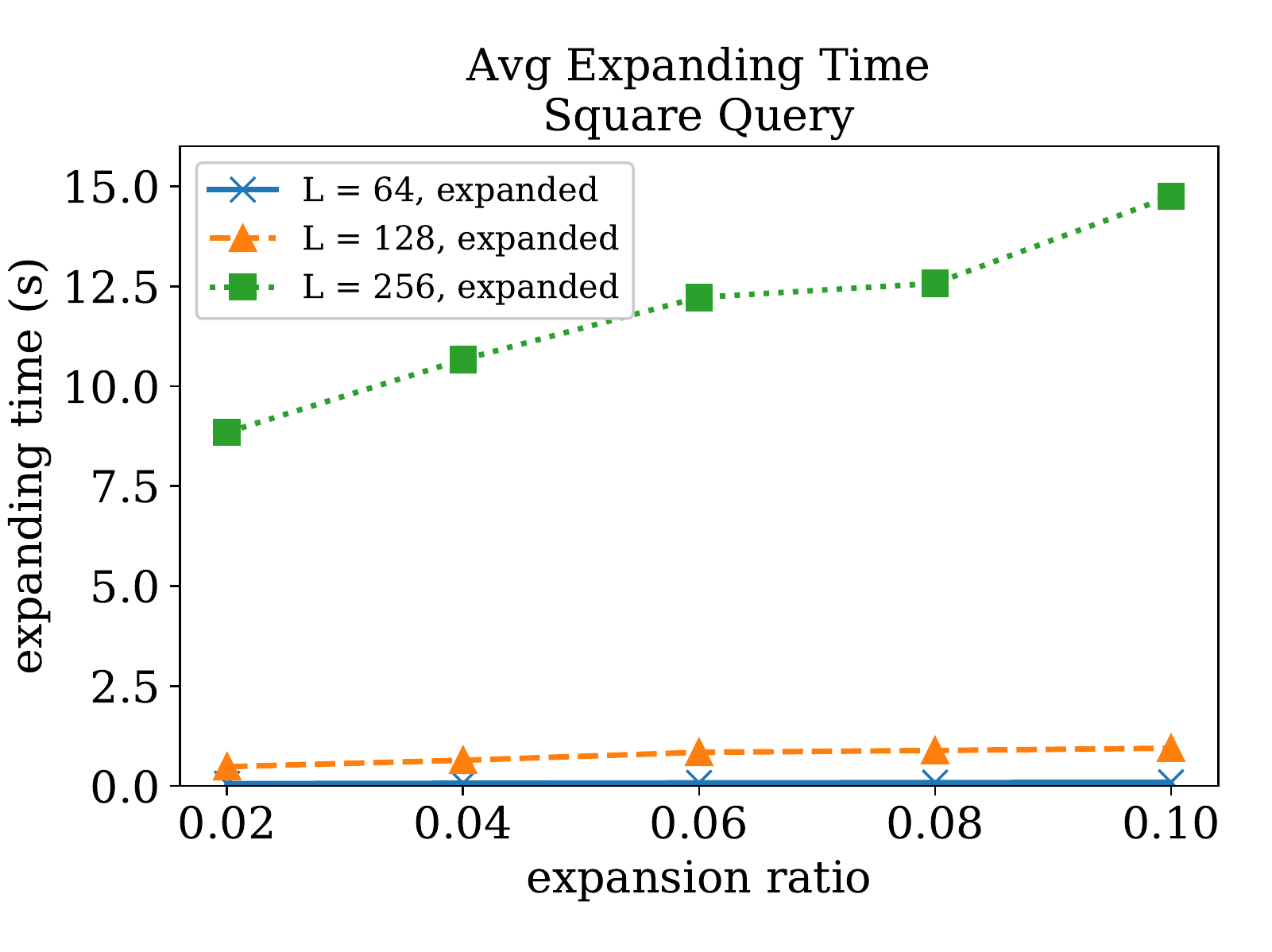}
		\label{fig:expansion_expanding_time:expansionratio:rect}
	}
	\subfloat[Circular Zone]{
		\centering
		\includegraphics[width=0.32\textwidth]{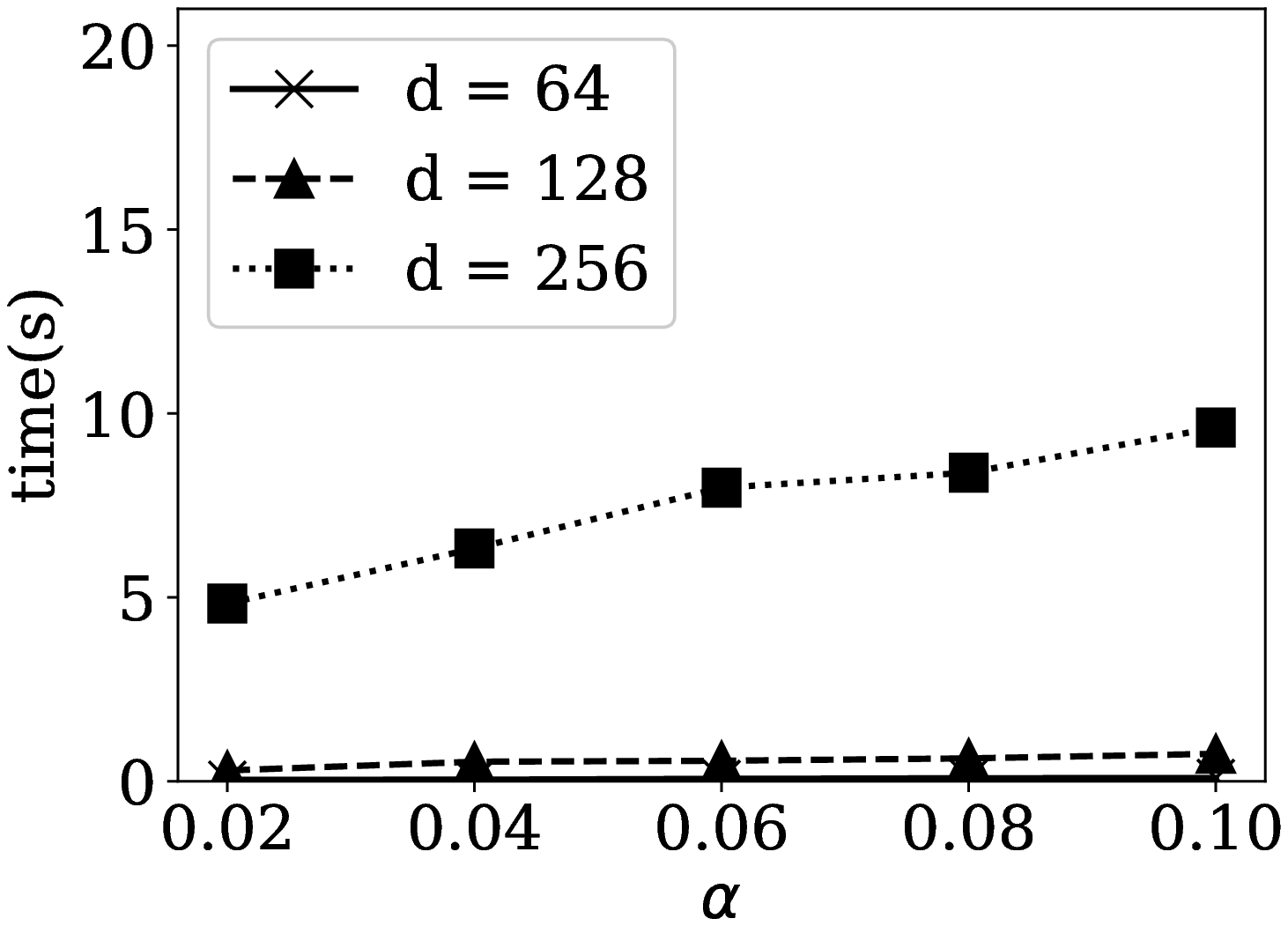}
		\label{fig:expansion_expanding_time:expansionratio:circle}
	}
	\figspb
	\caption{Zone expansion time vs expansion factor $\alpha$}
	\vspace{-20pt}
	\label{fig:timevaralpha}
\end{figure}

In our final experiment, we measure the execution time of the zone enlargement heuristic. Fig.~\ref{fig:timevarzone} and Fig.~\ref{fig:timevaralpha} show the time required to compute the enlarged zone when varying initial alert zone size and enlargement factor $\alpha$, respectively. An interesting trade-off emerges: as the granularity of the grid increases (i.e., finer grained grids), the improvement in token matching time increases (as seen in previous experiments), but at the same time the computation time for the enlarged zone grows. Furthermore, we emphasize that the token matching computational overhead can be parallelized, whereas zone enlargement computation is sequential in nature. The main reason why the zone enlargement computation is high for finer granularities is the quadratic increase in patch candidates, coupled with the relatively slow binary expression minimization step. 
Among different zone shapes, the circular shape takes the longest, due to the fact that it considers the most patch candidates within the given enlargement threshold $\alpha$.   

Nevertheless, we note that for coarser and moderate granularities ($d=64$ and $d=128$), the enlargement process is fast (less than half a second). Coupled with the significant improvement factors (ranging from $2$ to $4$ for granularities coarser than $d=256$, as can be observed from Figs.~\ref{fig:varzone} and \ref{fig:varalpha}), the heuristic can lead to very good overall execution time improvements. Furthermore, the enlargement cost is done once per zone, and remains the same regardless of the number of mobile users (i.e., ciphertexts to match against). Hence, as the user population grows, the performance gain of the heuristic (which is always a factor of the original zone evaluation time) will lead to linear gains in the number of users, whereas the enlargement computation overhead stays constant. We conclude that, overall, the zone enlargement heuristic is effective in reducing the matching overhead, even for small values of enlargement (i.e., only a small amount of privacy needs to be traded off for significant performance gains).

\figspb
\section{Related work}\label{sec:related}

\vspace{-10pt}

{\bf Location Privacy.}
A significant amount of research focused on the problem of private location-based 
queries, where users send their coordinates to obtain nearby points of interest.  Early work attempted to protect locations of real users by generating fake locations.
For instance, in \cite{KYS05} the querying user sends to the server $k-1$ fake locations
to reduce the likelihood of identifying the actual user position.
However, fake locations can be detected using filtering techniques,
which leaves the real users vulnerable. 

A new direction of research started by \cite{gruteser03} and continued by \cite{GL05,kgmp07,mca06}
relies on the concept of {\em Cloaking Regions (CRs)}. CR-based solutions  implement the
spatial $k$-anonymity (SKA) \cite{kgmp07} paradigm. For each query, 
a trusted anonymizer service generates
CRs that contain at least $k$ {\em real} user locations.
If the resulting CRs are {\em reciprocal} \cite{kgmp07}, 
SKA guarantees privacy for snapshots of user locations.
However, supporting continuous queries \cite{CM07} requires
generating large-sized CRs. In \cite{sec_priv,PROBE}, the objective is
to prevent the association between users and sensitive locations. Users define
privacy profiles \cite{PROBE} that specify their sensitivity with respect to certain {\em feature types}
(e.g., hospitals, bars, etc.), and every CRs must cover a diverse set of sensitive and non-sensitive features.
In \cite{KS07}, the set of POI is first encoded according to a secret transformation
by a trusted entity. A Hilbert-curve mapping (with secret parameters) transforms 2-D points to 1-D.
Users (who know the transformation key) map their queries to 1D, and the processing
is performed in the 1-D space. However, the mapping can decrease result accuracy,
and the transformation may be vulnerable to reverse-engineering.

The problem with CR-based methods is that the underlying $k$-anonymity paradigm is vulnerable
to background knowledge attacks. This is particularly a problem in the case of moving users,
since trajectory information can be used to derive the identities behind reported locations. More recently,
{\em differential privacy} \cite{dmns06}, a provably secure model for semantic privacy, has been used for
spatial data in \cite{cpssy12}. However, differential privacy is only suitable for aggregate releases
of data, and cannot handle processing of individual updates, as required by an alert system.

Closer to our work, a Private Information Retrieval (PIR) protocol is proposed in \cite{GKKST08}
for nearest-neighbor queries. The protocol is provably secure, and also uses cryptography.
However, it considers a 'pull-based' approach, and assumes that the user already knows the location
s/he wants to retrieve points of interest from. In contrast, our focus is on a 'push-based' notification service, where
the PIR solution cannot be applied since the user is not aware of where the alert zones are.

Since the publication of~\cite{codaspy14}, several works addressed processing on encrypted location data. In~\cite{range}
and~\cite{mixgeo}, two solutions are proposed for search on encrypted location data hosted at a cloud server. 
Both approaches
rely on {\em symmetric searchable encryption} (SSE), where the client has access to the secret key of the transformation. 
The FastGeo system~\cite{fastgeo} builds upon the concepts introduced in~\cite{range} and supports faster search under the same trust assumptions.
However, the SSE setting is not appropriate in our problem setting, where large populations of mobile users subscribe to
location-based alerts. If a single user colludes with the service provider, the security of the entire set of locations is compromised. This is
a strong trust assumption, suitable for cases where there are relatively few clients, who can be throughly vetted. 
Our solution relies on asymmetric encryption, and mobile users only have access to the public key, which is used for encryption.
No user is able to compromise the privacy of other participants.

Furthermore, all the above solutions build an index on encrypted data to speed up search performance. As shown in~\cite{SNN},
the index structure can leak a lot of sensitive details about the data, even when fully encrypted (e.g., data distribution,
or relative distance order among users). A similar approach that builds an R-tree
on location data protected using homomorphic encryption is proposed in~\cite{iot}, with emphasis on IoT data, and on parallelizing computation in big data environments.
The work in~\cite{biomed} is a position paper that looks at how some concepts similar to search on encrypted locations can be used for biomedical data, and also identifies other interesting type of queries that may be of interest, such as skyline queries.

A significant body of research focused on nearest-neighbor (NN) queries on encrypted data~\cite{Hashem10,ESBJ13, Hu11}, culminating with the work in~\cite{SNN} which showed that the most secure and efficient way to answer NN queries on encrypted data is through materialization of results and encryption of the resulting structure. All these works consider the symmetric encryption setting, hence they rely on trust assumptions that are too strong for our proposed location alert system.

{\bf Searchable Encryption.} One of the earliest works that coined the concept of searchable encryption was \cite{SWP00},
which proposed provably secure cryptographic techniques for keyword search. Only exact matches of keywords were supported.
Later in \cite{Boneh06}, the set of search predicates supported was extended to comparison queries. However, the resulting solution
could not be easily extended to conjunctions of conditions, without a considerable increase in ciphertext and token size.
The work in \cite{Boneh07} further extended the set of supported predicates to subset queries, as well as conjunctions of 
equality, comparison and subset queries with small ciphertext and token size. The authors of \cite{Boneh07} also introduced
HVE, which we employ as a building block in our solutions for private location-based alert systems.
HVE protects the privacy of the encrypted messages received from users, but assumes that the token information (e.g., alert zones) is public.
The more recent work in \cite{Blundo09} extends HVE to also protect the tokens. However, the solution is more expensive.

\section{Conclusion}\label{sec:conclusion}
We proposed a system for secure location-based alerts which utilizes searchable encryption. We introduced two alternate data encodings that allow efficient application of cryptographic primitives for search on encrypted data (namely HVE). Furthermore, we devised performance optimizations that reduce the overhead of searchable encryption, which is notoriously expensive. We also devised a heuristic that enlarges alert zones by a small factor in order to reduce matching time, thus achieving a tunable performance-privacy trade-off. 
The experimental evaluation results show that searchable encryption can be made practical with careful system design and optimizations.

In future work, we plan to investigate more advanced data and query encoding techniques (beyond regular grids) that will allow us to securely alert users with even lower overhead. We also plan to study other types of matching semantics beyond range queries (e.g., nearest-neighbors, top-$k$).

\noindent {\bf Acknowledgments.}
This research has been funded in part by NSF grants IIS-1910950 and  IIS-1909806, the USC Integrated Media Systems Center (IMSC), and unrestricted cash gifts from Microsoft and Google. Any opinions, findings, and conclusions or recommendations expressed in this material are those of the author(s) and do not necessarily reflect the views of the sponsors.

\bibliographystyle{abbrv}
\bibliography{PBC}

\noindent
\\

\vspace{-20pt}
\appendix
\section{Primer on HVE Encryption}
\label{sec:app}

HVE is built on top of a symmetrical bilinear map of composite order \cite{Boneh05}, which is a function $e : \mathbb{G} \times \mathbb{G} \rightarrow \mathbb{G}_T$ such that $\forall a,b \in G$ and $ \forall u,v \in \mathbb{Z}$ it holds that $e(a^u,b^v)=e(a,b)^{uv}$. $\mathbb{G}$ and $\mathbb{G}_T$ are cyclic multiplicative groups of composite order $N=P\cdot Q$ where $P$ and $Q$ are large primes of equal bit length. We denote by $\mathbb{G}_p$, $\mathbb{G}_q$ the subgroups of $\mathbb{G}$ of orders $P$ and $Q$, respectively. Let $l$ denote the HVE {\em width}, which is the bit length of the attribute, and consequently that of the search predicate. HVE consists of the following phases:

{\bf Setup.} The $TA$ generates the public/secret ($PK$/$SK$) key pair and shares $PK$ with the users. $SK$ has the form:
$$SK = ( g_q \in \mathbb{G}_q,\quad a \in \mathbb{Z}_p,\quad \forall i \in [1..l]: u_i,h_i, w_i, g, v \in \mathbb{G}_p )$$
To generate $PK$, the $TA$ first chooses at random elements \(R_{u,i}, R_{h,i}\), \(R_{w,i} 
\in \mathbb{G}_q, \forall i \in [1..l]\)  and \(R_v \in \mathbb{G}_q\). Next, $PK$ is determined as:
$$PK = (g_q,\quad V=vR_v,\quad A=e(g,v)^a,\quad$$
$$\forall i \in [1..l]: U_i=u_iR_{u,i},\quad  H_i=h_iR_{h,i},\quad  W_i=w_iR_{w,i})$$

{\bf Encryption} uses $PK$ and takes as parameters index attribute $I$ and message $M \in \mathbb{G}_T$. The following random elements are generated: \(Z, Z_{i,1}, Z_{i,2} \in \mathbb{G}_q\) and \(s \in \mathbb{Z}_n\). Then, the ciphertext is: 
$$C = (C^{'}= MA^s,\quad C_0=V^sZ, \quad $$
$$\forall i \in [1..l]: C_{i,1} = (U^{I_i}_iH_i)^sZ_{i,1}, \quad C_{i,2} = W^{s}_iZ_{i,2} )$$

{\bf Token Generation.} Using $SK$, and given a search predicate encoded as pattern vector \(I_{*}\), the TA generates 
a search token $TK$ as follows: let \(J\) be the set of all indices $i$ where \(I_{*}[i] \neq *\).
TA randomly generates \(r_{i,1}\) and \(r_{i,2} \in \mathbb{Z}_p, \forall i \in J\). 
Then
$$TK=(I_*, K_0 = g^a\prod_{i \in J}(u^{I_{*}[i]}_ih_i)^{r_{i,1}}w^{r_{i,2}}_i, \quad$$
$$ \forall i \in [1..l]: K_{i,1} = v^{r_i,1},\quad K_{i,2} = v^{r_i,2})$$

{\bf Query} is executed at the server, and evaluates if the predicate represented by $TK$ holds for ciphertext $C$. The server attempts to determine the value of \(M\) as 
\begin{equation}
M = C^{'}{/} (e(C_0,K_0) {/} \prod_{i \in J} e(C_{i,1},K_{i,1}) e(C_{i,2},K_{i,2}) \label{eq:query}
\end{equation}
If the index $I$ based on which $C$ was computed satisfies $TK$, then the actual value of \(M\) is returned, otherwise a special number which is not in the valid message domain (denoted by $\bot$) is obtained.

\end{document}